# Interfacial Polarization and Pyroelectricity in Antiferrodistortive Structures Induced by a Flexoelectric Effect and Rotostriction

Anna N. Morozovska [1,2*], Eugene A. Eliseev[1], Maya D. Glinchuk[1], Long-Qing Chen[3], and Venkatraman Gopalan [3†]

[1] Institute for Problems of Materials Science, National Academy of Science of Ukraine,
3, Krjijanovskogo, 03142 Kiev, Ukraine

[2] Institute of Semiconductor Physics, National Academy of Science of Ukraine,
41, pr. Nauki, 03028 Kiev, Ukraine

[3] Department of Materials Science and Engineering, Pennsylvania State University,
University Park, Pennsylvania 16802, USA

Abstract

Theoretical analysis based on the Landau-Ginzburg-Devonshire (LGD) theory is used to show that the combined effect of flexoelectricity and rotostriction can lead to a spontaneous polarization and pyroelectricity in the vicinity of antiphase boundaries, structural twin walls, surfaces and interfaces in the octahedrally tilted phase of otherwise non-ferroelectric perovskites such as $CaTiO_3$, $SrTiO_3$, and $EuTiO_3$. As an example, we numerically demonstrate a spontaneous polarization and pyroelectric response at the $SrTiO_3$ antiphase and twin boundaries at temperatures lower than the antiferrodistortive structural phase transition temperature of $T_S \sim 105$ K in agreement with previously unexplained experimental results.

At temperatures lower than effective Curie temperature $T_C^*$ (~25 K for twins and ~50 K for antiphase boundaries) biquadratic coupling between oxygen octahedron tilt and polarization vectors essentially enhances the polarization induced by the combined flexoelectric and rotostriction effects near the hard domain wall. Biquadratic coupling cannot induce polarization inside easy twins and antiphase boundaries, their polarization and pyroelectricity originates below $T_S$ from the built-in flexoelectric field. The spontaneous polarization reaches the values ~0.1-5μC/cm$^2$ at the $SrTiO_3$ antiphase boundaries and twins without free charges and ~1-5μC/cm$^2$ allowing for free charges. A principal difference between the influence of biquadratic and flexoelectric couplings on the interfacial polarization is the following: the biquadratic coupling induces bistable ferroelectric polarization inside hard antiphase boundaries and hard twins below $T_C^*$, while the flexoelectric coupling induces improper spontaneous polarization via the flexoelectric field below $T_S$.

--------
[*] Corresponding author: morozo@i.com.ua

[†] Corresponding author: vxg8@psu.edu

## 1. Introduction

Unique multifunctional properties of oxide interfaces are currently of widespread interest. These include such as 2-dimensional electron gas, superconductivity [1, 2, 3], charged domain walls [4], magnetism [5, 6] and multiferroicity at oxide interfaces [7] and thin strained films [8]. Interfaces by nature possess gradients of various order parameters such as strain, octahedral rotations, polarization, and magnetization, which can couple to induce new phenomena not present in the relevant bulk materials [9]. The influence of strain [10, 8] and strain gradients [11, 12, 13] in inducing ferroelectric polarization is well known. Recently, improper ferroelectricity induced by coupling to octahedral rotations has been predicted in a number of oxides (e.g. YMnO$_3$ [14], Ca$_3$Mn$_2$O$_7$ [15], CaTiO$_3$ [16]) and their multilayers [17].

Interfaces in antiferrodistorted perovskite oxides can possess both gradients in strain $u_{ij}$ and in oxygen octahedral rotations, characterized by spontaneous octahedral tilt angles, which in turn can be described by an axial vector $\Phi_i$ ($i$=1, 2, 3) [18]. As a consequence, both direct flexoelectric effect, namely the creation of a ferroelectric polarization due to a strain gradient, as well as rotostriction, namely a quadratic coupling between octahedral rotations and strain, exist at such interfaces. The coupling between these two phenomena can thus lead to a ferroelectric polarization at an interface across which the octahedral rotation varies, which is the subject of this paper. It has been previously predicted that a spontaneous polarization vector $P_i$ can appear inside structural walls due to biquadratic coupling term $\eta_{ijkl} P_i P_j \Phi_k \Phi_l$ [19, 20], but it is absent in the bulk. The biquadratic coupling term was later regarded as Houchmandazeh-Laizerowicz-Salje (HLS) coupling [21]. The coupling was considered as the reason of magnetization appearance inside the ferromagnetic domain wall in non-ferromagnetic media [22]. Biquadratic coupling leads to a polarization appearance inside antiphase boundaries in SrTiO$_3$ below 50 K [20]. However Zubko et al [23] experimentally observed strong changes of the apparent flexoelectric coefficient in SrTiO$_3$ at much higher temperatures, namely below the antiferrodistortive structural phase transition temperature (105 K), and supposed its reason in the polarization appearance at the domain walls between twins.

To the best of our knowledge, the flexoelectricity-induced polarization appearance across the structural twin boundaries (**TB**), antiphase boundaries (**APB**) and interfaces has not been previously addressed. However, the flexoelectric coupling, which is nonzero in any material and strong enough in many perovskites [11, 13, 23, 24, 25, 26, 27], should lead to the spontaneous polarization appearance across the structural domains walls of otherwise non-ferroelectric perovskites. Direct gradient coupling between the order parameters could lead to the oscillatory solutions and non-uniform pattern formation [28, 29]. This motivates us to perform calculations, based on the LGD free energy, to study the impact of flexoelectric coupling on the spontaneous polarization in the vicinity of structural domain walls in non-ferroelectric tilted perovskites such as $SrTiO_3$, $CaTiO_3$, and $EuTiO_3$. We present results for 90-degree TB and 180-degree APB in bulk $SrTiO_3$.

## 2. Theoretical formalism

We analyze the domain wall energy using approximate free energy functional corresponding to Tailor expansion on the polar and structural order parameter components. In the parent high temperature phase above the structural phase transition, the free energy density has the form:

$$F_b = a_i(T)P_i^2 + a_{ij}^u P_i^2 P_j^2 + ... + \frac{g_{ijkl}}{2}\left(\frac{\partial P_i}{\partial x_j}\frac{\partial P_k}{\partial x_l}\right) - P_i\left(\frac{E_i^d}{2} + E_i^{ext}\right) - q_{ijkl}u_{ij}P_k P_l + \frac{c_{ijkl}}{2}u_{ij}u_{kl}$$

$$+ b_i(T)\Phi_i^2 + b_{ij}^u \Phi_i^2 \Phi_j^2 - \eta_{ijkl}^u P_i P_j \Phi_k \Phi_l + \frac{v_{ijkl}}{2}\left(\frac{\partial \Phi_i}{\partial x_j}\frac{\partial \Phi_k}{\partial x_l}\right) \quad (1)$$

$$- r_{ijkl}^{(\Phi)} u_{ij}\Phi_k \Phi_l + \frac{f_{ijkl}}{2}\left(\frac{\partial P_k}{\partial x_l}u_{ij} - P_k \frac{\partial u_{ij}}{\partial x_l}\right) - \Phi_i \tau_i^d$$

$\Phi_i$ is the components ($i$=1, 2, 3) of an axial **tilt vector** corresponding to the octahedral rotation angles [18] (see **Scheme 1**), $\tau_i^d$ is de-elastification torque [18]; $u_{ij}(\mathbf{x})$ is the strain tensor. The summation is performed over all repeated indices. Coefficients $a_i(T)$ and $b_i(T)$ depend on temperature in accordance with Barrett law for quantum paraelectrics [30]: $a_1(T) = \alpha_T T_q^{(E)}\left(\coth\left(T_q^{(E)}/T\right) - \coth\left(T_q^{(E)}/T_0^{(E)}\right)\right)$ and $b_1(T) = \beta_T T_q^{(\Phi)}\left(\coth\left(T_q^{(\Phi)}/T\right) - \coth\left(T_q^{(\Phi)}/T_S\right)\right)$. Gradients coefficients $g_{ij}$ and $v_{ij}$ are regarded positive for commensurate ferroics. $f_{ijkl}$ is the forth-rank tensor of flexoelectric coupling, $q_{ijkl}$ is the forth-rank electrostriction tensor, $r_{ijkl}^{(\Phi)}$ is the rotostriction tensor. The biquadratic coupling between $\Phi_i$ and polarization components $P_i$ are defined by the constants $\eta_{ijkl}$. The flexoelectric effect tensor $f_{ijkl}$ and rotostriction tensor $r_{ijkl}^{(\Phi)}$ have nonzero components in all phases and for any symmetry of the system. Tensors form for cubic $m3m$ symmetry is well-known; in particular $f_{12}$, $f_{11}$ and $f_{44}$ are nonzero [23] similarly to elastic constants and electrostriction tensors [31]. Note, that the inclusion of the flexoelectric Lifshitz term in the free energy is critical for all effects discussed below.

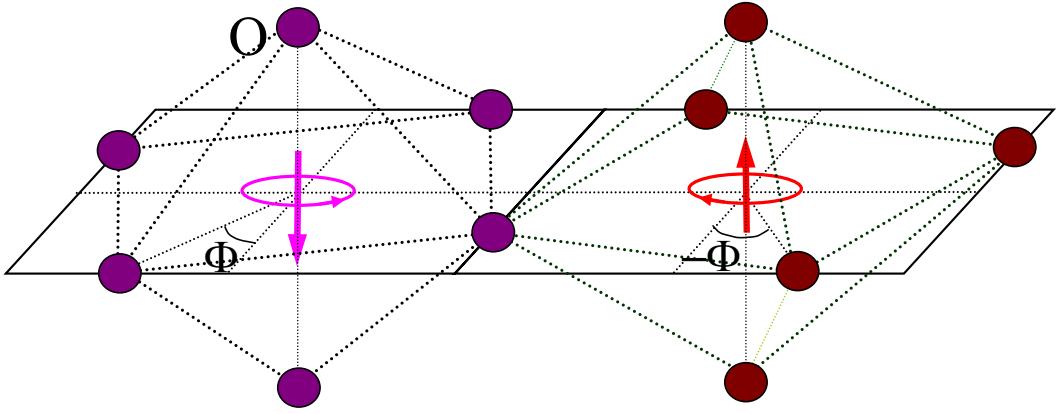

**Scheme 1.** The tilt value is typically opposite for the neighboring oxygen octahedrons far from the domain boundaries [18]. For the case free energy (1) considers the quasi-continuum tilt behavior in the next-nearest octahedral [20, 29].

External field is $E_i^{ext}$. In general case polarization distribution $P_i(x_i)$ can induce the depolarization field $E_i^d$ inside the wall. In the dielectric limit $E_i^d$ obeys electrostatic equation:

$$\varepsilon_0 \varepsilon_b \frac{\partial E_i^d}{\partial x_i} = -\frac{\partial P_i}{\partial x_i}, \qquad (i=1, 2, 3) \qquad (2)$$

where $\varepsilon_0 = 8.85 \times 10^{-12}$ F/m is the universal dielectric constant, $\varepsilon_b$ is the "base" isotropic lattice permittivity, different from the ferroelectric soft mode permittivity [32, 33, 34, 35]. Semiconductor case will be considered in the section 4.

Euler-Lagrange equations of state are obtained from the minimization of the free energy (1) as

$$\frac{\partial F_b}{\partial \Phi_i} - \frac{\partial}{\partial x_j}\left(\frac{\partial F_b}{\partial(\partial \Phi_i / \partial x_j)}\right) = 0, \qquad (3a)$$

$$\frac{\partial F_b}{\partial P_i} - \frac{\partial}{\partial x_j}\left(\frac{\partial F_b}{\partial(\partial P_i / \partial x_j)}\right) = 0, \qquad (3b)$$

$$\frac{\partial F_b}{\partial u_{ij}} - \frac{\partial}{\partial x_k}\left(\frac{\partial F_b}{\partial(\partial u_{ij} / \partial x_k)}\right) = \sigma_{ij}. \qquad (3c)$$

Where $\sigma_{ij}(\mathbf{x})$ is the stress tensor that satisfies mechanical equilibrium equation $\partial \sigma_{ij}(\mathbf{x})/\partial x_j = 0$. Note, that the stress tensor, polarization and tilt gradients vanish far from the domain walls.

Equation of state (3c) could be rewritten via the strains $u_{ij}(\mathbf{x})$ as follows:

$$u_{mn} = s_{mnij}\sigma_{ij} + R^{(\Phi)}_{mnkl}\Phi_k\Phi_l + Q_{mnkl}P_k P_l - F_{mnkl}\frac{\partial P_k}{\partial x_l}. \tag{4}$$

Where $s_{mnij}$ is the elastic compliances tensor; $R^{(\Phi)}_{ijkl} = s_{ijmn}r^{(\Phi)}_{mnkl}$ is the rotostriction strain tensor; $Q_{ijkl} = s_{ijmn}q_{mnkl}$ is the electrostriction strain tensor; $F_{ijkl} = s_{ijmn}f_{mnkl}$ is the flexoelectric strain tensor. The latter term corresponds to inverse flexoelectric effect.

The inhomogeneous strain $u_{ij}(\mathbf{x})$ given by Eq.(4) induces the polarization variation $\delta P_i(\mathbf{x})$ across the structural APB and TB, domain walls, defects and interfaces due to the direct flexoelectric effect:

$$\delta P_i(\mathbf{x}) = a_{iv}^{-1}f_{mnvl}\frac{\partial u_{mn}}{\partial x_l} \sim -a_{iv}^{-1}f_{mnvl}R^{(\Phi)}_{mnpq}\frac{\partial(\Phi_p\Phi_q)}{\partial x_l}. \tag{5}$$

The term $f_{mnvl}\frac{\partial u_{mn}}{\partial x_l}$ denotes direct flexoelectric effect. The proportionality in Eq.(5) suggests that the product of the flexoelectric $f_{mnvl}$ and rotostriction $R^{(\Phi)}_{mnpq}$ coefficients leads to the appearance of spontaneous polarization, which will *be abbreviated* in this study as **flexo-roto-effect**. To the best of our knowledge, the flexoelectric contribution to the interfacial polarization has not been considered earlier.

In the octahedral tilted phase at $T < T_S$, the spontaneous tilt vector appears and fluctuate subjected to small external stimulus: $\Phi_i(\mathbf{x}) = \Phi_i^S + \delta\Phi_i(\mathbf{x})$, where $\Phi_i^S = \sqrt{-b_i(T)/(2b_{11})}$. For the case substitution of $\Phi_i(\mathbf{x}) = \Phi_i^S + \delta\Phi_i(\mathbf{x})$ into Eq.(5) and linearization of Eq.(5) with respect to $\delta\Phi_i(\mathbf{x})$ gives $\delta P_i(\mathbf{x}) \sim W^{(\Phi)}_{ilq}\frac{\partial(\delta\Phi_q)}{\partial x_l}$, where the tensor of *apparent linear flexo-roto-effect* is introduced as $W^{(\Phi)}_{ilq} = -2a_{iv}^{-1}f_{mnvl}R^{(\Phi)}_{mnpq}\Phi_p^S$. However the effects, linear on tilt could be possible only as the result of linearization of the quadratic terms. In turn, the linearization could be possible if there exist some macroscopic external field, cojugated with tilt (like torque $\tau_i^d$). Below we will consider tilted perovskites without such external field, so we could not see any manifestation of linear on tilt efffects.

The gradient in the octahedral tilts across APB or TB may lead to a rather strong interfacial polarization due to the derivatives of the rotation angle in Eq.(5). In the next section we will consider the concrete example of SrTiO$_3$ with known numerical values of $R^{(\Phi)}_{mnkj}$ and $F_{mnvl}$ to check the validity of this supposition.

### 3. Flexo-roto-effect contribution to the interfacial polarization and pyroelectricity

Below we consider several one-dimensional problems, which follow from general results of the previous section, namely a typical 180-degree APB and 90-degree TB. Stable solutions of the coupled Euler-Lagrange equations (3) were obtained numerically by iteration method. We set initial distributions of the tilt and polarization vectors, which satisfy the boundary conditions. Special attention was paid to the parity of the obtained polarization distributions, namely we consider both odd and even initial polarization distributions with respect to the domain wall plane. Iterations were stopped when the relative tolerance reaches the value $10^{-4}$. Higher accuracy was required for the calculations of the walls energies.

Note, that the flexoelectric coupling was not included into the calculations performed in Ref.[20], while mentioned in the paper as giving rise to the renormalization of the gradient terms. Probably that is why Tagantsev et al [20] obtained nonzero polarization across SrTiO$_3$ APB only below effective Curie temperature renormalized by elastic stresses, i.e. below $T_C^* \sim 40$ K, but not below $T_S \sim 105$ K. However Zubko et al [23] attributed experimentally observed strong dependence of the SrTiO$_3$ flexoelectric response on applied elastic stress with the twins' motion in supposition of polarization appearance across the elastic walls exactly below 105 K. One of the most important results we obtained in the present research is the fact that the flexoelectric effect primary leads to the appearance of the strong built-in electric fields across the wall, besides the renormalization of the polarization gradient term also considered in Ref.[26]. We obtained that the flexoelectric effect can induce the polarization across TB and APB walls over the entire temperature range of the structural phase in complete agreement with the experiment [23].

One may ask a reasonable question about the experimental observation of an odd polarization distribution across an interface, because the mean polarization of the domain wall is zero in this case, and most of the macroscopic methods may detect it. In contrast scanning probe experiments, such as scanning piezoelectric force microscopy with sharp probes, allow local registration of the polarization over nanometer resolution (see e.g. [36] and refs therein). Kelvin probe and current scanning probe microscopy allows determination of the local electric potential and current distribution across the domain walls [4, 37]. Atomically resolved mapping of the octahedral tilts and polarization across a structural domain walls becomes possible from the combination of z-contrast Scanning Transmission

Electron Microscopy and atomic column shape image analysis [38]. Below we show that all elastic domain walls (including odd ones) possess noticeable pyroelectric response, which can be revealed by the novel scanning probe pyroelectric microscopy [39]. Note, that pyroelectric response exists in semiconducting solids also [40].

Thus below we consider the influence of the flexoelectric and rotostriction effect on the both odd and even polarization distributions across TB and APB and calculate the energy of such walls in order to answer the fundamental questions: (a) what domain walls are more thermodynamically preferable and (b) to know exactly the effective surface energy values at different temperatures. While it is expected that easy TB and APB should have essentially lower energies than the hard TB and APB (consistent with their classification), exact values of their effective surface energy are necessary for the calculations of the walls effective pressure as a driving force in kinetics of nucleation and formation process.

### *3.1. Flexo-roto-effect manifestation at the antiphase boundaries (APB)*

In the octahedral tilted phase at $T < T_S$, the one-component spontaneous tilt, $\Phi_S$, appear in a bulk SrTiO$_3$, other components, $\Phi_1$ and $\Phi_2$, can be nonzero in the vicinity of APB.

"Easy" APB with $\Phi_3(x_3) \neq 0, \Phi_2 \equiv 0, \Phi_1 \equiv 0$ (see **Fig. 1a**) induces nonzero odd or even distribution of polarization $P_3(x_3)$, while $P_1 \equiv 0, P_2 \equiv 0$. "Hard" APB with $\Phi_1(x_1) \neq 0, \Phi_3(x_1) \neq 0, \Phi_2 \equiv 0$ (see **Fig. 1b**) induces nonzero odd or even distributions of polarization $P_1(x_1)$ and even distribution of $P_3(x_1)$, while $P_2 \equiv 0$. Classification "easy" and "hard" APB comes from Ref.[20], but only even solutions $P_3(x_3)$ were considered in that work.

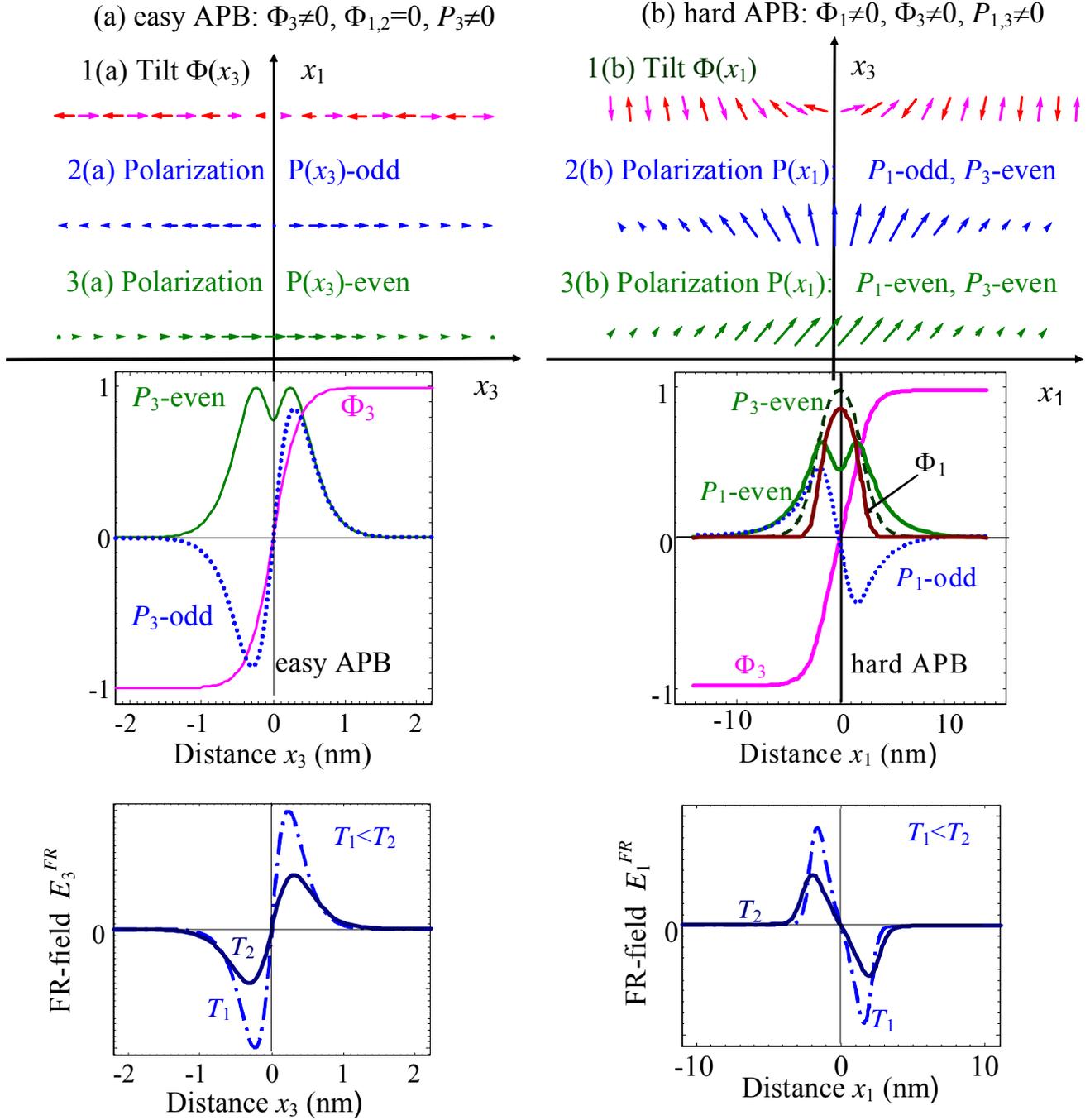

**Figure 1.** Schematics of the polarization appearance inside easy (a) and hard (b) APB. $x_1$=[100], $x_3$=[001] and $x_2$=[010] (not shown), are crystallographic axes directions of SrTiO$_3$. Flexo-roto field, which induces the polarization component parallel to the wall, is shown at the bottom plots at two different temperatures $T_1 < T_2$. Note, that Vasudevarao et al [8] observed and calculated by phase-field various orientations of the ferroelastic APB in SrTiO$_3$.

For the case of **easy APB** ($x_3$-dependent solution) one could easily derive the stress field that satisfies mechanical equilibrium equation $\partial \sigma_{ij}(\mathbf{x})/\partial x_j = 0$ and vanishes far from the domain walls as

$$\sigma_{11}(x_3) = \sigma_{22}(x_3) = \frac{U(x_3)}{s_{11} + s_{12}}, \quad \sigma_{33} = \sigma_{13} = \sigma_{12} = \sigma_{23} = 0. \tag{6}$$

The function:

$$U(x_3) = R_{12}^{(\Phi)}\left(\Phi_S^2 - \Phi_3^2\right) - \left(Q_{12}P_3^2 - F_{12}\frac{\partial P_3}{\partial x_3}\right) \tag{7}$$

Substitution of the elastic stresses (6) into Eqs.(3c) leads to a closed system for the polarization and tilt vector components:

$$2\left(b_1 - \eta_{11}P_3^2 - \frac{2R_{12}^{(\Phi)}U}{s_{11} + s_{12}}\right)\Phi_3 + 4b_{11}\Phi_3^3 - v_{11}\frac{\partial^2 \Phi_3}{\partial x_3^2} = 0, \tag{8a}$$

$$2\left(a_1 - \eta_{11}\Phi_3^2 - \frac{2Q_{12}U}{s_{11} + s_{12}}\right)P_3 + 4a_{11}P_3^3 - g_{11}\frac{\partial^2 P_3}{\partial x_3^2} = E_3^{ext} + E_3^d + \frac{2F_{12}}{s_{11} + s_{12}}\frac{\partial U}{\partial x_3}. \tag{8b}$$

Boundary conditions for rotations are $\Phi_3(x_3 \to \pm\infty) = \pm\Phi_S$, $\Phi_3(x_3 = 0) = 0$, where $\Phi_S = \sqrt{-b_1(T)/(2b_{11})}$ is the spontaneous tilt component of the single-domain tilted phase. Boundary conditions for polarization are $P_3(x_3 \to \pm\infty) = 0$ and $P_3(0) = 0$ (odd solution) or $\partial P_3(0)/\partial x_3 = 0$ (even solution).

Note, that tilt profile could be well approximated as $\Phi_3 = \Phi_S \tanh(x_3/l_\Phi)$, where $l_\Phi = \sqrt{-v_{11}/b_1(T)}$ is the correlation length of the correponding mode fluctuations. The distribution of $\Phi_3(x_3)$ is schematically shown in **Fig. 1a** (right). $\Phi_3(x_3)$ is rather weakly dependent on the polarization vector distribution. In contrast, polarization $P_3$ strongly depends on the tilt vector as proportional to the flexoelectric-rotostriction field (abbreviated as **flexo-roto field** below). Actually, substitution of the Eq.(7) into Eq.(8b) leads to the flexo-roto field appearance in the right-hand-side of the equation $E_3^{FR} = -2F_{12}R_{12}^{(\Phi)}(s_{11} + s_{12})^{-1}\partial(\Phi_3^2)/\partial x_3 \sim \sinh(x_3/l_\Phi)/\cosh^3(x_3/l_\Phi)$, which is odd function with respect to $x_3$ (see bottom **Fig.1a**). It is very important for further consideration that the trivial solution $P_3 \equiv 0$ does not exist in the vicinity of APB due to nonzero flexo-roto field $E_3^{FR} \neq 0$. Numerical estimations for SrTiO$_3$ material parameters listed in the **Table 1** proved that renormalized coefficient $a_3^R = a_1 + 1/(\varepsilon_0\varepsilon_b) - \eta_{11}\Phi_3^2 - 2Q_{12}U/(s_{11} + s_{12})$ in Eq.(8b) is positive at all temperatures due to the strong depolarization field $E_3^d(x_3) = -P_3(x_3)/(\varepsilon_0\varepsilon_b)$ [41], that is calculated in the dielectric limit in this section. So the spontaneous polarization component $P_3$ perpendicular to the APB plane is induced by the flexo-roto-field $E_3^{FR}$ only. Naturally, the depolarization field $E_3^d(x_3)$ strongly reduces the component $P_3$ value. Since $a_3^R > 0$ the "true" ferroelectricity, i.e. polarization hysteresis is absent in external field $E_3^{ext}$, but pyroelectric r, i.e. nonzero pyroelectric response $\Pi_3 = dP_3/dT$, should exist, since the polarization component is temperature dependent due to the temperature dependence of $E_3^{FR}$ and $a_3^R$, which in turn depend on temperature due to the temperature dependence of

$$b_1(T) \sim \left( \coth\left(\frac{T_q^{(\Phi)}}{T}\right) - \coth\left(\frac{T_q^{(\Phi)}}{T_S}\right) \right), \quad a_1(T) \sim \left( \coth\left(\frac{T_q^{(E)}}{T}\right) - \coth\left(\frac{T_q^{(E)}}{T_0^{(E)}}\right) \right) \text{ and } \Phi_3(x_3).$$ Substitution of

Eq.(7) into Eq.(8b) shows that the flexoelectric coupling also leads to the gradient terms renormalization like $g_{11}^R = \left(g_{11} + 2F_{12}^2/(s_{11} + s_{12})\right)$ similar to reported earlier [26] (note that we used coefficients at fixed strains in Ref. [26], so signs of renormalization are different).

Eqs.(8) were solved numerically by iteration method using SrTiO$_3$ material parameters listed in the **Table 1.** The results of numerical solution of Eqs.(8) is analyzed below. The spontaneous polarization across easy APB is induced by the flexo-roto field shown in **Figs.2a** and **2b**. $P_3$-odd and $P_3$-even spontaneous polarization distributions inside easy APB are shown in **Figs. 2c,d** at two different temperatures below (c) and well above (d) the temperature $T_0^{(E)} \sim 30$ K. It is seen that polarization is identically zero under the absence of the flexoelectric field (see curves 1,2 calculated at $F_{ij} \equiv 0$ and $\eta_{ij} \neq 0$): biquadratic coupling does not induce any polarization for the case. Naturally, flexoelectric coefficient sign determines the appearance of head-to-head or tail-to-tail easy $P_3$-odd polarization distributions; for SrTiO$_3$ with positive $F_{12}$ value only head-to-head distributions are realized. Polarization amplitude decreases with the temperature increase, while the shape of the profile appears almost the same. Note, that double maximums in $P_3$-even distributions originate from the internal flexo-roto electric field $E_3^{FR}$; they are not related with oscillatory solutions [28].

**Table 1.** SrTiO$_3$ material parameters and LGD free energy (1) coefficients collected from Refs. [20, 23, 30, 42, 43, 44, 45, 46]

| Parameter | SI units | Value | Source and notes |
|---|---|---|---|
| $\varepsilon_b$ | dimensionless | 43 | [45] |
| $\alpha_T$ | $10^6 \times$ m/(F K) | 0.75 | [20, 43] |
| $T_0^{(E)}$ | K | 30 | ibidem |
| $T_q^{(E)}$ | K | 54 | ibidem |
| $a_{ij}$ | $10^9 \times$ m$^5$/(C$^2$F) | $a_{11}^u$=2.025, $a_{12}^u$=1.215, $a_{11}^\sigma$=0.820, $a_{12}^\sigma$=1.396 | ibidem |
| $q_{ij}$ | $10^{10} \times$ m/F | $q_{11}$=1.251, $q_{12}$= −0.108, $q_{44}$=0.243 | [43] |
| $Q_{ijkl}$ | m$^4$/C$^2$ | $Q_{11}$=0.051, $Q_{12}$= −0.016, $Q_{44}$=0.020 | [20] |
| $g_{ijkl}$ | $10^{-11} \times$ V·m$^3$/C | $g_{11}$=$g_{44}$=1, $g_{12}$=0.5 | Estimation [46] |
| $\beta_T$ | $10^{26} \times$ J/(m$^5$ K) | 9.1 | [43] |
| $T_S$ | K | 105 | [43] |
| $T_q^{(\Phi)}$ | K | 145 | [43] |
| $b_{ij}$ | $10^{50} \times$ J/m$^7$ | $b_{11}^u$=1.94, $b_{12}^u$=3.96, $b_{11}^\sigma$=0.93, $b_{12}^\sigma$=3.88 | [43] |
| $r_{ij}$ | $10^{30} \times$ J/(m$^5$) | $r_{11}$=1.3, $r_{12}$= −2.5, $r_{44}$=-2.3 | [43] |
| $R_{ij}$ | $10^{19} \times$ m$^{-2}$ | $R_{11}$=0.882, $R_{12}$= −0.777, $R_{44}$= −1.811 | calculated from $r$ |
| $\eta_{ijkl}$ | $10^{29}$ (F m)$^{-1}$ | $\eta_{11}^u$=−3.366, $\eta_{12}^u$= 0.135, $\eta_{44}^u$=6.3 $\eta_{11}^\sigma$=−2.095, $\eta_{12}^\sigma$= −0.849, $\eta_{44}^\sigma$=5.860 | [43] calculated from $\eta_{ij}^u$ |
| $v_{ijkl}$ | $10^{10} \times$ J/m$^3$ | $v_{11}$=0.28, $v_{12}$= −7.34, $v_{44}$=7.11 | [20] |
| $c_{ij}$ | $10^{11} \times$ J/m$^3$ | $c_{11}$=3.36, $c_{12}$=1.07, $c_{44}$=1.27 | [20, 43] |
| $s_{ij}$ | $10^{-12} \times$ m$^3$/J | $s_{11}$=3.52, $s_{12}$= −0.85, $s_{44}$=7.87 | calculated from $c_{ij}$ |
| $F_{ijkl}$ | $10^{-12} \times$ m$^3$/C | $F_{11}$= − 13.80, $F_{12}$= 6.66, $F_{44}$= 8.48 | calculated from $f_{ij}$ from [23] at given stress |
| $\Phi_S$ | radian | 0.0235 | at low temperatures |
| $n_e$ | m$^{-3}$ | $10^{22} - 5 \cdot 10^{26}$ | [47, 48], strongly dependent on the oxygen vacancies concentration |

**\*)** Superscripts "u" and "σ" denote coefficients at given strain (clamped sample) and given stress (free sample) respectively

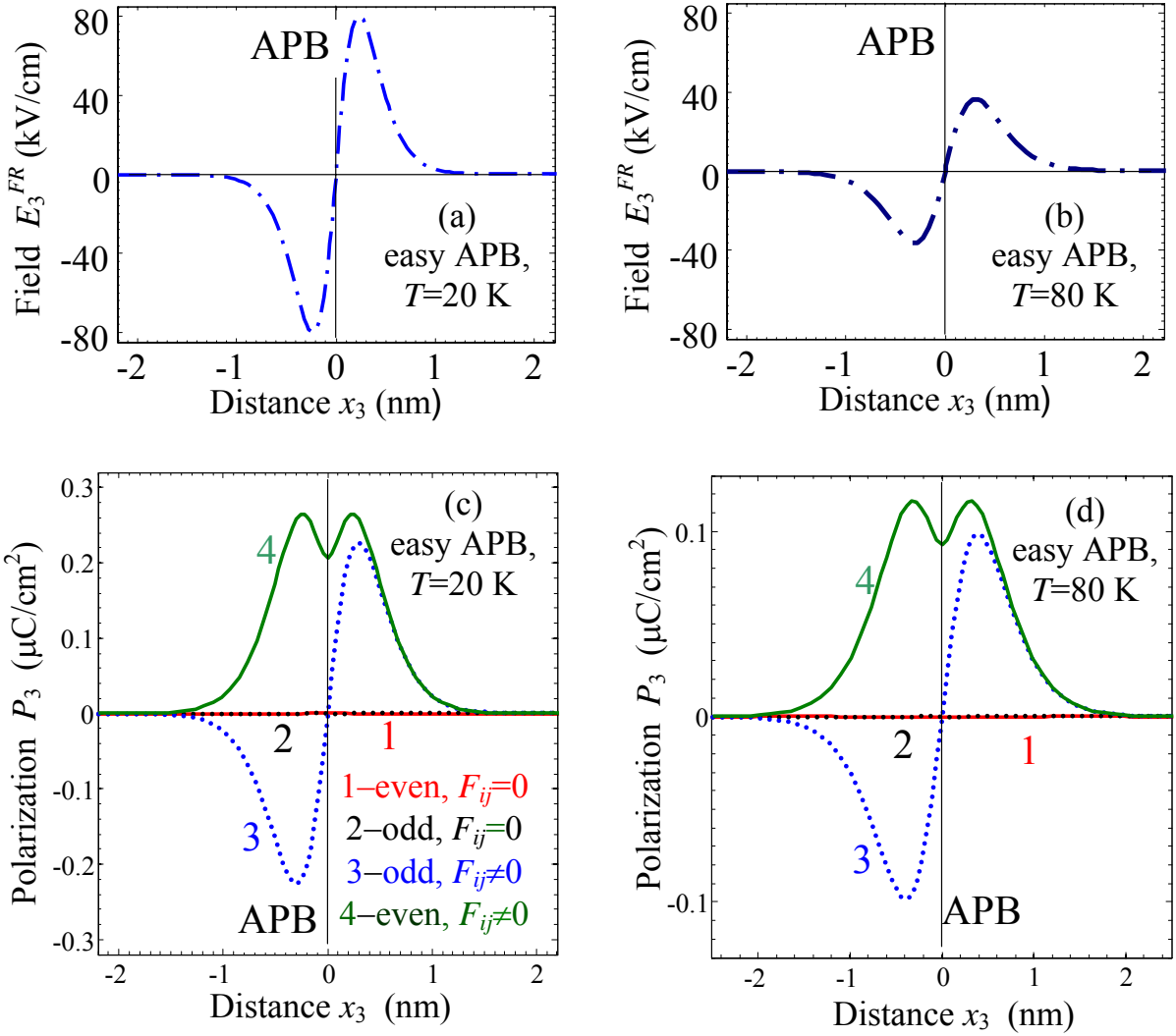

**Figure 2.** Flexo-roto field (a, b) induces spontaneous polarization (c,d) distributions across easy APB. Plots are calculated at temperatures $T=20$ K (a, c) and 80 K (b, d) Perpendicular to APB component $P_3$-odd (dotted curves) and $P_3$-even (solid curves) are calculated for nonzero flexoelectric effect $F_{ij} \neq 0$ and biquadratic coupling $\eta_{ij} \neq 0$ (curves 3, 4) and for the case of nonzero biquadratic coupling $\eta_{ij} \neq 0$ and zero flexoelectric effect $F_{ij} \equiv 0$ (curves 1, 2). Curves 1-4 style and color coding for plots (c, d) are the same and described in the legend to plot (c).

Temperature dependences of the maximal spontaneous polarization values and pyroelectric coefficient induced by the flexo-roto-effect and biquadratic coupling inside easy APB are shown in **Figs. 3a** and **3b**. Under the absence of the flexo-roto field polarization and pyroelectric coefficient are zero (see curves 1,2 calculated at $F_{ij} \equiv 0$ and $\eta_{ij} \neq 0$). It is seen from the **Fig.3a** that spontaneous polarization component appears below $T_S = 105$ K at $F_{12} \neq 0$, then quasi-linearly increases with temperature decrease and then saturates at temperatures $T < T_0^{(E)} \sim 30$ K due to the Barrett law for $b_1(T)$ and $a_1(T)$. Actually, $P_3 \sim E_3^{FR} \sim \partial(\Phi_3^2)/\partial x_3 \sim \Phi_S^2/l_\Phi \sim (-b_1(T))^{3/2} \sim (T_S - T)^{3/2}$ at temperatures near $T_S$. It is seen from the **Fig.3b** that pyroelectric coefficient $\Pi_3$ appears below $T_S = 105$ K at $F_{12} \neq 0$, then rapidly increases with temperature decrease, reaches maximum at the polarization inflection point ~80 K and then strongly decreases with further temperatures decrease due to the spontaneous polarization $P_3$ saturation at temperatures $T \ll T_q \sim 50$ K in accordance with the Barrett law for $b_1(T)$ and $a_1(T)$. Actually, in the range $T \ll T_q$ polarization becomes almost temperature independent and its temperature derivative vanishes.

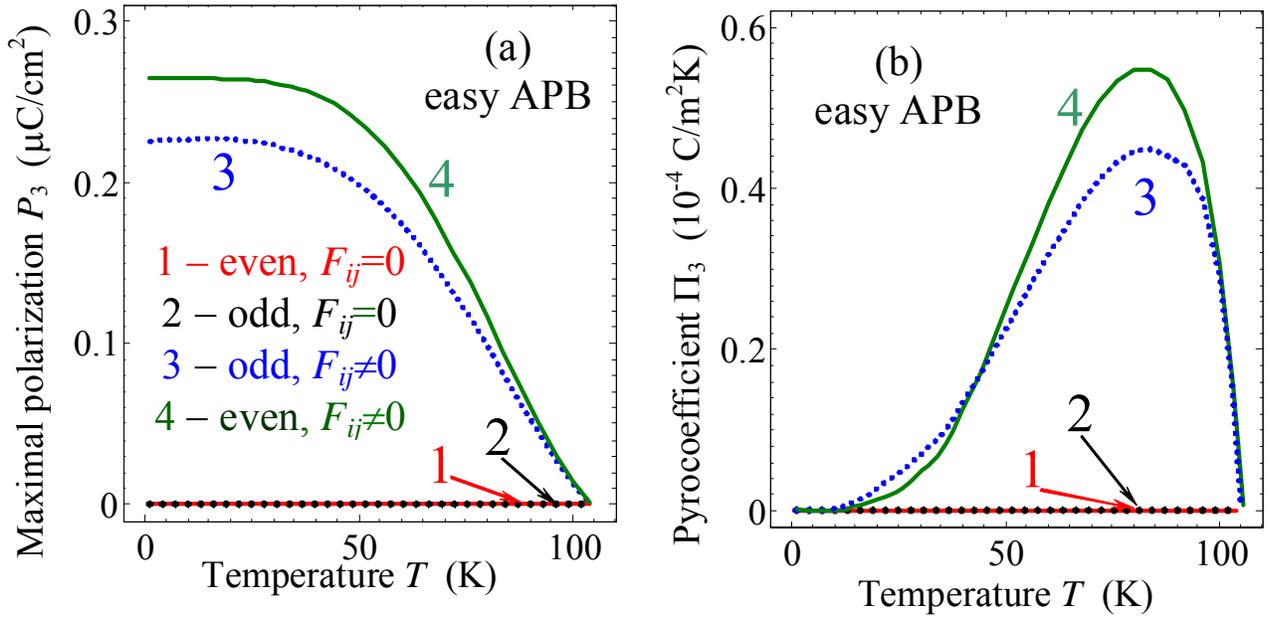

**Figure 3.** Temperature dependences of (a) the spontaneous polarization $P_3$ maximal value and (b) corresponding pyroelectric coefficient $\Pi_3$ calculated for easy APB in SrTiO$_3$ without free carriers. Temperature dependences are calculated for nonzero flexoelectric effect $F_{ij} \neq 0$ and biquadratic coupling $\eta_{ij} \neq 0$ (curves 3, 4) and for the case of nonzero biquadratic coupling $\eta_{ij} \neq 0$ and zero flexoelectric effect $F_{ij} \equiv 0$ (curves 1, 2). Curves 1-4 style and color coding for plots (a, b) are the same and described in the legend to plot (a). Solid and dotted curves correspond to $P_3$-even and $P_3$-odd solutions respectively.

For the case of **hard APB** ($x_1$-dependent solution) we derive the stress field that satisfies mechanical equilibrium equation $\partial \sigma_{ij}(\mathbf{x})/\partial x_j = 0$ and vanishes far from the domain walls as

$$\sigma_{11} = \sigma_{13} = \sigma_{12} = \sigma_{23} = 0, \quad \sigma_{33} = \frac{s_{11}U_3 - s_{12}U_2}{s_{11}^2 - s_{12}^2}, \quad \sigma_{22} = \frac{s_{11}U_2 - s_{12}U_3}{s_{11}^2 - s_{12}^2}. \tag{9}$$

Functions:

$$U_2(x_1) = R_{12}^{(\Phi)}(\Phi_S^2 - \Phi_3^2) - \left(Q_{12}(P_1^2 + P_3^2) + R_{12}^{(\Phi)}\Phi_1^2 - F_{12}\frac{\partial P_1}{\partial x_1}\right), \tag{10a}$$

$$U_3(x_1) = R_{11}^{(\Phi)}(\Phi_S^2 - \Phi_3^2) - \left(Q_{11}P_3^2 + Q_{12}P_1^2 + R_{12}^{(\Phi)}\Phi_1^2 - F_{12}\frac{\partial P_1}{\partial x_1}\right). \tag{10b}$$

Substitution of the elastic stress (9) into Eqs.(3) leads to a closed system for the polarization and tilt vector components:

$$2\left(b_1 + b_{12}\Phi_3^2 - \eta_{11}P_1^2 - \eta_{12}P_3^2 - R_{12}^{(\Phi)}\left(\frac{U_2 + U_3}{s_{11} + s_{12}}\right)\right)\Phi_1 + 4b_{11}\Phi_1^3 - \eta_{44}P_1P_3\Phi_3 - v_{11}\frac{\partial^2 \Phi_1}{\partial x_1^2} = 0, \tag{11a}$$

$$2\left(b_1 + b_{12}\Phi_1^2 - \eta_{11}P_3^2 - \eta_{12}P_1^2 - \frac{s_{11}U_3 - s_{12}U_2}{s_{11}^2 - s_{12}^2}R_{11}^{(\Phi)} - \frac{s_{11}U_2 - s_{12}U_3}{s_{11}^2 - s_{12}^2}R_{12}^{(\Phi)}\right)\Phi_3$$
$$- \eta_{44}P_1P_3\Phi_1 - v_{44}\frac{\partial^2 \Phi_3}{\partial x_1^2} = 0 \tag{11b}$$

$$2\left(a_1 - \eta_{11}\Phi_1^2 - \eta_{12}\Phi_3^2 - Q_{12}\left(\frac{U_2 + U_3}{s_{11} + s_{12}}\right) + a_{12}P_3^2\right)P_1 + 4a_{11}P_1^3 - g_{11}\frac{\partial^2 P_1}{\partial x_1^2}$$
$$= E_1^{ext} + E_1^d + F_{12}\frac{\partial}{\partial x_1}\left(\frac{U_2 + U_3}{s_{11} + s_{12}}\right) + \eta_{44}\Phi_1\Phi_3 P_3 \tag{11c}$$

$$2\left(a_1 + a_{12}P_1^2 - \eta_{11}\Phi_3^2 - \eta_{12}\Phi_1^2 - \frac{s_{11}U_3 - s_{12}U_2}{s_{11}^2 - s_{12}^2}Q_{11} - \frac{s_{11}U_2 - s_{12}U_3}{s_{11}^2 - s_{12}^2}Q_{12}\right)P_3$$
$$+ 4a_{11}P_3^3 - g_{44}\frac{\partial^2 P_3}{\partial x_1^2} = E_3^{ext} + \eta_{44}\Phi_1\Phi_3 P_1 \tag{11d}$$

Boundary conditions are: $\Phi_3(x_1 \to \pm\infty) = \pm\Phi_S$, $\Phi_3(x_1 = 0) = 0$, $\Phi_1(x_1 \to \pm\infty) = 0$, $\partial\Phi_1/\partial x_1|_{x_1=0} = 0$ for the rotation vector; polarization $P_3(x_1 \to \pm\infty) = 0$, $\partial P_3/\partial x_1|_{x_1=0} = 0$, $P_1(x_1 \to \pm\infty) = 0$ and either $\partial P_1/\partial x_1|_{x_1=0} = 0$ for even solution, or $P_1(x_1 = 0) = 0$ for odd solution.

Distributions $\Phi_{1,3}(x_3)$ are schematically shown in **Fig. 1b** (right). Appeared that $\Phi_{1,3}(x_3)$ are rather weakly dependent on the polarization vector. Nonzero odd flexo-roto-field $E_1^{FR} = -F_{12}(s_{11} + s_{12})^{-1}\partial\left((R_{11}^{(\Phi)} + R_{12}^{(\Phi)})\Phi_3^2 + 2R_{12}^{(\Phi)}\Phi_1^2\right)/\partial x_1$ exists only for the polarization component $P_1(x_1)$ perpendicular to APB plane (see the right-hand-side of Eq.(11c) and bottom **Fig.1b**). Thus the component $P_1$ appears just below $T_S$ and strongly depends on the tilt vector as proportional to the field $E_3^{FR}$. Corresponding coefficient $a_1^R = a_1 + 1/(\varepsilon_0 \varepsilon_b) - \eta_{11}\Phi_1^2 - \eta_{12}\Phi_3^2 - Q_{12}(U_2 + U_3)/(s_{11} + s_{12}) + a_{12}P_3^2$ in Eq.(11c) is positive at all temperatures due to the strong depolarization field $E_1^d(x_1) = -P_1(x_1)/(\varepsilon_0 \varepsilon_b)$. Therefore hysteresis loop for polarization component $P_1$ is absent in external field $E_1^{ext}$.

Nonzero component $P_1$ perpendicular to the APB should induce the component $P_3$ parallel to the APB just below $T_S$ due to the biquadratic coupling term $-\eta_{44}\Phi_1\Phi_3 P_1$ in Eq.(11d). Thus the trivial solution $P_1 \equiv P_3 \equiv 0$ does not exist in the vicinity of hard APB due to nonzero flexo-roto field $E_1^{FR} \neq 0$ and the coupling term $-\eta_{44}\Phi_1\Phi_3 P_1$ (compare with Ref.[20], where the stability of the trivial solution $P_3 \equiv 0$ was studied without flexo-effect). Polarization component $P_3$ depends on the tilt vector due to the tilt-dependent coefficient $a_3^R = a_1 - \eta_{11}\Phi_3^2 - \eta_{12}\Phi_1^2 - \frac{s_{11}U_3 - s_{12}U_2}{s_{11}^2 - s_{12}^2}Q_{11} - \frac{s_{11}U_2 - s_{12}U_3}{s_{11}^2 - s_{12}^2}Q_{12}$ that becomes negative for SrTiO$_3$ at temperatures $T < T_C^{APB}$ due to the biquadratic coupling terms $-\eta_{11}\Phi_3^2 - \eta_{12}\Phi_1^2$ and elastic stresses [20], where the effective Curie temperature $T_C^{APB} \approx 50$ K is introduced for hard APB. So, ferroelectric hysteresis for polarization component $P_3$ should exist in external field $E_3^{ext}$ in the temperature range $T < T_C^{APB}$.

Below $T_C^{APB}$ the ferroelectric parallel component $P_3$ strongly enhances the perpendicular one $P_1$ via the biquadratic coupling term $-\eta_{44}\Phi_1\Phi_3 P_3$ in Eq.(11c), in the region $P_1$ is induced to the sum of the $-\eta_{44}\Phi_1\Phi_3 P_3$ and $E_3^{FR}$. The depolarization field strongly reduces the component $P_1$ in comparison with ferroelectric polarization $P_3$ below $T_C^{APB}$. Pyroelectric coefficients $\Pi_3 = dP_3/dT$ and $\Pi_1 = dP_1/dT$ should be nonzero in the temperature range $T < T_S$, but their temperature behavior are rather different at $T_C^{APB} < T < T_S$ and $T < T_C^{APB}$ as shown below.

Eqs.(11) were solved numerically by iteration method for SrTiO$_3$ material parameters listed in the **Table 1**. $P_1$-odd and $P_1$-even and $P_3$-even spontaneous polarization distributions induced by the flexoelectric effect and biquadratic coupling inside the hard APB are shown in **Figs. 4** at two temperatures below (a,b) and above (c,b) the effective Curie temperature $T_C^{APB}$. Polarization components $P_1$ and $P_3$ are identically zero above $T_C^{APB}$ under the absence of the flexoelectric coupling, since biquadratic coupling does not induce any polarization for the case (see curves 1,2 calculated at $F_{ij} \equiv 0$ and $\eta_{ij} \neq 0$). Polarization component $P_3$ is almost independent on the flexoelectric effect below $T_C^{APB}$, since the component is determined by the biquadratic coupling in the temperature range (see curves 1-4 in **Figs. 4b**). Polarization amplitude strongly decreases with the temperature increase, while the shape of the profile appears almost the same for $P_1$, but not for the $P_3$. $P_3$ has a single maximum at the wall at low temperatures, which splits into the double maximums with the temperature increase (compare **Figs. 4b** and **4d**). A very small component $P_3$ exists at temperatures $T_C^{APB} < T < T_S$ due to the biquadratic coupling term $-\eta_{44}\Phi_1\Phi_3 P_1$ in Eq.(14d). The double maximums at $P_1$-even distributions originate from the internal electric field $E_1^{FR}$ and exist at all temperatures below $T_S$. The double maximums at $P_3$-even distributions originate from the coordinate dependence of $a_3^R$ via the functions $U_{2,3}(x_1)$; the maximums exist at temperatures above $T_C^{APB}$.

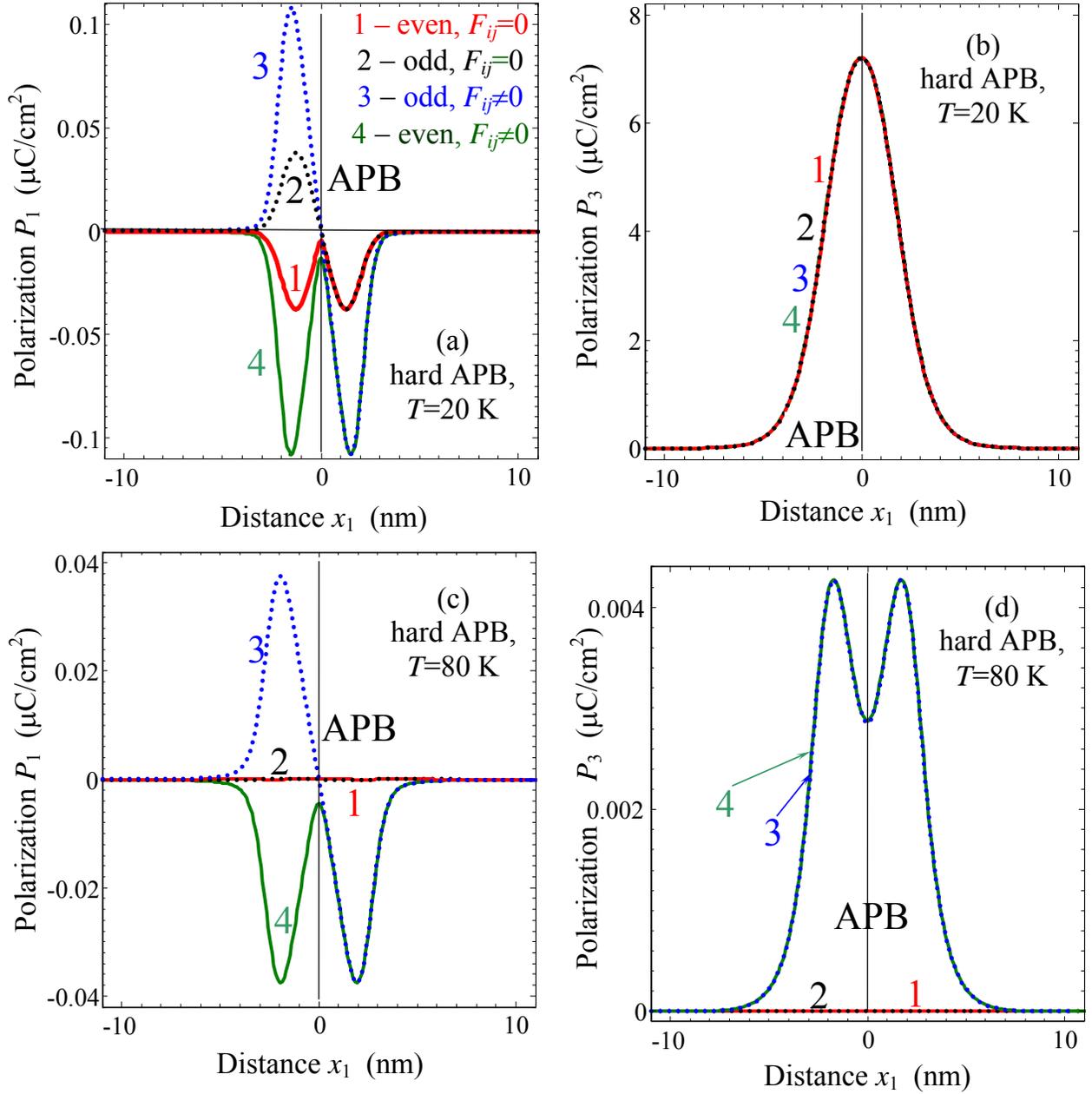

**Figure 4.** Spontaneous polarization distribution across hard APB calculated at temperatures $T$=20 K (a, b) and 80 K (c, d) for SrTiO$_3$ without free carriers. Polarization components perpendicular ($P_1$) and parallel ($P_3$) to APB are shown. $P_1$-odd (dotted curves) and $P_1$-even (solid curves) are calculated for nonzero flexoelectric effect $F_{ij} \neq 0$ and biquadratic coupling $\eta_{ij} \neq 0$ (curves 3, 4) and for the case of nonzero biquadratic coupling $\eta_{ij} \neq 0$ and zero flexoelectric effect $F_{ij} \equiv 0$ (curves 1, 2). Curves 1-4 style and color coding for plots (a-d) are the same and described in the legend to plot (a).

Temperature dependences of the maximal spontaneous polarization values and pyroelectric response induced by the flexo-roto-effect and biquadratic coupling inside hard APB are shown in

**Figs. 5**. Under the absence of the flexoelectric field the spontaneous polarization and pyroelectric coefficient are zero at temperatures higher than the effective Curie temperature $T_C^{APB}$ (see curves 1,2 calculated at $F_{ij} \equiv 0$ and $\eta_{ij} \neq 0$). The flexoelectric field rather weakly influences on the polarization component $P_3$ (compare curves 1, 2 with curves 3, 4 in **Fig.5a**). However the flexoelectric field $E_1^{FR}$ strongly increases the component $P_1$ below $T_S$, since $P_1 \sim E_1^{FR}$ (compare curves 1, 2 with curves 3, 4 in **Fig.5b**). Actually, for the case $F_{ij} \neq 0$ the component $P_1$ appears below $T_S$, firstly quasi-linearly increases with temperature decrease, then has a pronounced jump at $T_C^{APB}$ and then saturates at temperatures $T \ll T_q$ due to the Barrett law for $b_1(T)$. The break at $T_C^{APB}$ originates from the appearance of reversible polarization component $P_3$ below $T_C^{APB}$. The component $P_1 \sim E_1^{FR} \sim \dfrac{\partial \Phi^2}{\partial x_1} \sim \Phi_S^2/l_\Phi \sim (-b_1(T))^{3/2} \sim (T_S - T)^{3/2}$ in the vicinity of $T_S$. It is seen from **Fig. 5a-b** that the maximal values of polarization are very close for odd and even types of solutions in the dielectric limit. Note, that Tagantsev et al [20] analytically predicted spontaneous polarization about 4 μC/cm$^2$ at hard APB below 35-40 K without considering flexo-roto-effect contribution. Allowing for the flexo-roto-effect we obtained $P_3 \sim 8$ μC/cm$^2$ and $P_1 \sim 0.1$ μC/cm$^2$ at hard APB below $T_C^{APB} \approx 50$ K. At temperatures $T < T_C^{APB}$ the amplitude of $P_1$ is much smaller than the amplitude of $P_3$ due to the strong depolarization field $E_1^d(x_1)$.

It is seen from the **Figs.5c-d** that pyroelectric coefficients $\Pi_3$ and $\Pi_1$ appears below $T_S = 105$K only at nonzero flexoelectric coefficient $F_{12} \neq 0$. Pyroelectric coefficient $\Pi_3$ has the sharp maximum at $T_C^{APB}$ corresponding to the second order ferroelectric phase transition (appearance of the ferroelectric polarization $P_3$). Pyroelectric coefficient $\Pi_1$ has two maximums: a smooth maximum at the polarization inflection point ~80 K and the sharp maximum at $T_C^{APB}$ originated from $P_3$ appearance, since $P_3$ enhances $P_1$ via the biquadratic coupling term $-\eta_{44}\Phi_1\Phi_3 P_3$ in Eq.(14c). Pyroelectric coefficients monotonically decrease below $T_C^{APB}$ with the temperatures decrease due to the spontaneous polarization components saturation at temperatures $T \ll T_q$ in accordance the Barrett law for $b_1(T)$ and $a_1(T)$. Actually, in the range $T \ll T_q$ polarization becomes almost temperature independent and its temperature derivative vanishes. Allowing for the flexo-roto-effect contribution we obtained maximal pyroelectric coefficients $\Pi_3 \sim 6 \times 10^{-3}$ C/m$^2$K at $T_C^{APB}$ and $\Pi_1 \sim (1-3) \times 10^{-5}$ C/m$^2$ at hard APB.

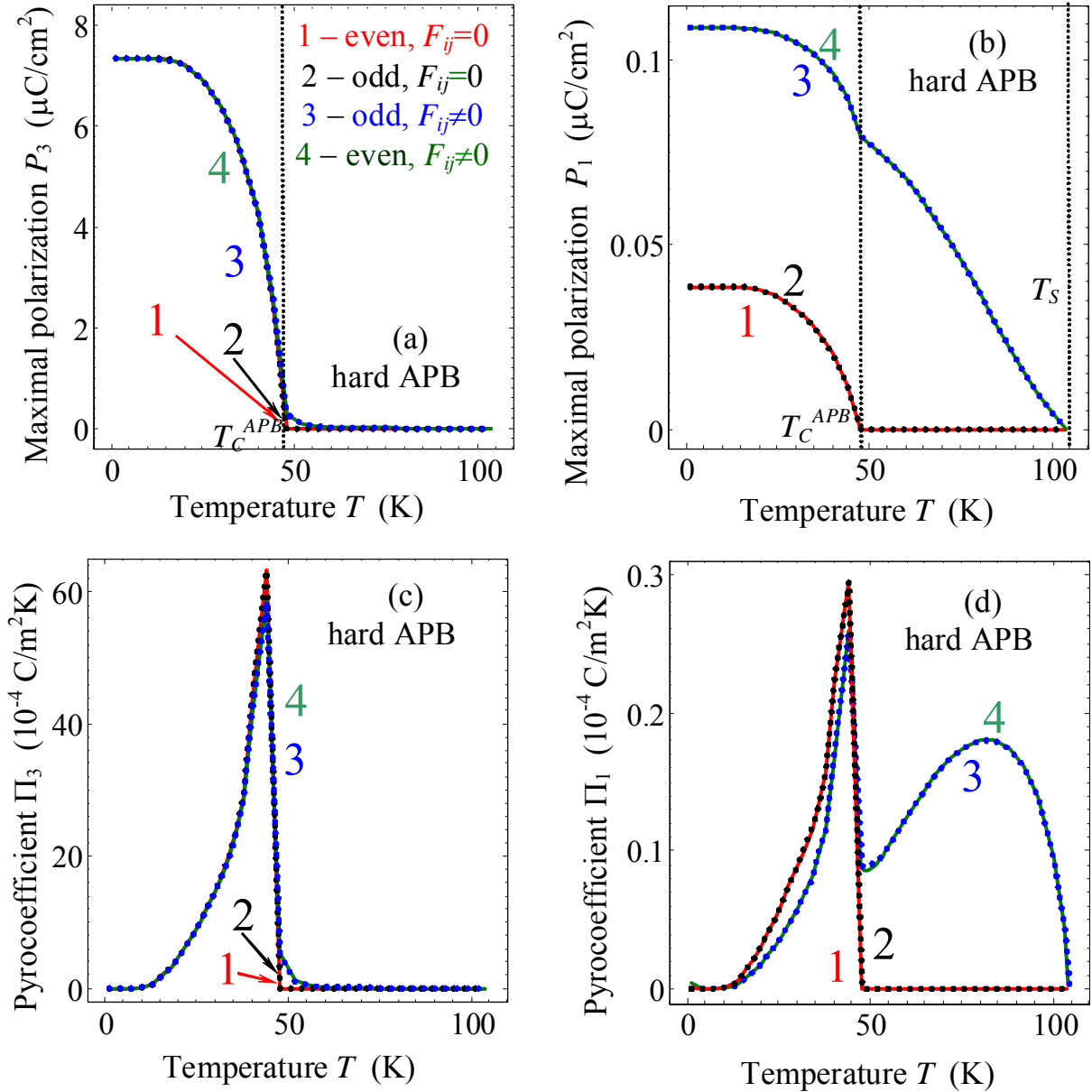

**Figure 5.** Temperature dependences of spontaneous polarization components $P_3$ and $P_1$ maximal values (a,b) and corresponding pyro-coefficient components $\Pi_3$ and $\Pi_1$ (c,d) calculated for hard APB in SrTiO$_3$ without free screening charges. Temperature dependences are calculated for nonzero flexoelectric effect $F_{ij} \neq 0$ and biquadratic coupling $\eta_{ij} \neq 0$ (curves 3, 4) and for the case of nonzero biquadratic coupling $\eta_{ij} \neq 0$ and zero flexoelectric effect $F_{ij} \equiv 0$ (curves 1, 2). Curves 1-4 style and color coding for plots (a-d) are the same and described in the legend to plot (a). Solid and dotted curves correspond to $P_3$-even and $P_3$-odd solutions respectively.

The principal difference between biquadratic coupling and flexo-roto effect influence on the polarization is the fact that the biquadratic coupling can induce bistable ferroelectric polarization inside antiphase boundaries, while the flexo-roto effect induces improper built-in polarization via the flexoelectric field (see **Fig. 6**). At temperatures lower than effective Curie temperature $T_C^{APB}$ polarization component $P_3(x_1)$ can be reversed by the localized electric field $E_3(x_1, x_3 = 0) = E_0 \exp(-x_1^2/2\rho_0^2)$ applied to the hard APB. The hysteresis disappears at higher temperatures. Loops are insensitive to the value of flexoelectric coefficient $F_{ij}$. Polarization component $P_1(x_1)$ does not show any bistable or hysteresis behavior entire the temperature range, since it is induced by the flexo-roto field $E_1^{FR}$, but not by the biquadratic coupling.

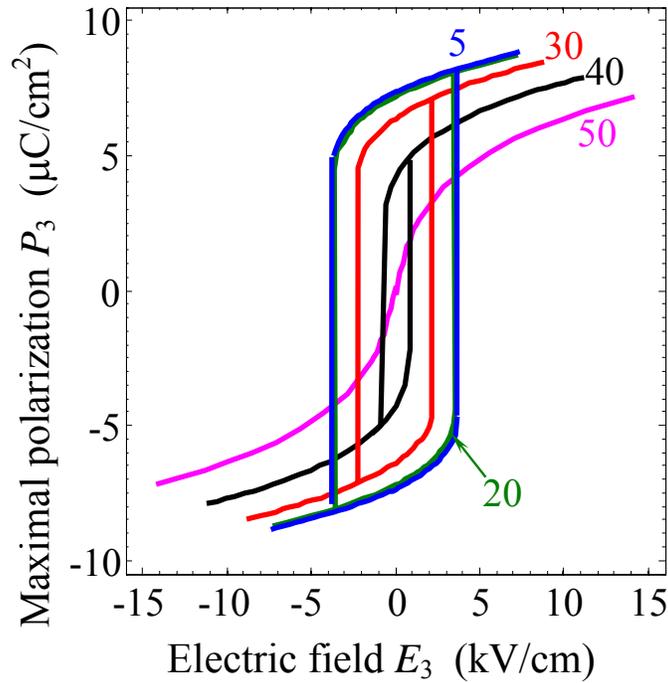

**Figure 6.** Polarization $P_3(x_1 = 0)$ vs. electric field $E_3(x_1, x_3 = 0) = E_0 \exp(-x_1^2/2\rho_0^2)$ applied to the hard APB in SrTiO$_3$ at different temperatures $T$= 5, 20, 30, 30, 40, 50 K (numbers near the curves). Contact radius $\rho_0 = 7$nm.

### *3.2. Flexo-roto-effect manifestation at 90-degree TB*

90-degree twins can have structure with rotation vector parallel (**Fig.7a**) or perpendicular (**Fig.7b**) to the domain wall plane in the immediate vicinity of the plane. Far from the wall the tilt vectors of twins are perpendicular. We will regard parallel twins as "**hard**" TB, since they have higher energy and perpendicular twins as "**easy**" TB, since they have lower energy as will be demonstrated in the section 3.3.

Let us consider domains "1" and "2" with different orientation of structural order parameters very far from the wall, $\mathbf{\Phi}^{(1)} = (\pm\Phi_S, 0, 0)$ and $\mathbf{\Phi}^{(2)} = (0, \Phi_S, 0)$, where different signs correspond to different orientation of the tilt arrows near domain walls. The spontaneous tilt component $\Phi_S(\mathbf{x}) \sim \sqrt{-b_1(T)}$ appears in the tilted phase at $T < T_S$ of bulk SrTiO$_3$. Introducing 45°-rotated coordinate system [49] with one axis, perpendicular to domain wall, $\tilde{z} \equiv x_3$, and two other rotated axes, $\tilde{x}_1 = (x_1 + x_2)/\sqrt{2}$, $\tilde{x}_2 = (-x_1 + x_2)/\sqrt{2}$, we write the tilt in the rotated system as $\tilde{\Phi}_1 = (\Phi_1 + \Phi_2)/\sqrt{2}$, $\tilde{\Phi}_2 = (-\Phi_1 + \Phi_2)/\sqrt{2}$ and $\tilde{\Phi}_3 \equiv 0$. Very far from the DW: $\tilde{\Phi}_1 = \pm\Phi_S/\sqrt{2}$ and $\tilde{\Phi}_2 = \mp\Phi_S/\sqrt{2}$ for **domain (1)**; $\tilde{\Phi}_1 = \Phi_S/\sqrt{2}$ and $\tilde{\Phi}_2 = \Phi_S/\sqrt{2}$ for **domain (2).** In the vicinity of the wall the tilt $\tilde{\mathbf{\Phi}} = \{\tilde{\Phi}_1(\tilde{x}_1), \tilde{\Phi}_2(\tilde{x}_1), 0\}$ induces nonzero distributions of polarization $\tilde{P}_1 = (P_1 + P_2)/\sqrt{2}$ and $\tilde{P}_2 = (-P_1 + P_2)/\sqrt{2}$, while $\tilde{P}_3 \equiv P_3 \equiv 0$ for the considered geometry of the problem.

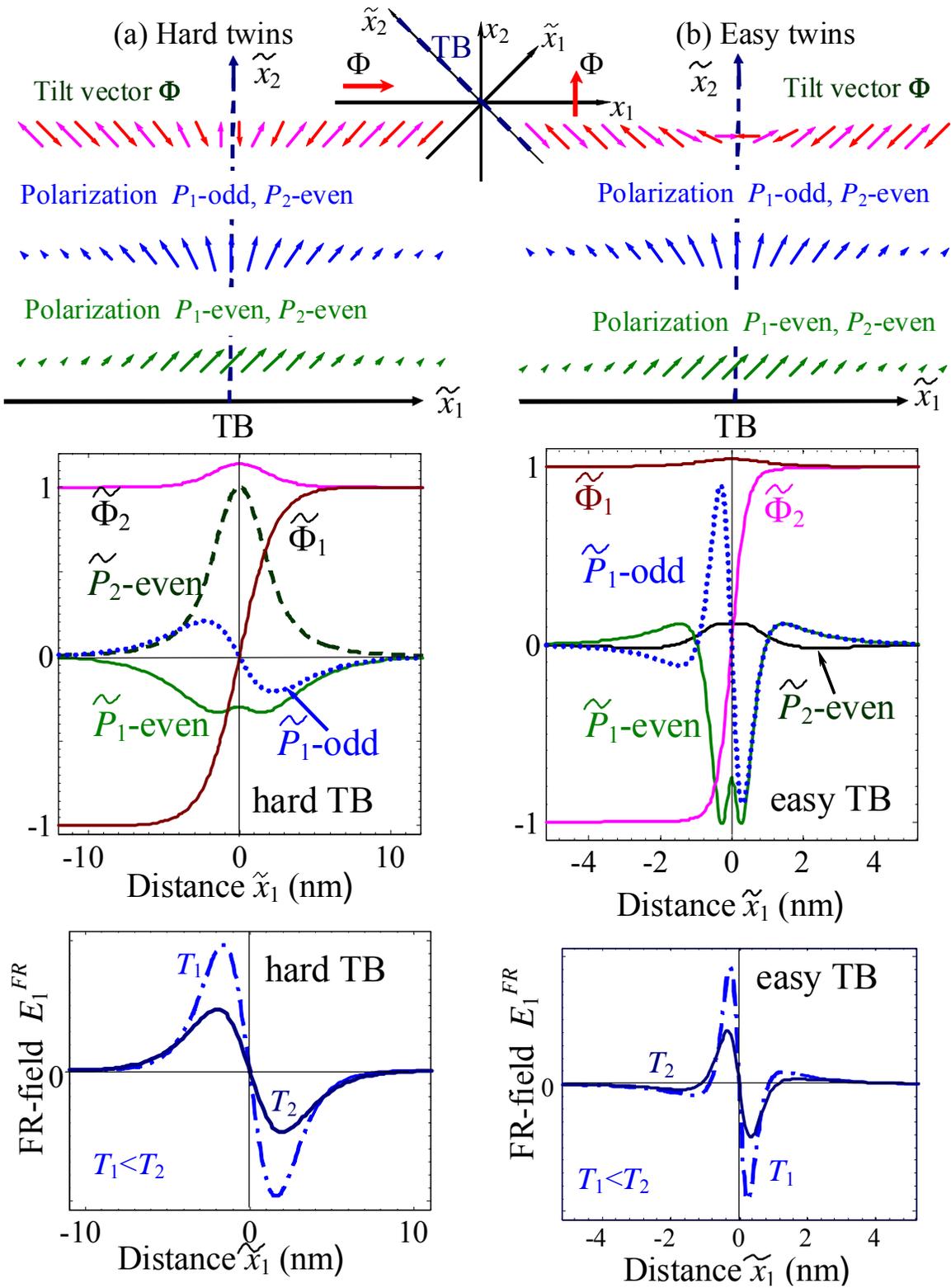

**Figure 7.** Schematics of 90-degree TB: rotation vector $\Phi$ is parallel (a) or perpendicular (b) to the domain wall plane in the immediate vicinity of the wall plane. Polarization appears inside the twins. TB plane $\tilde{x} = 0$ (denoted as TB-plane) is in the centre. Flexo-roto fields are shown at the bottom plots (calculated across $SrTiO_3$ twins for temperatures T=15 K and 80 K).

Using compatibility and mechanical equilibrium conditions the case of $\tilde{x}_1$-dependent solution we obtained the evident form of elastic strains and stresses in rotated system:

$$\tilde{\sigma}_{22} = \frac{s_{11}U_2 - s_{12}U_3}{s_{11}\tilde{s}_{11} - s_{12}^2}, \qquad \tilde{\sigma}_{33} = \frac{\tilde{s}_{11}U_3 - s_{12}U_2}{s_{11}\tilde{s}_{11} - s_{12}^2}, \qquad \tilde{\sigma}_{11} = \tilde{\sigma}_{13} = \tilde{\sigma}_{12} = \tilde{\sigma}_{23} = 0, \qquad (12a)$$

$$\tilde{u}_{22} = \left(\tilde{R}_{11}^{(\Phi)} + \tilde{R}_{12}^{(\Phi)}\right)\frac{\Phi_S^2}{2}, \qquad \tilde{u}_{33} = R_{12}^{(\Phi)}\Phi_S^2, \qquad (12b)$$

$$\tilde{u}_{11} = -\tilde{F}_{11}\frac{\partial \tilde{P}_1}{\partial \tilde{x}_1} + \tilde{Q}_{11}\tilde{P}_1^2 + \tilde{Q}_{12}\tilde{P}_2^2 + \tilde{R}_{11}^{(\Phi)}\tilde{\Phi}_1^2 + \tilde{R}_{12}^{(\Phi)}\tilde{\Phi}_2^2 + \frac{(\tilde{s}_{12}s_{11} - s_{12}^2)U_2 + (s_{12}\tilde{s}_{11} - \tilde{s}_{12}s_{12})U_3}{s_{11}\tilde{s}_{11} - s_{12}^2}, \qquad (12c)$$

$$\tilde{u}_{12} = \frac{1}{2}\left(-\tilde{F}_{66}\frac{\partial \tilde{P}_2}{\partial \tilde{x}_1} + \tilde{Q}_{66}\tilde{P}_1\tilde{P}_2 + \tilde{R}_{66}^{(\Phi)}\tilde{\Phi}_1\tilde{\Phi}_2\right), \quad \tilde{u}_{13} \equiv 0, \quad \tilde{u}_{23} \equiv 0. \qquad (12d)$$

Functions:

$$U_2 = \tilde{F}_{12}\frac{\partial \tilde{P}_1}{\partial \tilde{x}_1} + \tilde{R}_{11}^{(\Phi)}\left(\frac{\Phi_S^2}{2} - \tilde{\Phi}_2^2\right) + \tilde{R}_{12}^{(\Phi)}\left(\frac{\Phi_S^2}{2} - \tilde{\Phi}_1^2\right) - \tilde{Q}_{12}\tilde{P}_1^2 - \tilde{Q}_{11}\tilde{P}_2^2 \qquad (13a)$$

$$U_3 = F_{12}\frac{\partial \tilde{P}_1}{\partial \tilde{x}_1} + R_{12}^{(\Phi)}\left(\Phi_S^2 - \tilde{\Phi}_2^2 - \tilde{\Phi}_1^2\right) - Q_{12}\left(\tilde{P}_2^2 + \tilde{P}_1^2\right) \qquad (13b)$$

In Eqs.(12)-(13) we used the tensor components in the new reference frame for elastic compliances $\tilde{s}_{11} = (s_{11} + s_{12} + s_{44}/2)/2$, $\tilde{s}_{12} = (s_{11} + s_{12} - s_{44}/2)/2$, $\tilde{s}_{66} = 2(s_{11} - s_{12})$; rotostriction $\tilde{R}_{11}^{(\Phi)} = (R_{11}^{(\Phi)} + R_{12}^{(\Phi)} + R_{44}^{(\Phi)}/2)/2$, $\tilde{R}_{12}^{(\Phi)} = (R_{11}^{(\Phi)} + R_{12}^{(\Phi)} - R_{44}^{(\Phi)}/2)/2$, $\tilde{R}_{66}^{(\Phi)} = 2(R_{11}^{(\Phi)} - R_{12}^{(\Phi)})$; electrostriction $\tilde{Q}_{11} = (Q_{11} + Q_{12} + Q_{44}/2)/2$, $\tilde{Q}_{12} = (Q_{11} + Q_{12} - Q_{44}/2)/2$, $\tilde{Q}_{66} = 2(Q_{11} - Q_{12})$; flexoelectric coefficients $\tilde{F}_{11} = (F_{11} + F_{12} + F_{44})/2$, $\tilde{F}_{12} = (F_{11} + F_{12} - F_{44})/2$, $\tilde{F}_{66} = F_{11} - F_{12}$. Since polarization components are tends to zero far from the twin wall, the stresses (12a) vanish far from the wall; the strains (12b-d) tend to the spontaneous strains.

Substitution of the elastic solution (12) in Eqs.(3) leads the system of four coupled equations for the tilt and polarization vectors components:

$$2\left(b_1 - \tilde{R}_{12}\frac{s_{11}U_2 - s_{12}U_3}{s_{11}\tilde{s}_{11} - s_{12}^2} - R_{12}\frac{\tilde{s}_{11}U_3 - s_{12}U_2}{s_{11}\tilde{s}_{11} - s_{12}^2} + \tilde{b}_{12}\tilde{\Phi}_2^2 - \tilde{\eta}_{11}\tilde{P}_1^2 - \tilde{\eta}_{12}\tilde{P}_2^2\right)\tilde{\Phi}_1$$
$$+ 4\tilde{b}_{11}\tilde{\Phi}_1^3 - \tilde{\eta}_{66}\tilde{P}_1\tilde{P}_2\tilde{\Phi}_2 - \tilde{v}_{11}\frac{\partial^2 \tilde{\Phi}_1}{\partial \tilde{x}_1^2} = 0 \qquad (14a)$$

$$2\left(b_1 - \tilde{R}_{11}\frac{s_{11}U_2 - s_{12}U_3}{s_{11}\tilde{s}_{11} - s_{12}^2} - R_{12}\frac{\tilde{s}_{11}U_3 - s_{12}U_2}{s_{11}\tilde{s}_{11} - s_{12}^2} + \tilde{b}_{12}\tilde{\Phi}_1^2 - \tilde{\eta}_{11}\tilde{P}_2^2 - \tilde{\eta}_{12}\tilde{P}_1^2\right)\tilde{\Phi}_2$$
$$+ 4\tilde{b}_{11}\tilde{\Phi}_2^3 - \tilde{\eta}_{66}\tilde{P}_1\tilde{P}_2\tilde{\Phi}_1 - \tilde{v}_{66}\frac{\partial^2\tilde{\Phi}_2}{\partial \tilde{x}_1^2} = 0$$
(14b)

$$2\left(a_1 - \tilde{Q}_{12}\frac{s_{11}U_2 - s_{12}U_3}{s_{11}\tilde{s}_{11} - s_{12}^2} - Q_{12}\frac{\tilde{s}_{11}U_3 - s_{12}U_2}{s_{11}\tilde{s}_{11} - s_{12}^2} + \tilde{a}_{12}\tilde{P}_2^2 - \tilde{\eta}_{11}\tilde{\Phi}_1^2 - \tilde{\eta}_{12}\tilde{\Phi}_2^2\right)\tilde{P}_1 + 4\tilde{a}_{11}\tilde{P}_1^3$$
$$- \tilde{\eta}_{66}\tilde{\Phi}_2\tilde{\Phi}_1\tilde{P}_2 - \tilde{g}_{11}\frac{\partial^2\tilde{P}_1}{\partial \tilde{x}_1^2} = \tilde{E}_1^{ext} + \tilde{E}_1^d + F_{12}\frac{\partial}{\partial \tilde{x}_1}\left(\frac{s_{11}U_2 - s_{12}U_3}{s_{11}\tilde{s}_{11} - s_{12}^2}\right) + F_{12}\frac{\partial}{\partial \tilde{x}_1}\left(\frac{\tilde{s}_{11}U_3 - s_{12}U_2}{s_{11}\tilde{s}_{11} - s_{12}^2}\right)$$
(14c)

$$2\left(a_1 + \tilde{a}_{12}\tilde{P}_1^2 - \tilde{\eta}_{11}\tilde{\Phi}_2^2 - \tilde{\eta}_{12}\tilde{\Phi}_1^2 - \tilde{Q}_{11}\frac{s_{11}U_2 - s_{12}U_3}{s_{11}\tilde{s}_{11} - s_{12}^2} - Q_{12}\frac{\tilde{s}_{11}U_3 - s_{12}U_2}{s_{11}\tilde{s}_{11} - s_{12}^2}\right)\tilde{P}_2$$
$$+ 4\tilde{a}_{11}\tilde{P}_2^3 - \tilde{\eta}_{66}\tilde{\Phi}_1\tilde{\Phi}_2\tilde{P}_1 - \tilde{g}_{66}\frac{\partial^2\tilde{P}_2}{\partial \tilde{x}_1^2} = E_2^{ext}$$
(14d)

Here we introduced the tensor components in the new reference frame: $\tilde{b}_{11} = \frac{1}{4}(2b_{11} + b_{12})$,
$\tilde{b}_{12} = \frac{1}{2}(6b_{11} - b_{12})$, $\tilde{a}_{11} = \frac{1}{4}(2a_{11} + a_{12})$, $\tilde{a}_{12} = \frac{1}{2}(6a_{11} - a_{12})$, $\tilde{\eta}_{11} = (\eta_{11} + \eta_{12} + \eta_{44}/2)/2$,
$\tilde{\eta}_{12} = (\eta_{11} + \eta_{12} - \eta_{44}/2)/2$, $\tilde{\eta}_{66} = 2(\eta_{11} - \eta_{12})$, $\tilde{v}_{11} = (v_{11} + v_{12} + 2v_{44})/2$, $\tilde{v}_{66} = (v_{11} - v_{12})/2$,
$\tilde{g}_{11} = (g_{11} + g_{12} + 2g_{44})/2$, $\tilde{g}_{66} = (g_{11} - g_{12})/2$.

Polarization vector components are zero far from the TB

$$\tilde{P}_1(\tilde{x}_1 \to \pm\infty) = 0, \quad \tilde{P}_2(\tilde{x}_1 \to \pm\infty) = 0 \quad (15a)$$

Polarization component $\tilde{P}_2$ has zero derivatives at the TB (even type solution). Polarization component $\tilde{P}_1$ either has zero derivatives at the TB (even type solution) or is zero at the TB (odd type solution):

$$\frac{\partial \tilde{P}_1}{\partial \tilde{x}_1}(\tilde{x}_1 = 0) = 0, \quad \frac{\partial \tilde{P}_2}{\partial \tilde{x}_1}(\tilde{x}_1 = 0) = 0, \quad (\tilde{P}_1\text{-even and } \tilde{P}_2\text{-even}) \quad (15b)$$

$$\tilde{P}_1(\tilde{x}_1 = 0) = 0, \quad \frac{\partial \tilde{P}_2}{\partial \tilde{x}_1}(\tilde{x}_1 = 0) = 0. \quad (\tilde{P}_1\text{-odd and } \tilde{P}_2\text{-even}) \quad (15c)$$

Note that $\tilde{P}_2$-odd solution appeared unstable.

Boundary conditions for the tilt vector at **hard** twins (with rotation vector parallel to the wall plane as shown in **Fig.7a**) are $\tilde{\Phi}_1(\tilde{x}_1 = 0) = 0$, $\frac{\partial \tilde{\Phi}_2}{\partial \tilde{x}_1}(\tilde{x}_1 = 0) = 0$ at the TB and very far from TB:

$$\tilde{\Phi}_1(\tilde{x}_1 \to +\infty) = \mp \frac{\Phi_S}{\sqrt{2}}, \quad \tilde{\Phi}_1(\tilde{x}_1 \to -\infty) = \pm \frac{\Phi_S}{\sqrt{2}}, \quad \tilde{\Phi}_2(\tilde{x}_1 \to +\infty) = \frac{\Phi_S}{\sqrt{2}}, \quad \tilde{\Phi}_2(\tilde{x}_1 \to -\infty) = \frac{\Phi_S}{\sqrt{2}}, \quad (16a)$$

Boundary conditions for **easy** twins (with rotation vector perpendicular to the wall plane as shown in **Fig.7b**) are $\frac{\partial \tilde{\Phi}_1}{\partial \tilde{x}_1}(\tilde{x}_1 = 0) = 0$, $\tilde{\Phi}_2(\tilde{x}_1 = 0) = 0$ at the TB and very far from for TB:

$$\tilde{\Phi}_1(\tilde{x}_1 \to +\infty) = \tilde{\Phi}_1(\tilde{x}_1 \to -\infty) = \pm \frac{\Phi_S}{\sqrt{2}}, \quad \tilde{\Phi}_2(\tilde{x}_1 \to +\infty) = \pm \frac{\Phi_S}{\sqrt{2}}, \quad \tilde{\Phi}_2(\tilde{x}_1 \to -\infty) = \mp \frac{\Phi_S}{\sqrt{2}} \quad (16b)$$

The roto-flexo field

$$\tilde{E}_1^{FR} = \frac{\tilde{F}_{12}\left(s_{12} R_{12}^{(\Phi)} - s_{11} \tilde{R}_{11}^{(\Phi)}\right) + F_{12}\left(s_{12} \tilde{R}_{11}^{(\Phi)} - \tilde{s}_{11} R_{12}^{(\Phi)}\right)}{s_{11}\tilde{s}_{11} - s_{12}^2} \frac{\partial(\tilde{\Phi}_2^2)}{\partial \tilde{x}_1} + \frac{\tilde{F}_{12}\left(s_{12} R_{12}^{(\Phi)} - s_{11} \tilde{R}_{12}^{(\Phi)}\right) + F_{12}\left(s_{12} \tilde{R}_{12}^{(\Phi)} - \tilde{s}_{11} R_{12}^{(\Phi)}\right)}{s_{11}\tilde{s}_{11} - s_{12}^2} \frac{\partial(\tilde{\Phi}_1^2)}{\partial \tilde{x}_1} \quad (17)$$

appears in the right-hand-side of Eq.(14c). $\tilde{E}_1^{FR}(\tilde{x}_1)$ it is an odd function but has much more complex structure than the one for APBs, (see bottom **Figs.7a** and **7b** and compare them with the bottom **Figs.1a** and **1b**).

Coefficient $a_1^R = a_1 - \tilde{Q}_{12} \frac{s_{11} U_2 - s_{12} U_3}{s_{11}\tilde{s}_{11} - s_{12}^2} - Q_{12} \frac{\tilde{s}_{11} U_3 - s_{12} U_2}{s_{11}\tilde{s}_{11} - s_{12}^2} + \tilde{a}_{12} \tilde{P}_2^2 - \tilde{\eta}_{11} \tilde{\Phi}_1^2 - \tilde{\eta}_{12} \tilde{\Phi}_2^2 + \frac{1}{\varepsilon_0 \varepsilon_b}$ in

Eq.(14c) is positive at all temperatures due to the strong depolarization field $\tilde{E}_1^d(\tilde{x}_1) = -\tilde{P}_1(\tilde{x}_1)/(\varepsilon_0 \varepsilon_b)$ considered in the dielectric limit. Therefore polarization hysteresis for $\tilde{P}_1$ component is absent in external field $\tilde{E}_1^{ext}$. Depolarization field $\tilde{E}_1^d(\tilde{x})$ strongly reduces the component $\tilde{P}_1$ value.

Nonzero component $\tilde{P}_1$ perpendicular to the TB induces the component $\tilde{P}_2$ parallel to the TB just below $T_S$ due to the biquadratic coupling term $-\tilde{\eta}_{66} \tilde{\Phi}_1 \tilde{\Phi}_2 \tilde{P}_1$ in Eq.(14d). The trivial solution $\tilde{P}_1 \equiv \tilde{P}_2 \equiv 0$ does not exist in the vicinity of hard TB due to nonzero flexo-roto field $\tilde{E}_1^{FR} \neq 0$ and the coupling term $-\tilde{\eta}_{66} \tilde{\Phi}_1 \tilde{\Phi}_2 \tilde{P}_1$. Polarization component $\tilde{P}_2$ depends on the tilt vector due to the tilt-dependent coefficient $a_2^R = a_1 + \tilde{a}_{12} \tilde{P}_1^2 - \tilde{\eta}_{11} \tilde{\Phi}_2^2 - \tilde{\eta}_{12} \tilde{\Phi}_1^2 - \tilde{Q}_{11} \frac{s_{11} U_2 - s_{12} U_3}{s_{11}\tilde{s}_{11} - s_{12}^2} - Q_{12} \frac{\tilde{s}_{11} U_3 - s_{12} U_2}{s_{11}\tilde{s}_{11} - s_{12}^2}$ that becomes negative for SrTiO$_3$ at temperatures $T < T_C^{TB}$ due to the biquadratic coupling terms $-\tilde{\eta}_{11} \tilde{\Phi}_2^2 - \tilde{\eta}_{12} \tilde{\Phi}_1^2$ and elastic fields (12), where the effective Curie temperature $T_C^{TB} \approx 20$ K exists for hard twins. Therefore ferroelectric hysteresis for polarization component $\tilde{P}_2$ should exist in the temperature range $T < T_C^{TB}$. Below $T_C^{TB}$ the ferroelectric parallel component $\tilde{P}_2$ strongly enhances the perpendicular one $\tilde{P}_1$ via the biquadratic coupling term $\tilde{\eta}_{66}\tilde{\Phi}_2\tilde{\Phi}_1\tilde{P}_2$ in Eq.(14c), in the region $\tilde{P}_1$ is induced to the sum of the $\tilde{\eta}_{66}\tilde{\Phi}_2\tilde{\Phi}_1\tilde{P}_2$ and $\tilde{E}_1^{FR}$. Pyroelectric coefficients $\tilde{\Pi}_2 = d\tilde{P}_2/dT$ and $\tilde{\Pi}_1 = d\tilde{P}_1/dT$ should be nonzero in the temperature range $T < T_S$ as shown below.

Below we analyze numerical solutions of Eq.(14) obtained for SrTiO$_3$ material parameters and without free charges. Generally speaking polarization spatial distribution and its temperature behavior across TB is similar to the ones calculated for APB (compare **Figs.8-12** with **Figs.2-6**) and all comments made to **Figs.2-6** are qualitatively valid for TB. However the numerical values of polarization and pyroelectric coefficient for TBs are typically much smaller than for APBs (namely more than in 3 times for hard TB and up to 30 times for easy TB). The difference originated from the smaller effective flexoelectric field, which in turn originate from smaller stress gradients. Differences in the stress gradients originate from the different orientation of the tilt vector $\Phi$ inside the TB and APB.

The spontaneous polarization component $\tilde{P}_1$ inside easy TB are induced by the flexo-roto fields shown in **Figs.8a** and **8b** at temperatures below (c) and above (d) $T_0^{(E)}$. The component $\tilde{P}_1$ distributions inside the easy TB are shown in **Figs. 8c** and **8d** correspondingly. Polarization components are identically zero under the absence of the flexo-roto coupling (see curves 1,2 calculated at $F_{ij} \equiv 0$ and $\eta_{ij} \neq 0$). Polarization distributions calculated allowing for flexo-roto coupling (curves 3, 4) have another shape (extra sign change, smaller double maxima) and much lower maximum values ~5 nC/cm$^2$ that the ones calculated for easy APB (compare with **Figs.2**). Polarization amplitude strongly decreases with the temperature increase, but the shape of the curves remains almost the same.

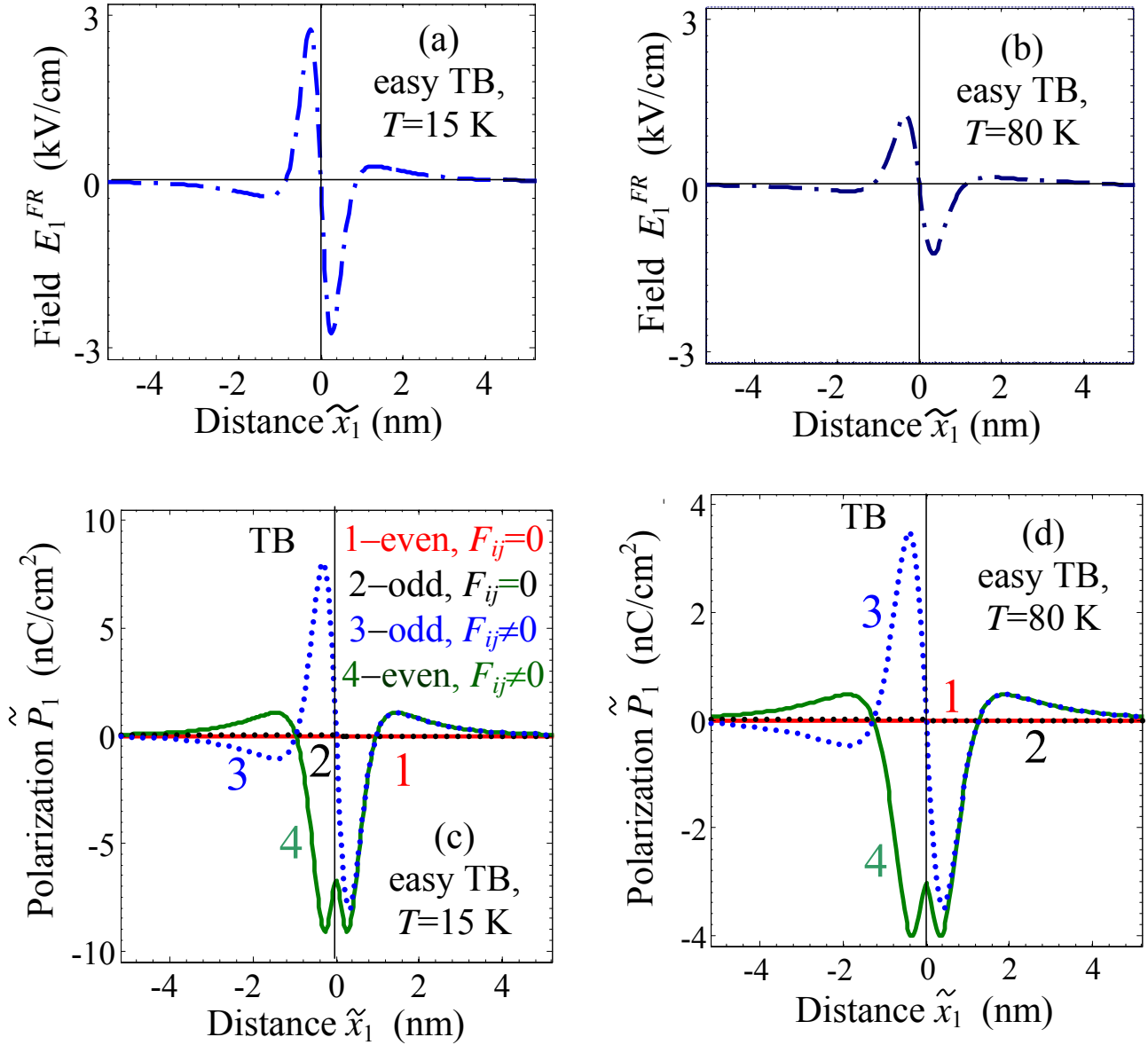

**Figure 8.** Flexo-roto field (a, b) induces spontaneous polarization (c,d) distributions across easy TB. Plots are calculated at temperatures $T=15$ K (a, c) and 80 K (b, d) for dielectric $SrTiO_3$ without free carriers. Perpendicular to APB component $\widetilde{P}_1$-odd (dotted curves) and $\widetilde{P}_1$-even (solid curves) are calculated for nonzero flexoelectric effect $F_{ij} \neq 0$ and biquadratic coupling $\eta_{ij} \neq 0$ (curves 3, 4) and for the case of nonzero biquadratic coupling $\eta_{ij} \neq 0$ and zero flexoelectric effect $F_{ij} \equiv 0$ (curves 1, 2). Curves 1-4 style and color coding for plots (c, d) are the same and described in the legend to plot (d).

Spontaneous polarization components $\widetilde{P}_1$ and $\widetilde{P}_2$ distributions inside hard TB are shown in **Figs. 9** at different temperatures. It is seen from the **Figs. 9** that polarization are identically zero above $T_C^{TB}$ under the absence of the flexo-roto coupling (see curves 1,2 calculated at $F_{ij} \equiv 0$ and $\eta_{ij} \neq 0$). Polarization amplitude decreases with the temperature increase, while the shape of the curves remains almost the same for $\tilde{P}_1$ (compare **Figs. 9a** and **9c**). Hard TB always has a significant polarization component $\tilde{P}_2$ parallel to the wall plane that is insensitive to the depolarization field. Polarization component $\tilde{P}_2$ is almost independent on the flexoelectric effect below $T_C^{TB}$, since the component is determined by the biquadratic coupling in the temperature range (curves 1-4 coincide in **Figs. 9b**). Component $\tilde{P}_2$ has a single maximum at the TB at low temperatures, which splits into the double maximums with the temperature increase (compare **Figs. 9b** and **9c**). A very small component $\tilde{P}_2$ exists at temperatures $T_C^{TB} < T < T_S$ due to the biquadratic coupling term. Similarly to the hard APB case, the double maximums at $\tilde{P}_2$-even distributions originate from the internal electric field $E_1^{FR}$ and exist at all temperatures below $T_S$. Polarization values induced by the flexo-roto effect and biquadratic coupling inside TB are essentially smaller than the one inside APB (compare **Figs.9** with **Figs.4**).

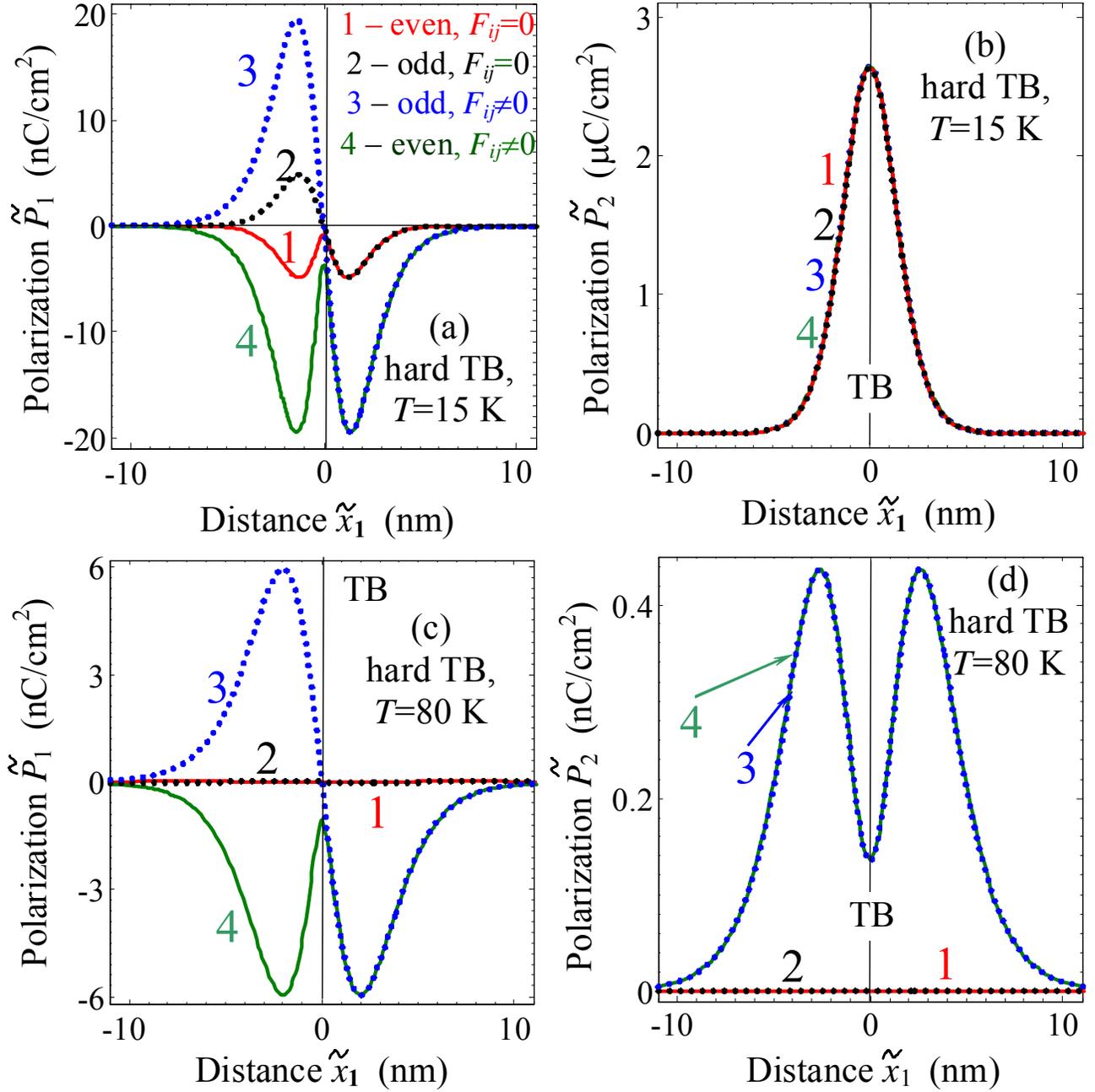

**Figure 9.** Spontaneous polarization distribution across hard TB calculated at temperatures $T$=20 K (a, b) and 80 K (c, d) for SrTiO$_3$ parameters without free carriers. Polarization components perpendicular ($\widetilde{P}_1$) and parallel ($\widetilde{P}_2$) to TB are shown. $\widetilde{P}_1$-odd (dotted curves) and $\widetilde{P}_1$-even (solid curves) are calculated for nonzero flexoelectric effect $F_{ij} \neq 0$ and biquadratic coupling $\eta_{ij} \neq 0$ (curves 3, 4) and for the case of nonzero biquadratic coupling $\eta_{ij} \neq 0$ and zero flexoelectric effect $F_{ij} \equiv 0$ (curves 1, 2). Curves 1-4 style and color coding for plots (a-d) are the same and described in the legend to plot (a).

Temperature dependences of the maximal spontaneous polarization values and pyroelectric coefficient inside easy TB are shown in **Figs. 10**. Under the absence of the flexo-roto field $\tilde{E}_1^{FR}$ spontaneous polarization and pyroelectric coefficient are zero (see curves 1,2 calculated at $F_{ij} \equiv 0$ and $\eta_{ij} \neq 0$). Curves temperature behavior is very similar to the ones calculated for easy APB (compare **Figs.10** with **Figs.3**) and all comments made to **Figs.3** are relevant to **Figs.10**. However the numerical values of polarization and pyroelectric coefficient across easy TB are essentially smaller (more than in 10 times) than across easy APB.

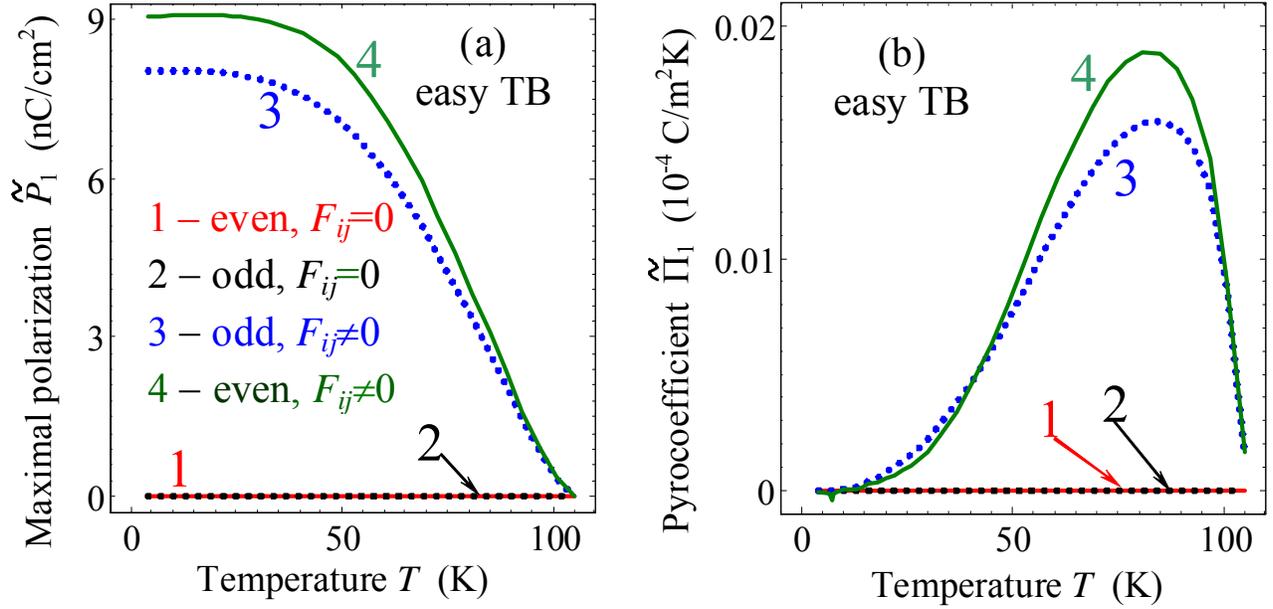

**Figure 10.** Temperature dependences of (a) the spontaneous polarization component $\tilde{P}_1$ maximal value and (b) corresponding pyroelectric coefficient $\tilde{\Pi}_1$ calculated for easy TB in SrTiO$_3$ without free carriers. Temperature dependences are calculated for nonzero flexoelectric effect $F_{ij} \neq 0$ and biquadratic coupling $\eta_{ij} \neq 0$ (curves 3, 4) and for the case of nonzero biquadratic coupling $\eta_{ij} \neq 0$ and zero flexoelectric effect $F_{ij} \equiv 0$ (curves 1, 2). Curves 1-4 style and color coding for plots (a, b) are the same and described in the legend to plot (a). Solid and dotted curves correspond to $\tilde{P}_1$-even and $\tilde{P}_1$-odd solutions respectively.

Temperature dependences of the maximal spontaneous polarization values at hard TB are shown in **Figs. 11a-b**. Under the absence of the flexoelectric field spontaneous polarization and pyroelectric coefficient are zero at temperatures higher than the effective Curie temperature $T_C^{TB}$ (see curves 1, 2 calculated at $F_{ij} \equiv 0$ and $\eta_{ij} \neq 0$). The flexo-roto effect rather weakly influences on the polarization component $\widetilde{P}_2$. For the case $F_{ij} \neq 0$ the component $\widetilde{P}_1 \sim E_1^{FR}$ appears below $T_S$, firstly quasi-linearly increases with temperature decrease, then non-linearly increases, then has a pronounced jump at $T_C^{TB}$ and then saturates at low temperatures $T \ll T_q$ due to the Barrett law for $b_1(T)$. The jump at $T_C^{TB}$ originates from the appearance of reversible ferroelectric polarization component $\widetilde{P}_2$ below $T_C^{TB}$. The maximal values of polarization are very close for odd and even types of solutions in the dielectric limit. Allowing for the flexo-roto-effect contribution we obtained $\widetilde{P}_2 \sim 2$ μC/cm² and $\widetilde{P}_1 \sim 0.02$ μC/cm² below $T_C^{TB}$. Without the flexo-roto-effect $\widetilde{P}_2$ is still ~2 μC/cm² at low temperatures, but $\widetilde{P}_1 < 0.005$ μC/cm².

Temperature dependences of the maximal pyroelectric coefficients $\widetilde{\Pi}_2$ and $\widetilde{\Pi}_1$ of hard TB are shown in **Figs. 11c-d**. Pyroelectric coefficients appear below $T_S$ only at nonzero flexoelectric coefficient $F_{ij} \neq 0$. Pyroelectric coefficient $\widetilde{\Pi}_1$ has two maximums: a smooth maximum at the polarization inflection point ~80 K and the sharp maximum at $T_C^{TB}$ originated from $\widetilde{P}_2$ appearance, since $\widetilde{P}_2$ enhances $\widetilde{P}_1$ via the biquadratic coupling term $\widetilde{\eta}_{66}\widetilde{\Phi}_2\widetilde{\Phi}_1\widetilde{P}_2$ in Eq.(14c). Pyroelectric coefficient $\widetilde{\Pi}_2$ has a single sharp maximum at $T_C^{TB}$ corresponding to the second order ferroelectric phase transition (appearance of the ferroelectric polarization $\widetilde{P}_2$).

Pyroelectric coefficients monotonically decrease below $T_C^{TB}$ with the temperatures decrease due to the spontaneous polarization components saturation at temperatures $T \ll T_q$ in accordance the Barrett law for $b_1(T)$ and $a_1(T)$. Allowing for the flexo-roto-effect contribution we obtained maximal pyroelectric coefficients $\widetilde{\Pi}_2 \sim 7 \times 10^{-3}$ C/m²K at $T_C^{TB}$ and $\widetilde{\Pi}_1 \sim (0.5\text{-}1.5) \times 10^{-3}$ C/m²K at hard TB below 80 K.

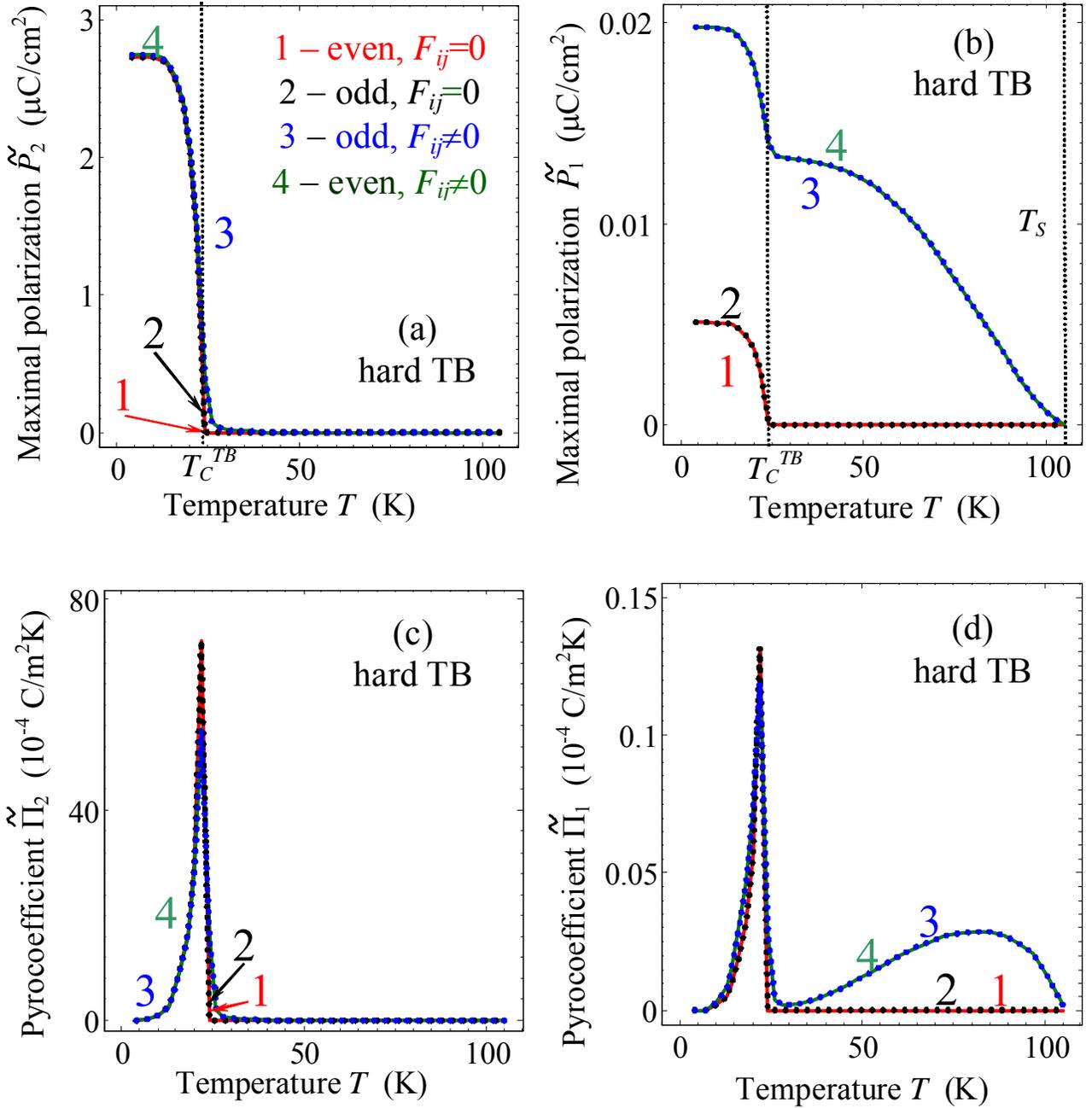

**Figure 11.** Temperature dependences of spontaneous polarization components $\tilde{P}_2$ and $\tilde{P}_1$ maximal values (a,b) and corresponding pyroelectric coefficient components $\tilde{\Pi}_2$ and $\tilde{\Pi}_1$ (c,d) calculated for hard TB in SrTiO$_3$ without free carriers. Temperature dependences are calculated for nonzero flexoelectric effect $F_{ij} \neq 0$ and biquadratic coupling $\eta_{ij} \neq 0$ (curves 3, 4) and for the case of nonzero biquadratic coupling $\eta_{ij} \neq 0$ and zero flexoelectric effect $F_{ij} \equiv 0$ (curves 1, 2). Curves 1-4 style and color coding for plots (a-d) are the same and described in the legend to plot (a). Solid and dotted curves correspond to $\tilde{P}_1$-even and $\tilde{P}_1$-odd solutions respectively.

Polarization component $\tilde{P}_2(\tilde{x}_1)$ can be reversed by the localized electric field $\tilde{E}_2(\tilde{x}_1) = E_0 \exp(-\tilde{x}_1^2/2\rho_0^2)$ applied to the TB only at temperatures lower than the effective Curie temperature $T_C^{TB}$ (see **Fig.12**). The hysteresis disappears at higher temperatures. Loops are insensitive to the value of flexoelectric coefficient $F_{ij}$. Polarization component $\tilde{P}_1(\tilde{x})$ does not show any bistable or hysteresis behavior entire the temperature range, since it is induced by the flexo-roto effect, but not by the biquadratic coupling.

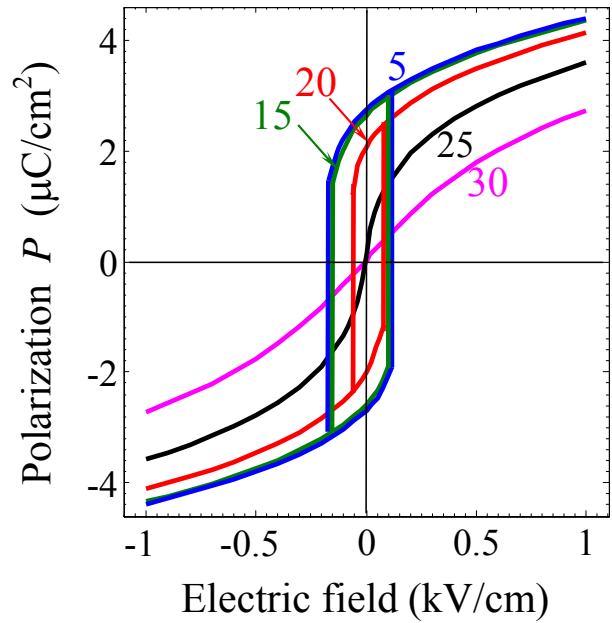

**Figure 12.** Polarization $\tilde{P}_2(\tilde{x}=0)$ vs. electric field $E_2(\tilde{x}, z=0) = E_0 \exp(-\tilde{x}^2/2\rho_0^2)$ applied to the hard TB of SrTiO$_3$ without free carriers. Numbers near the curves indicate different temperatures $T$= 5, 15, 20, 25, 30 K (numbers near the curves), $F_{ij} \neq 0$ and contact radius $\rho_0$ = 7nm.

To summarize the section 3, let us underline that pyroelectric response and polarization across TB and APB originate from the flexo-roto effect in the temperature range $T_0^{(E)} < T < T_S$. Since most likely 90-degree TBs appeared in SrTiO$_3$ just below $T_S$, our prediction of their spontaneous polarization appearance qualitatively explain the flexoelectric response features in the experiment of Zubko et al [23], and, more important, can be easy verified experimentally.

### *3.3. Energy of antiphase boundaries: elastic and electric contributions*

Temperature dependences of energies of easy and hard APB and TB are shown in **Fig.13.** APBs have higher energy than TBs (compare curves in **Fig.13a** and the ones in **Fig.13b**). Easy APB and TB are more energetically stable than the hard ones (compare upper and bottom curves in **Fig.13a,b**), but the energy difference decreases with the temperature increase and tends to zero at $T \to T_S$. Note, that existence of different types of antiphase boundaries in nature is not determined by energetic balance alone [20], it also depends on the spatial confinement as well as the nature of external stimulus.

The elastic energy contribution strongly dominates over the electric one as anticipated, since the tilt vector distribution across the wall is virtually independent on the polarization vector. Quantitatively, the tilt vector distribution determines more than 95% of the wall energy. The remainder ~5% depends on the electric energy, i.e. on the odd and/or even polarization components distribution. The total electric energy can be negative or positive, but anyway it only slightly changes the positive elastic energy. Appeared that odd distributions of the polarization component perpendicular to the wall are more energetically preferable than the even ones.

To show this, relative electric energies of the APBs counted from the lowest $P_3$-odd solution (curves 3 on plots (a) and (b)) are shown in **Figs. 13c** and **13d** for easy and hard APBs correspondingly. $P_3$-even polarization distributions have the highest energy, but the energy difference between all types of polarization solutions are very small; or TB (not shown) the difference is even smaller than for APB. However the electric energy difference between the odd and even distributions are small enough and disappear when approaching the structural phase transition temperature $T_S$. The fact can be realized from comparison of the absolute values of odd and even polarization distributions shown in **Figs.2, 4, 8, 9.** One could conclude from the figures that the difference between the absolute values of odd and even polarization distribution is very small (except the very enclosure of APB), so that the energies of these structures also should be close as determined mainly by the square of polarization components.

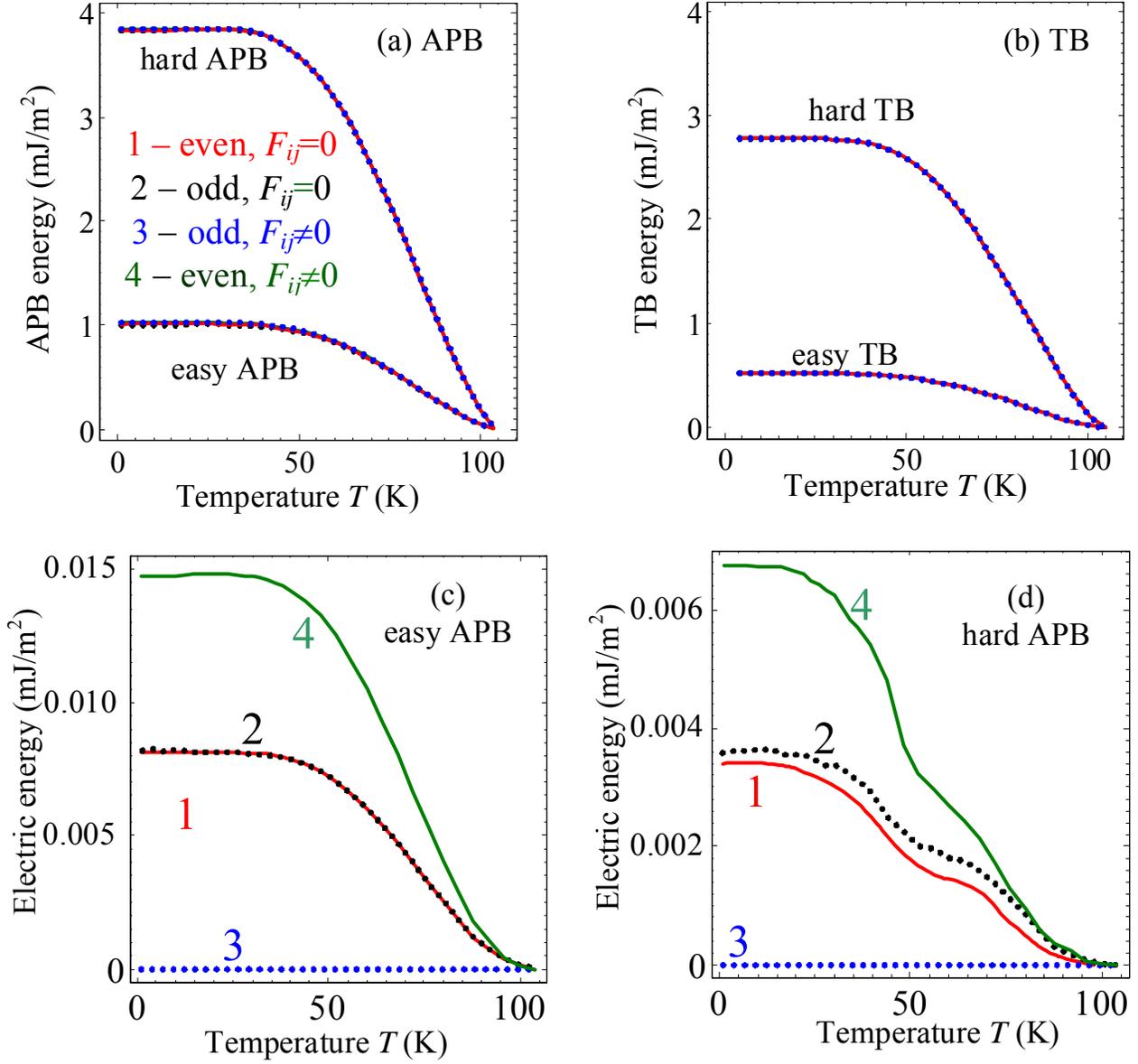

**Figure 13.** Energies of APB (a) and TB (b) boundaries vs. temperature calculated for SrTiO$_3$ without free carriers. Relative electric energies of easy (c) and hard (d) APB counted from the lowest $P_3$-odd solution (curves 3 on plots (a) and (b)). Temperature dependences are calculated for nonzero flexoelectric effect $F_{ij} \neq 0$ and biquadratic coupling $\eta_{ij} \neq 0$ (curves 3, 4) and for the case of nonzero biquadratic coupling $\eta_{ij} \neq 0$ and zero flexoelectric effect $F_{ij} \equiv 0$ (curves 1, 2). Solid and dotted curves correspond to $P_3$-even and $P_3$-odd solutions respectively. Curves 1-4 style and color coding for plots (a-d) are the same and described in the legend to plot (a).

## 4. Free carriers impact on the interfacial polarization and pyroelectric response

In the previous sections we consider $SrTiO_3$ without free carriers and so depolarization effects were maximal. However one can govern the free carriers concentration in $SrTiO_3$, as well as in other tilted perovskites, e.g. by varying concentration of the oxygen vacancies. For instance experimental data [47] report about weakly temperature dependent concentration of free carriers in $SrTiO_3$ that vary in the range $n_0=(10^{22} - 5\cdot10^{26})m^{-3}$ depending on the amount of oxygen vacancies. So the question about the influence of free carriers on the spontaneous polarization and pyroelectric response of the antiphase boundaries should be studied in details. In the section we perform numerical simulations of interfacial polarization and pyroelectric response for "relatively good" semiconductor $SrTiO_3$ with $n_0=(10^{24} - 10^{26})m^{-3}$ (but still not approaching the border of highly conductive media with $n_0>10^{28}m^{-3}$) and compare obtained dependences with ones calculated in the previous section for dielectric $SrTiO_3$.

For semiconductor $SrTiO_3$ the depolarization field $E_i^d = -\partial\varphi/\partial x_i$ inside the wall that should be determined not from Eq.(2), but from the Poisson equation:

$$\frac{\partial^2\varphi}{\partial x_i^2} = -\frac{e}{\varepsilon_0\varepsilon^b}\left(N_d^+(\varphi) - n(\varphi) + p(\varphi)\right) + \frac{\partial P_i}{\varepsilon_0\varepsilon^b \partial x_i}, \quad (18)$$

Ionized shallow donors $N_d^+(\varphi)$ (e.g. oxygen vacancies), and free electrons $n(\varphi)$ and holes $p(\varphi)$ concentrations $\varphi$-dependence were taken the same as in Ref.[50]. In Debye approximation $-\frac{e}{\varepsilon_0\varepsilon^b}\left(N_d^+ - n + p\right) \approx \frac{\varphi}{R_d^2}$, where $R_d = \sqrt{\varepsilon^b\varepsilon_0 k_B T/(2e^2 n_0)}$ is the "net" Debye screening radius, reflecting the screening by free carriers accumulated around the domain walls, $n_0$ is the equilibrium average concentration of the free carriers, $k_B=1.3807\times10^{-23}$ J/K, where $T$ is the absolute temperature. We checked the applicability of Debye approximation in Eq.(18) during the numerical simulations of Eqs.(3) and appeared that $|e\varphi| < k_B T$ up to 10 K, making Debye approximation self-consistent in the wide temperature range.

Spontaneous polarization distributions across easy and hard APB are shown in **Figs. 14** and **16** correspondingly. Temperature dependences of the spontaneous polarization maximal value and corresponding pyroelectric response coefficients are shown in **Figs. 15** and **17** correspondingly. Spontaneous polarization components distributions across easy and hard TB are shown in **Figs. 18** and **20** correspondingly. Temperature dependences of the polarization maximal values and corresponding pyroelectric coefficients are shown in **Figs. 19** and **21** correspondingly.

Interfacial polarization spatial distribution and its temperature behavior calculated for semiconductor SrTiO$_3$ with $n_0$=(10$^{24}$ – 10$^{26}$)m$^{-3}$ appeared semi-qualitatively similar to the ones calculated dielectric SrTiO$_3$. Naturally, free electrons and mobile oxygen vacancies effectively screen the depolarization field across polar interfaces and strongly enhance the components of interfacial polarization conjugated with depolarization field. Thus several other important differences exist:

1) Numerical values of polarization across easy APB and easy TB appeared much higher (up to 100 times for $n_0$=10$^{26}$ m$^{-3}$!) for semiconductor SrTiO$_3$ than for the dielectric one (compare **Figs. 14-15** with **Figs. 2-3**). The polarization component, that is perpendicular to the wall plane, increases with the carrier concentration increase due to much smaller depolarization field, which decrease comes from the screening carriers.

2) The values of polarization component parallel to the hard APB or hard TB plane also increase (up to 10 times at higher temperatures) with the carrier concentration increase. Since the component is not directly affected by the depolarization field the increase originated from the coupling with parallel component (compare **Figs.16c,d** and **20c,d** with **Figs.4b,d** and **9b,d** correspondingly).

3) The wall width of easy APB and TB (as well as the width of the polarization component perpendicular to the hard APB or TB plane) significantly increases in the semiconductor SrTiO$_3$, up to 5 times for the polarization component perpendicular to the wall plane (compare *x*-axes scale in **Figs. 14, 16, 18** with the ones in **Figs.2, 4, 8**). The reason for the width increase with free carriers concentration increase is the depolarization field decrease. The parallel component depends on the perpendicular one via the biquadratic coupling terms, therefore the wall width of the polarization component parallel to the hard APB and hard TB also increases with the carriers concentration increase, but the effect is weaker.

4) Temperature dependence of the maximal polarization at easy APB or TB (as well as the polarization component perpendicular to the hard APB or TB plane) has no saturation at low temperatures in semiconductor SrTiO$_3$, in contrast to the dielectric SrTiO$_3$ (see e.g. **Figs. 15, 17b, 19a, 21** and compare them with **Figs. 3, 5b, 10, 11b**). Moreover, maximal polarization of easy antiphase boundaries super-linearly increases with the temperature decrease. The increase originated from the decrease of the depolarization field with the temperature increase. Really, in accordance with Debye equation $\dfrac{\partial^2 \varphi}{\partial x_i^2} \approx \dfrac{\varphi}{R_d^2} + \dfrac{\partial P_i}{\varepsilon_0 \varepsilon^b \partial x_i}$ for depolarization field $E_i^d = -\partial\varphi/\partial x_i$, the field decreases with the temperature increase, since the Debye screening radius $R_d = \sqrt{\varepsilon^b \varepsilon_0 k_B T/(2e^2 n_0)}$ decreases.

5) Pyroelectric coefficients increases with free carrier concentration increase and demonstrate additional peculiarities (maxima and minima) at temperatures below the effective Curie temperatures. Pyroelectric coefficients of easy antiphase boundaries super-linearly increases with the temperature decrease, since the polarization increases with the temperature decrease via the decrease of the Debye screening radius.

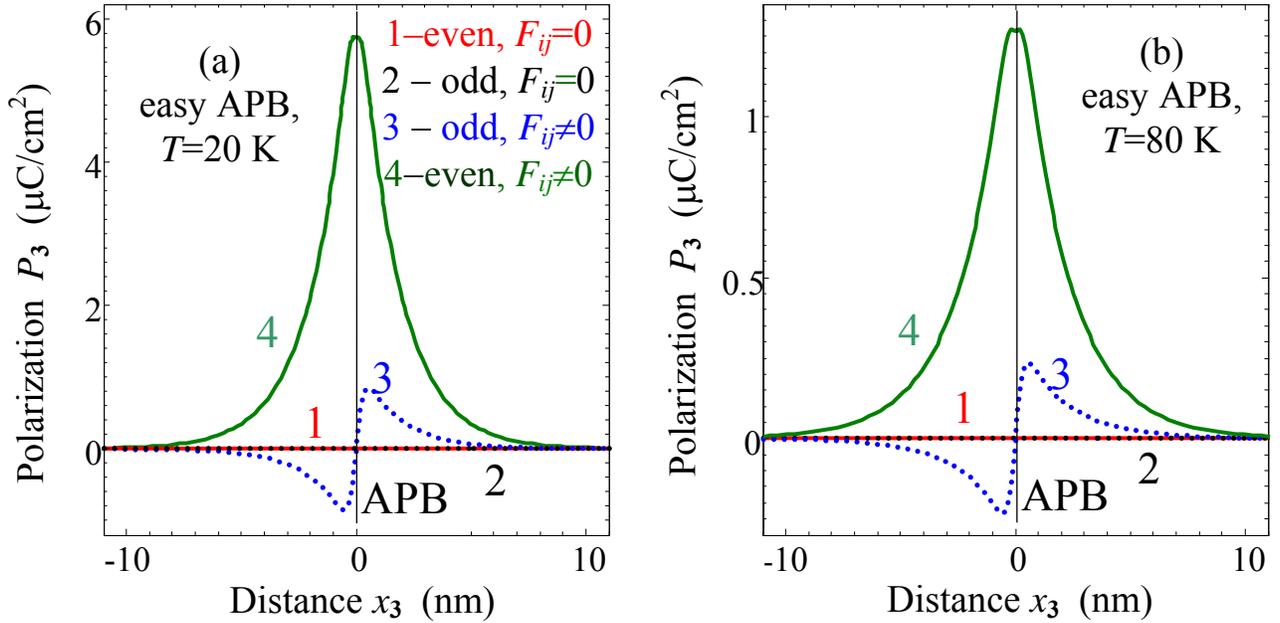

**Figure 14.** Spontaneous polarization distribution across easy APB calculated at temperatures $T$=20 K (a) and 80 K (b) for $SrTiO_3$ parameters; concentration of free carriers $n_0=10^{26}$ m$^{-3}$. Perpendicular to APB component $P_3$-odd (dotted curves) and $P_3$-even (solid curves) are calculated for nonzero flexoelectric effect $F_{ij} \neq 0$ and biquadratic coupling $\eta_{ij} \neq 0$ (curves 3, 4) and for the case of nonzero biquadratic coupling $\eta_{ij} \neq 0$ and zero flexoelectric effect $F_{ij} \equiv 0$ (curves 1, 2).

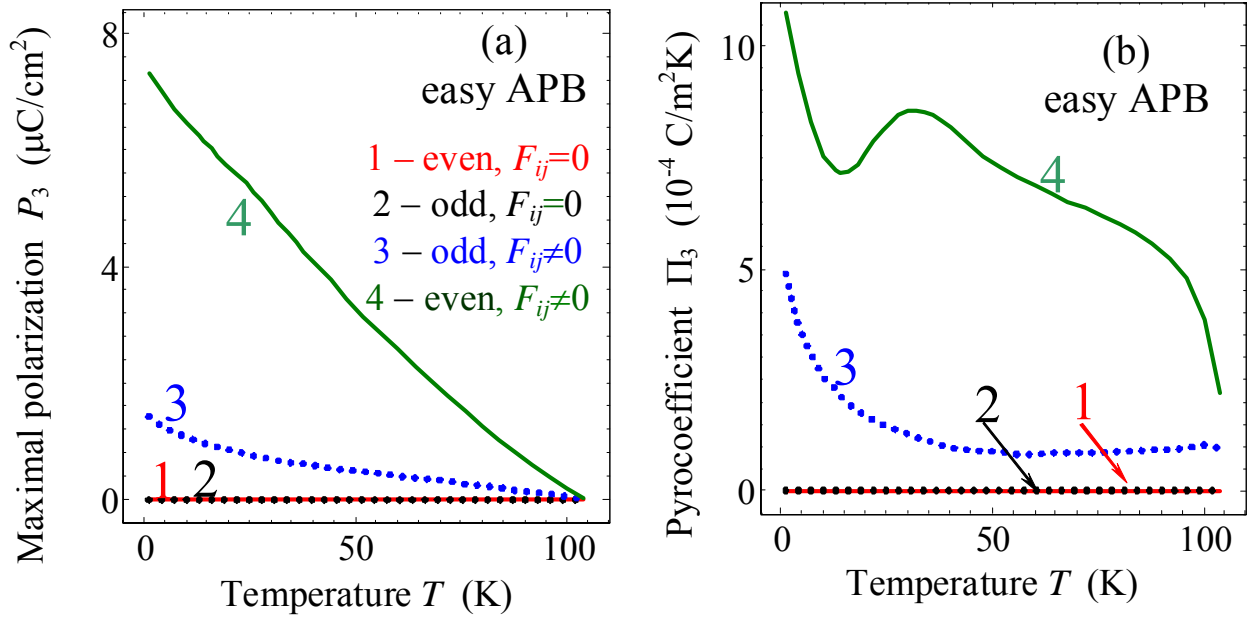

**Figure 15.** Temperature dependences of the spontaneous polarization maximal value (a) and (b) corresponding pyroelectric coefficient $\Pi_3$ calculated at easy APB in SrTiO$_3$ with concentration of free carriers $n_0=10^{26}$ m$^{-3}$. Temperature dependences are calculated for nonzero flexoelectric effect $F_{ij} \neq 0$ and biquadratic coupling $\eta_{ij} \neq 0$ (curves 3, 4) and for the case of nonzero biquadratic coupling $\eta_{ij} \neq 0$ and zero flexoelectric effect $F_{ij} \equiv 0$ (curves 1, 2). Curves 1-4 style and color coding for plots (a, b) are the same and described in the legend to plot (a). Solid and dotted curves correspond to $P_3$-even and $P_3$-odd solutions respectively.

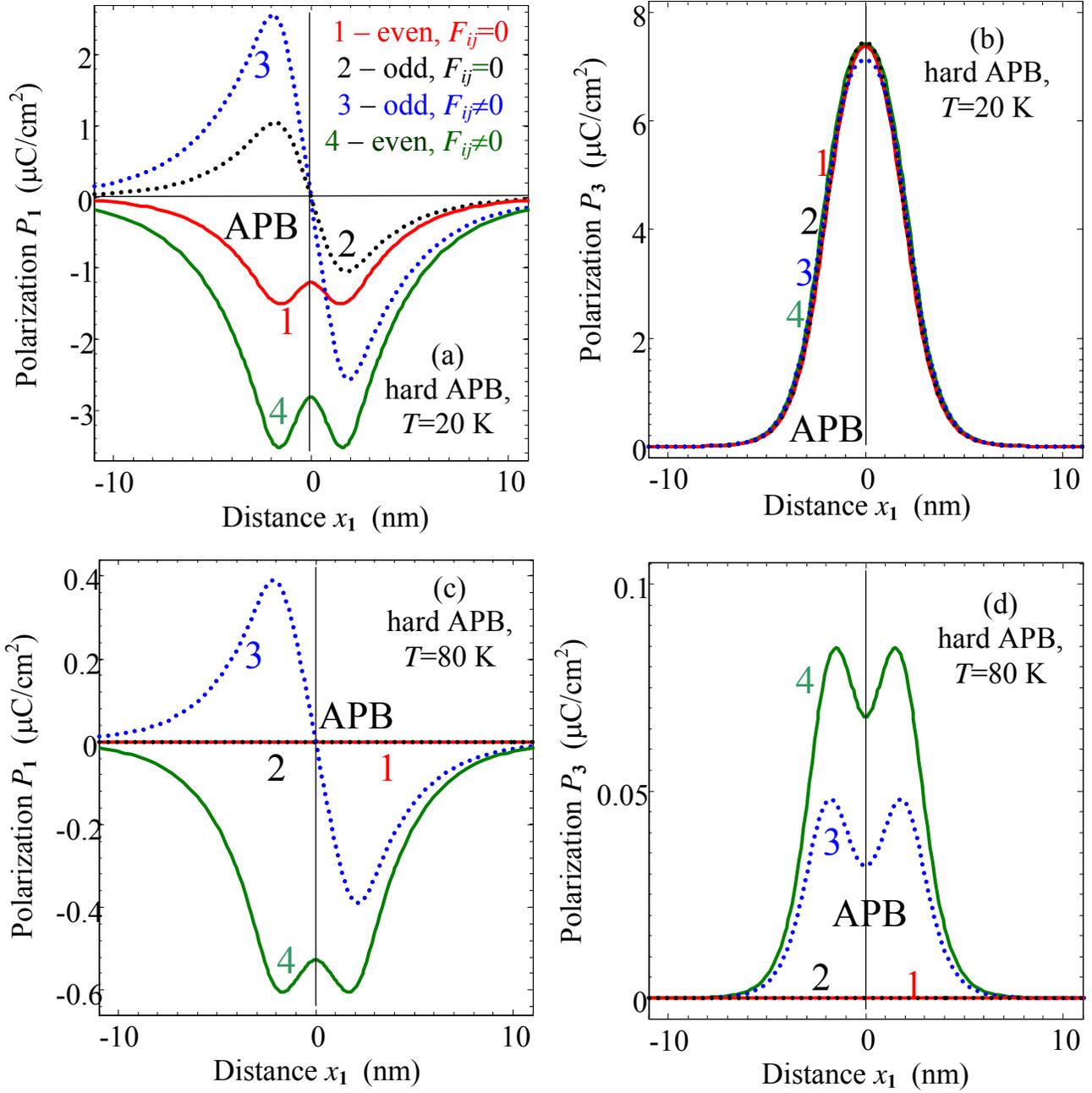

**Figure 16.** Spontaneous polarization distribution across hard APB calculated at temperatures $T$=20 K (a, b) and 80 K (c, d) for SrTiO$_3$ parameters, concentration of free carriers $n_0$=10$^{26}$ m$^{-3}$. Polarization components perpendicular ($P_1$) and parallel ($P_3$) to APB are shown. $P_1$-odd (dotted curves) and $P_1$-even (solid curves) are calculated for nonzero flexoelectric effect $F_{ij} \neq 0$ and biquadratic coupling $\eta_{ij} \neq 0$ (curves 3, 4) and for the case of nonzero biquadratic coupling $\eta_{ij} \neq 0$ and zero flexoelectric effect $F_{ij} \equiv 0$ (curves 1, 2). Curves 1-4 style and color coding for plots (a-d) are the same and described in the legend to plot (a).

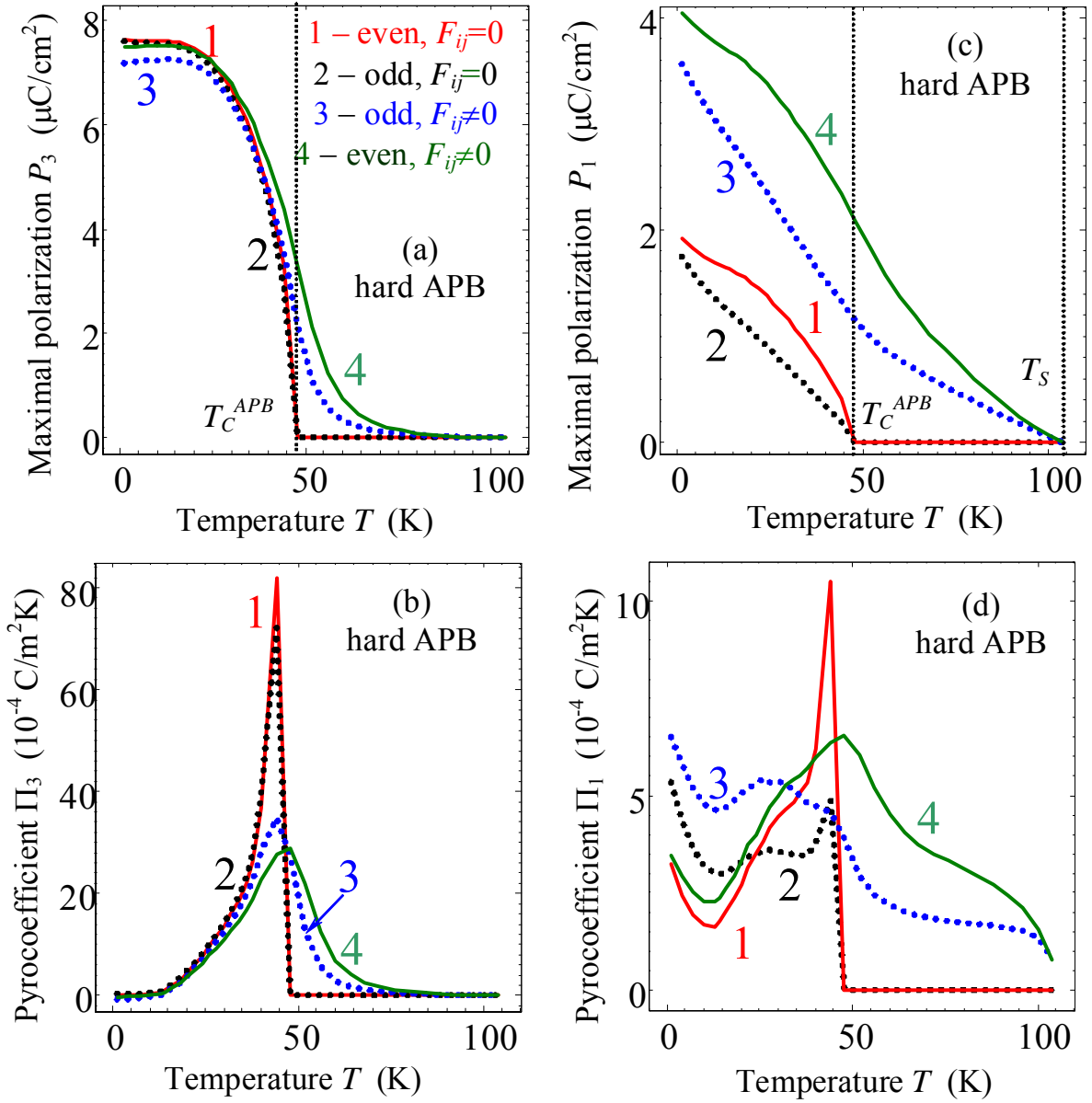

**Figure 17.** Temperature dependences of maximal values of spontaneous polarization components $P_3$ and $P_1$ (a,b) and corresponding pyroelectric coefficient components $\Pi_3$ and $\Pi_1$ (c,d) calculated at hard APB for SrTiO$_3$, concentration of free carriers $n_0=10^{26}$ m$^{-3}$. Temperature dependences are calculated for nonzero flexoelectric effect $F_{ij} \neq 0$ and biquadratic coupling $\eta_{ij} \neq 0$ (curves 3, 4) and for the case of nonzero biquadratic coupling $\eta_{ij} \neq 0$ and zero flexoelectric effect $F_{ij} \equiv 0$ (curves 1, 2). Curves 1-4 style and color coding for plots (a, b) are the same and described in the legend to plot (a). Solid and dotted curves correspond to $P_3$-even and $P_3$-odd solutions respectively.

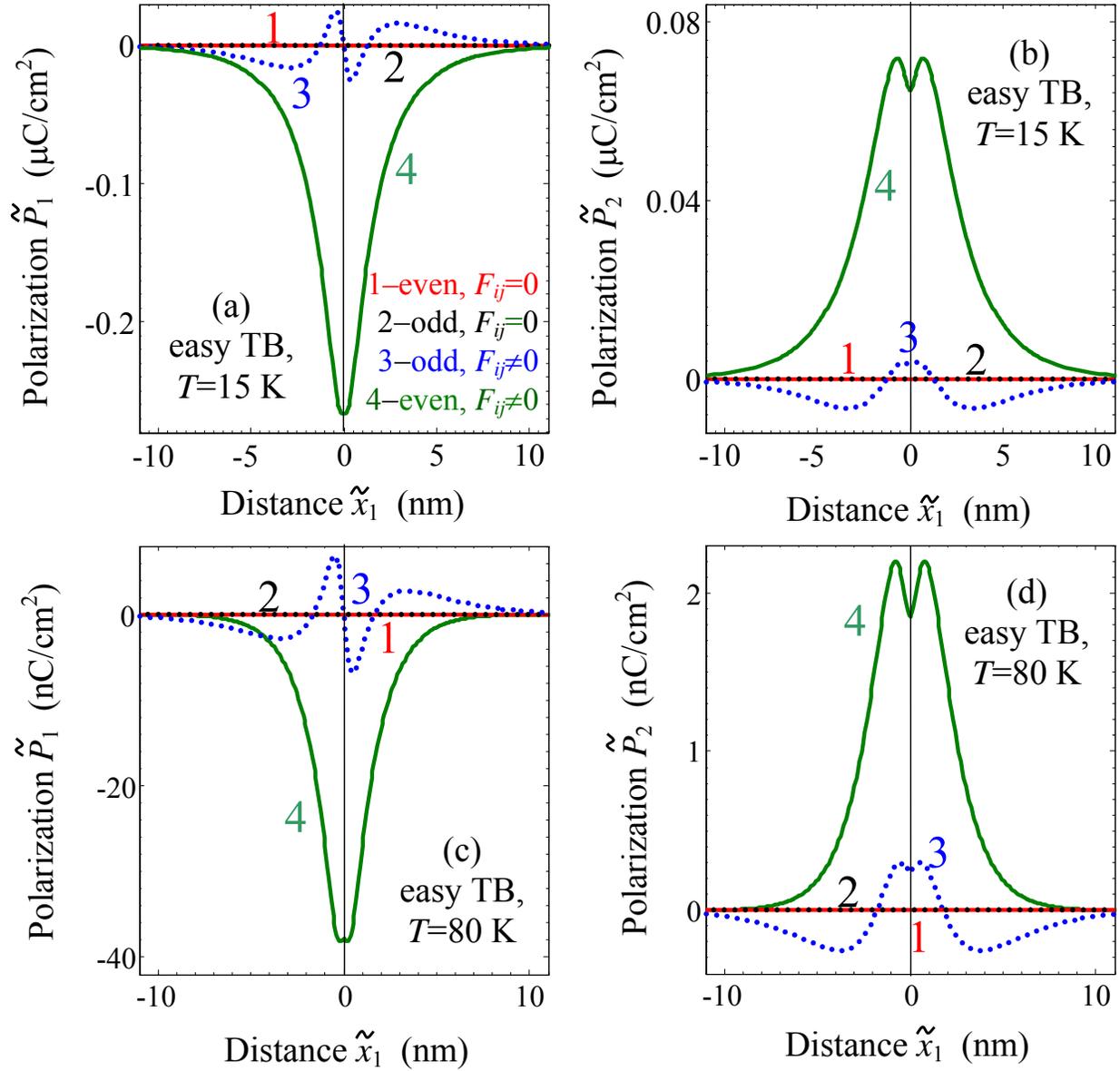

**Figure 18.** Spontaneous polarization distribution across easy TB calculated at temperatures $T$=15 K (a, b) and 80 K (c, d) for SrTiO$_3$ parameters, concentration of free carriers $n_0$=10$^{26}$ m$^{-3}$. Polarization components perpendicular ($\widetilde{P}_1$) and parallel ($\widetilde{P}_2$) to TB are shown. $\widetilde{P}_1$-odd (dotted curves) and $\widetilde{P}_1$-even (solid curves) are calculated for nonzero flexoelectric effect $F_{ij} \neq 0$ and biquadratic coupling $\eta_{ij} \neq 0$ (curves 3, 4) and for the case of nonzero biquadratic coupling $\eta_{ij} \neq 0$ and zero flexoelectric effect $F_{ij} \equiv 0$ (curves 1, 2). Curves 1-4 style and color coding for plots (a-d) are the same and described in the legend to plot (a).

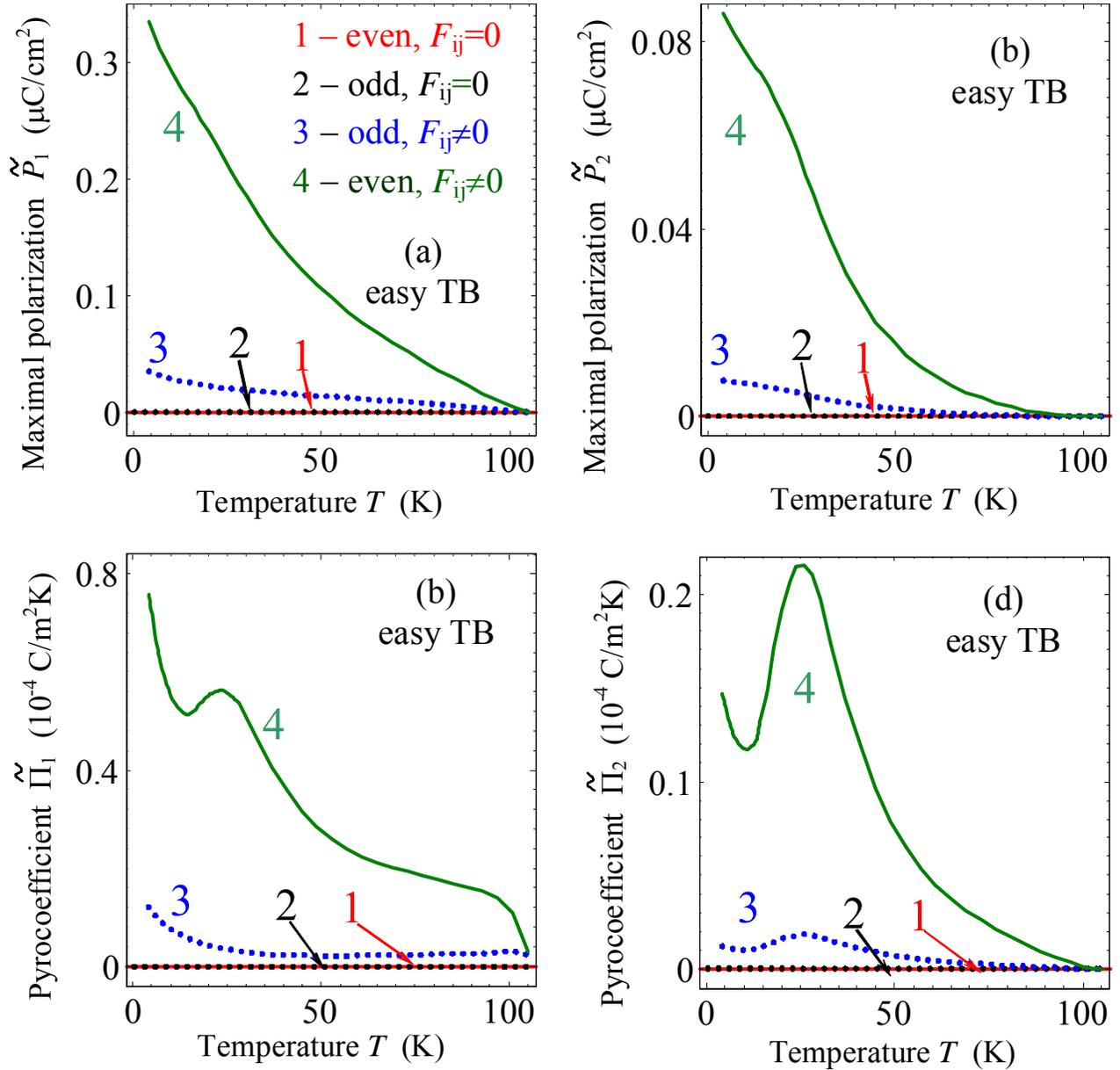

**Figure 19.** Temperature dependences of spontaneous polarization components $\tilde{P}_1$ and $\tilde{P}_2$ maximal values (a,b) and corresponding pyroelectric coefficient components $\tilde{\Pi}_1$ and $\tilde{\Pi}_2$ (c,d) calculated at easy TB for SrTiO$_3$ parameters, concentration of free carriers $n_0=10^{26}$ m$^{-3}$. Temperature dependences are calculated for nonzero flexoelectric effect $F_{ij} \neq 0$ and biquadratic coupling $\eta_{ij} \neq 0$ (curves 3, 4) and for the case of nonzero biquadratic coupling $\eta_{ij} \neq 0$ and zero flexoelectric effect $F_{ij} \equiv 0$ (curves 1, 2). Curves 1-4 style and color coding for plots (a, b) are the same and described in the legend to plot (a). Solid and dotted curves correspond to $\tilde{P}_1$-even and $\tilde{P}_1$-odd solutions respectively.

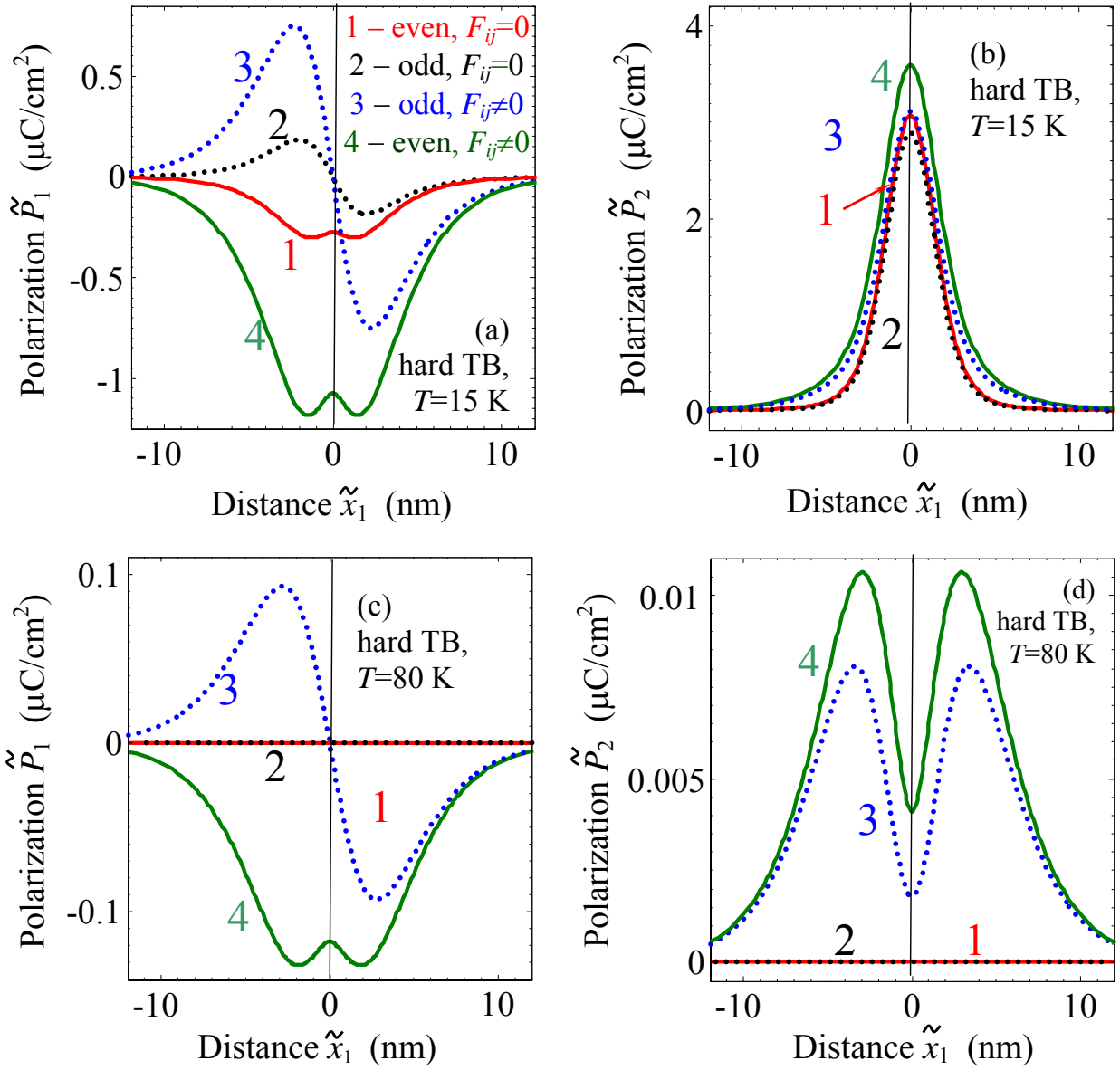

**Figure 20.** Spontaneous polarization distribution across hard TB calculated at temperatures $T=15$ K (a, b) and 80 K (c, d) for SrTiO$_3$ parameters, concentration of free carriers $n_0=10^{26}$ m$^{-3}$. Polarization components perpendicular ($\widetilde{P}_1$) and parallel ($\widetilde{P}_2$) to TB are shown. $\widetilde{P}_1$-odd (dotted curves) and $\widetilde{P}_1$-even (solid curves) are calculated for nonzero flexoelectric effect $F_{ij} \neq 0$ and biquadratic coupling $\eta_{ij} \neq 0$ (curves 3, 4) and for the case of nonzero biquadratic coupling $\eta_{ij} \neq 0$ and zero flexoelectric effect $F_{ij} \equiv 0$ (curves 1, 2). Curves 1-4 style and color coding for plots (a-d) are the same and described in the legend to plot (a).

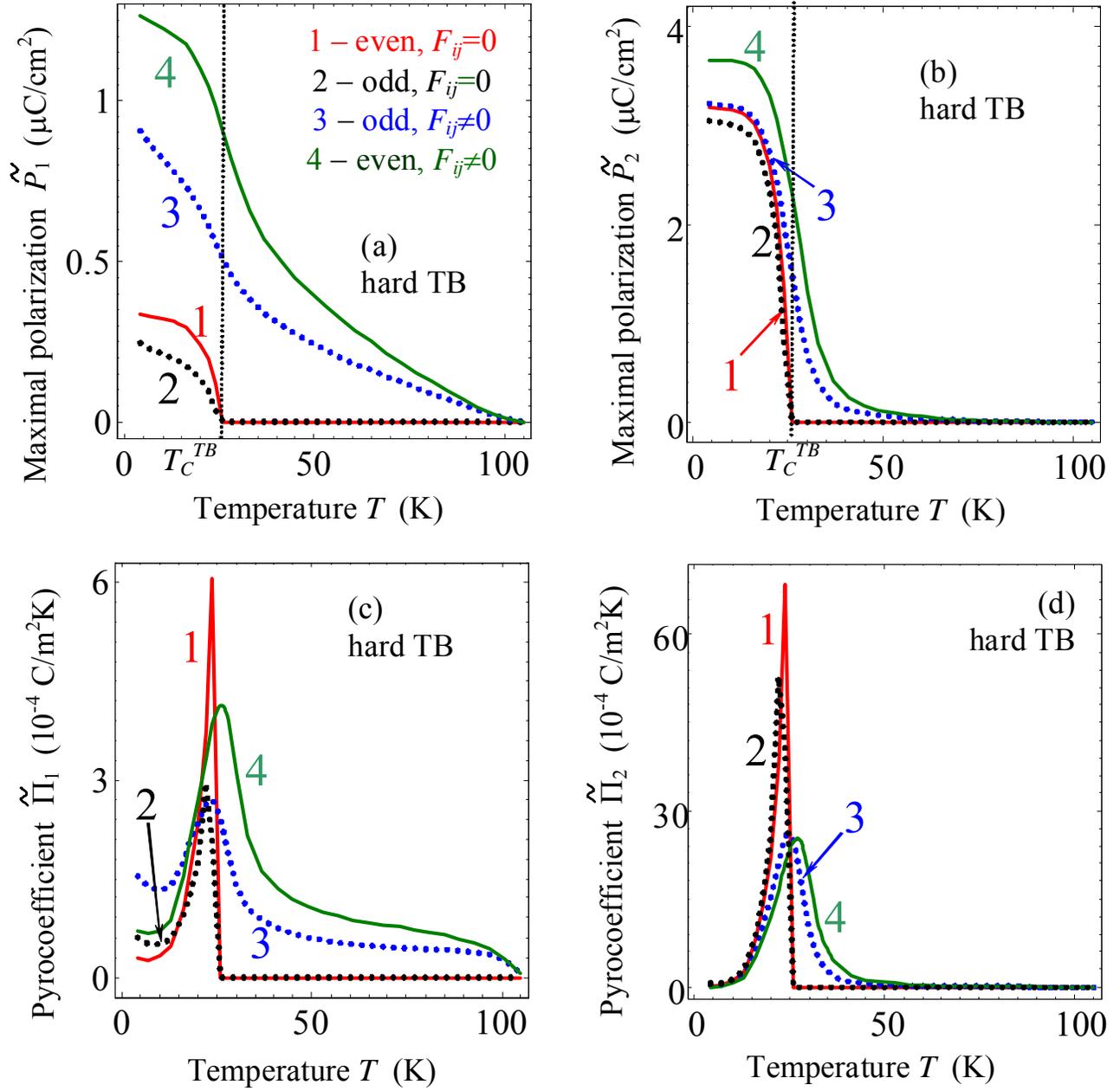

**Figure 21.** Temperature dependences of spontaneous polarization components $\tilde{P}_1$ and $\tilde{P}_2$ maximal values (a,b) and corresponding pyroelectric coefficient components $\tilde{\Pi}_1$ and $\tilde{\Pi}_2$ (c,d) calculated at hard TB for SrTiO$_3$ parameters, concentration of free carriers $n_0 = 10^{26}$ m$^{-3}$. Temperature dependences are calculated for nonzero flexoelectric effect $F_{ij} \neq 0$ and biquadratic coupling $\eta_{ij} \neq 0$ (curves 3, 4) and for the case of nonzero biquadratic coupling $\eta_{ij} \neq 0$ and zero flexoelectric effect $F_{ij} \equiv 0$ (curves 1, 2). Curves 1-4 style and color coding for plots (a, b) are the same and described in the legend to plot (a). Solid and dotted curves correspond to $\tilde{P}_1$-even and $\tilde{P}_1$-odd solutions respectively.

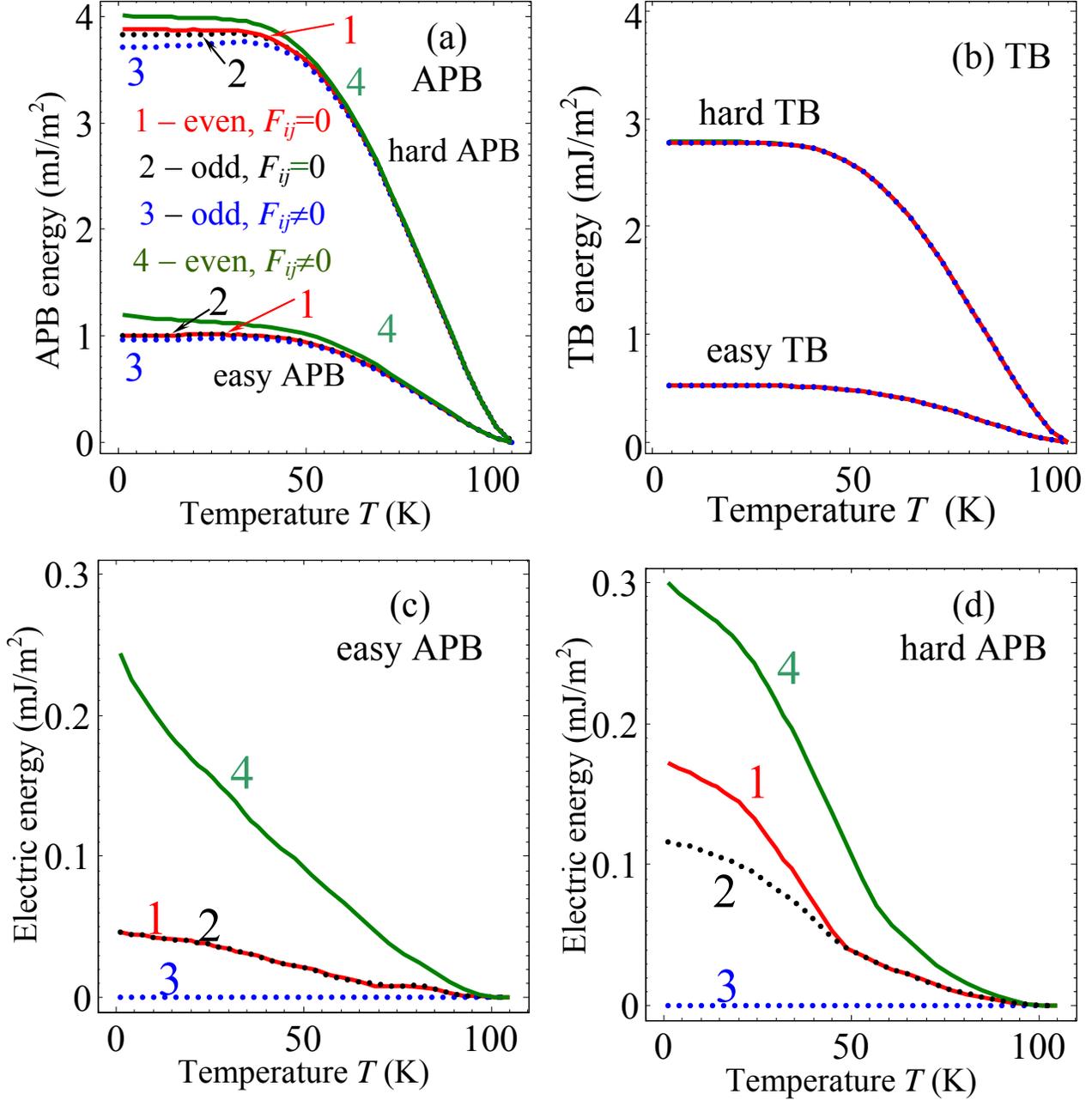

**Figure 22.** Energies of (a) antiphase and (b) twin boundaries vs. temperature calculated for SrTiO$_3$ with concentration of free carriers $n_0 = 10^{26}$ m$^{-3}$. Relative electric energies of easy (c) and hard (d) APBs counted from the lowest $P_3$-odd solution (curves 3 on plots (a) and (b)). Temperature dependences are calculated for nonzero flexoelectric effect $F_{ij} \neq 0$ and biquadratic coupling $\eta_{ij} \neq 0$ (curves 3, 4) and for the case of nonzero biquadratic coupling $\eta_{ij} \neq 0$ and zero flexoelectric effect $F_{ij} \equiv 0$ (curves 1, 2). Solid and dotted curves correspond to $P_3$-even and $P_3$-odd solutions respectively. Curves 1-4 style and color coding for plots (a-d) are the same and described in the legend to plot (a).

It is seen from **Figs. 22a** that APB energy is rather weakly dependent on the polarization distribution and its screening conditions. It is seen from **Figs. 22b** that TB energy looks almost independent on these factors, since the polarization component perpendicular to the TB plane is very small. The explanation is the weak dependence of the wall energy on the electric contribution. Relative electric energies of the APBs counted from the lowest $P_3$-odd solution (curves 3 on plots (a) and (b)) are shown in **Figs. 22c** and **22d** for easy and hard APBs correspondingly. Similarly to the case of dielectric SrTiO$_3$, $P_3$-even polarization distributions have the highest energy, but the electric energy difference between all types of polarization solutions are very small and super-linearly decreases with temperature increase; for TB (not shown) the difference is even smaller than for APB.

## Summary

In summary, we report a new mechanism, namely through the coupling of flexoelectric and rotostriction effects, that can give rise to the appearance of a significant improper spontaneous polarization and pyroelectricity across a structural antiphase boundary and twins, and by extension across interfaces in otherwise non-ferroelectric perovskites such as CaTiO$_3$, SrTiO$_3$, and EuTiO$_3$. In SrTiO$_3$, we show that this mechanism leads to a larger spontaneous polarization with an onset at a higher temperature than previously predicted through other coupling mechanisms (see **Table 2**). The result is in agreement with previously unexplained experimental results of Zubko et al [23], who attributed experimentally observed strong dependence of the SrTiO$_3$ flexoelectric response on applied elastic stress with the twins' motion allowing for polarization appearance across the elastic walls exactly below 105 K.

**Table 2**

| Polarization and pyro-coefficients | Hard 180-degree tilt APB and Hard 90-degree tilt TB | Easy 180-degree tilt APB and Easy 90-degree tilt TB |
|---|---|---|
| **Properties without free carries (dielectric limit)** | | |
| Polarization component $P_\parallel$ **parallel** to the domain wall plane | Ferroelectric hysteresis loop $P_\parallel(E_\parallel)$ exists at $0<T<T_C^*$ | Identically zero for easy APB  Negligibly small for easy TB |
| | Second order phase transition to ferroelectric phase occurs at $T=T_C^*$ | |
| | Hysteresis loop is absent at $T_C^*<T<T_S$. At these temperatures the amplitude $P_\parallel$ is proportional to $\eta\Phi_\perp\Phi_\parallel P_\perp$ | |

| Polarization component $P_\perp$ **perpendicular** to the domain wall plane | Amplitude $P_\perp$ is proportional to the local flexoelectric field $E^{FR}$ and $\eta\Phi_\perp\Phi_\parallel P_\parallel$ at $0<T<T_S$. Ferroelectricity and hysteresis loop for $P_\perp(E_\perp)$ is absent due to the strong depolarization field | Amplitude $P_\perp$ is proportional to the local flexoelectric field $E^{FR}$ at $0<T<T_S$. Ferroelectricity and hysteresis loop for $P_\perp$ is absent due to the strong depolarization field |
|---|---|---|
| Pyroelectric response **parallel** component $\Pi_\parallel = dP_\parallel/dT$ | Increases with $T$ increase at $0<T<T_C^*$ | Identically zero for APB |
|  | Sharp maximum occurs at $T=T_C^*$ and then response decreases with $T$ increase at $T_C^*<T<T_S$ | Negligibly small for TB |
| Pyroelectric response **perpendicular** component $\Pi_\perp = dP_\perp/dT$ | Increases with $T$ increase at $0<T<T_C^*$ | $\Pi_\perp$ is nonzero in the temperature range $0<T<T_S$, but vanishes at low temperatures $T\to 0$ and tends to zero at $T\to T_S$ |
|  | Sharp maximum occurs at $T=T_C^*$, since in the vicinity of $T_C^*$ $\Pi_\parallel \sim \Pi_\perp$ via the coupling term $\eta\Phi_\perp\Phi_\parallel P_\parallel$ |  |
|  | Smooth maximum exists at polarization inflection point located in the range $T_C^*<T<T_S$ | Smooth maximum exists at the polarization temperature dependence inflection point, that is located in the range $T_C^*<T<T_S$ |
| **Influence of the free carriers concentration $n_0$** | | |
| **Parallel** components $P_\parallel$ and $\Pi_\parallel$ | Increase with $n_0$ increase | Identically zero for easy APB Appear for easy TB and increases with $n_0$ increase |
| **Perpendicular** components $P_\perp$ and $\Pi_\perp$ | Strongly increase with $n_0$ increase due to the depolarization field decrease. | Strongly increase with $n_0$ increase due to the depolarization field decrease. |

\*) $T_C^*$ is effective Curie temperature that is different for APB and TB, namely $T_C^{APB} \approx 50$ K for hard APB and $T_C^{TB} \approx 25$ K for hard TB in SrTiO$_3$; $T_S$ is the temperature of the structural phase transition.

\*\*) $FR$ – product of flexoelectric and rotostriction coefficients; $\eta$ – biquadratic coupling coefficient.

Free electrons and mobile oxygen vacancies can effectively screen the depolarization field across polar interfaces and strongly enhance the components of interfacial polarization conjugated with depolarization field. The spontaneous polarization reaches the values ~0.1-5μC/cm$^2$ at the SrTiO$_3$ antidistortive and twin boundaries without free charges and ~1-5μC/cm$^2$ allowing for free charges. Since the induced polarizations are relatively large and since this effect is allowed at interfaces in all structures with static rotations, which are abundant in nature, it allows for an understanding of a large class of polar interfaces in nonpolar materials.

Authors gratefully acknowledge multiple discussions with Daniel Litvin, Behera K. Rakesh, and Sergei V. Kalinin; useful comments and suggestions from Alexander K. Tagantsev, and especially for his idea about interfacial pyroelectricity. National Science Foundation (DMR-0908718 and DMR-0820404) and user agreement with CNMS N UR-08-869 are acknowledged.

*Supplementary Materials to*

Interfacial Polarization and Pyroelectricity in Antiferrodistortive Structures Induced by a Flexoelectric Effect and Rotostriction

Anna N. Morozovska [1,2,*], Eugene A. Eliseev[1], Maya D. Glinchuk[1], Long-Qing Chen[3], and Venkatraman Gopalan [3,†]

[1] Institute for Problems of Materials Science, National Academy of Science of Ukraine, 3, Krjijanovskogo, 03142 Kiev, Ukraine

[2] Institute of Semiconductor Physics, National Academy of Science of Ukraine, 41, pr. Nauki, 03028 Kiev, Ukraine

[3] Department of Materials Science and Engineering, Pennsylvania State University, University Park, Pennsylvania 16802, USA

**Appendix A. Free energy, Euler-Lagrange equations of state and their transformations**

In the low temperature octahedrally tilted phase the free energy density has the form:

$$F_b = a_{ij}(T)P_i^2 + a_{ij}^u P_i^2 P_j^2 + ... + \frac{g_{ijkl}}{2}\left(\frac{\partial P_i}{\partial x_j}\frac{\partial P_k}{\partial x_l}\right) - P_i\left(\frac{E_i^d}{2} + E_i^{ext}\right) - q_{ijkl}u_{ij}P_k P_l + \frac{c_{ijkl}}{2}u_{ij}u_{kl}$$

$$+ b_i(T)\Phi_i^2 + b_{ij}^u \Phi_i^2 \Phi_j^2 - \eta_{ijkl}^u P_i P_j \Phi_k \Phi_l + \frac{v_{ijkl}}{2}\left(\frac{\partial \Phi_i}{\partial x_j}\frac{\partial \Phi_k}{\partial x_l}\right) \quad (A.1)$$

$$- r_{ijkl}^{(\Phi)} u_{ij}\Phi_k\Phi_l + \frac{f_{ijkl}}{2}\left(\frac{\partial P_k}{\partial x_l}u_{ij} - P_k \frac{\partial u_{ij}}{\partial x_l}\right)$$

Here $f_{ijkl}$ is the forth-rank tensor of flexoelectric stresses, $q_{ijkl}$ is the forth-rank tensor of electrostriction stress, $r_{ijkl}^{(\Phi)}$ is the tensor of rotostriction stress.

Euler-Lagrange equations of state are obtained from the minimization of the free energy (A.1) as

$$\frac{\partial F_b}{\partial \Phi_i} = 0 \quad \rightarrow \quad 2b_i\Phi_i + 4b_{ij}^u \Phi_j^2 \Phi_i - v_{ijkl}\frac{\partial^2 \Phi_k}{\partial x_j \partial x_l} - 2r_{mjki}u_{mj}\Phi_k - \eta_{klij}P_k P_l \Phi_j = 0, \quad (A.2a)$$

---

$$\frac{\partial F_b}{\partial u_{ij}} = \sigma_{ij} \quad \rightarrow \quad c_{ijkl} u_{kl} - r_{ijkl} \Phi_k \Phi_l + f_{ijkl} \frac{\partial P_k}{\partial x_l} - q_{ijkl} P_k P_l = \sigma_{ij}, \tag{A.2b}$$

$$\frac{\partial F_b}{\partial P_i} = 0 \rightarrow 2a_{ij} P_j + 4a_{ijkl} P_j P_k P_l - g_{ijkl} \frac{\partial^2 P_k}{\partial x_j \partial x_l} - 2q_{mjki} u_{mj} P_k - f_{mnil} \frac{\partial u_{mn}}{\partial x_l} - \eta_{ijkl} P_j \Phi_k \Phi_l = E_i^d. \tag{A.2c}$$

Where $\sigma_{ij}(\mathbf{x})$ is the stress tensor that satisfy mechanical equilibrium equation $\partial \sigma_{ij}(\mathbf{x})/\partial x_j = 0$ and so the application of the divergence operation to Eq.(A.2b) gives Lame-type equation:

$$c_{ijkl} \frac{\partial u_{kl}}{\partial x_j} - r_{ijkl} \frac{\partial}{\partial x_j}(\Phi_k \Phi_l) + f_{ijkl} \frac{\partial^2 P_k}{\partial x_j \partial x_l} - q_{ijkl} \frac{\partial}{\partial x_j}(P_k P_l) = 0.$$

It is seen from the equation of state (A.2b), that in the octahedrally tilted phase the strains $u_{ij}(\mathbf{x})$ are as:

$$u_{mn} = s_{mnij} \sigma_{ij} + R_{mnkl} \Phi_k \Phi_l + Q_{mnkl} P_k P_l - F_{mnkl} \frac{\partial P_k}{\partial x_l}. \tag{A.3}$$

Where $s_{mnij}$ is the elastic compliances tensor and $\sigma_{ij}$ is the stress tensor. $R_{ijkl} = s_{ijmn} r_{mnkl}$ is the rotostriction strain tensor; $Q_{ijkl} = s_{ijmn} q_{mnkl}$ is the electrostriction strain tensor; $F_{ijkl} = s_{ijmn} f_{mnkl}$ is the flexoelectric strain tensor.

Free energy expansion coefficients for SrTiO3
For the cubic m3m symmetry (A.1) could bewritten as

$$\begin{aligned}
F_b &= a_1(T)\left(P_1^2 + P_2^2 + P_3^2\right) + a_{11}^u\left(P_1^4 + P_2^4 + P_3^4\right) + a_{12}^u\left(P_1^2 P_2^2 + P_2^2 P_3^2 + P_3^2 P_1^2\right) + \frac{1}{2} g_{ijkl}\left(\frac{\partial P_i}{\partial x_j} \frac{\partial P_k}{\partial x_l}\right) \\
&- q_{11}\left(u_1 P_1^2 + u_2 P_2^2 + u_3 P_3^2\right) - q_{12}\left[u_1\left(P_2^2 + P_3^2\right) + u_2\left(P_3^2 + P_1^2\right) + u_3\left(P_1^2 + P_2^2\right)\right] \\
&- q_{44}\left(u_4 P_2 P_3 + u_5 P_3 P_1 + u_6 P_1 P_2\right) + \\
&+ \frac{1}{2} c_{11}\left(u_1^2 + u_2^2 + u_3^2\right) + c_{12}\left(u_1 u_2 + u_2 u_3 + u_3 u_1\right) + \frac{1}{2} c_{44}\left(u_4^2 + u_5^2 + u_6^2\right) + \\
&+ b_1(T)\left(\Phi_1^2 + \Phi_2^2 + \Phi_3^2\right) + b_{11}^u\left(\Phi_1^4 + \Phi_2^4 + \Phi_3^4\right) + b_{12}^u\left(\Phi_1^2 \Phi_2^2 + \Phi_2^2 \Phi_3^2 + \Phi_3^2 \Phi_1^2\right) - \\
&- r_{11}\left(u_1 \Phi_1^2 + u_2 \Phi_2^2 + u_3 \Phi_3^2\right) - r_{12}\left[u_1\left(\Phi_2^2 + \Phi_3^2\right) + u_2\left(\Phi_3^2 + \Phi_1^2\right) + u_3\left(\Phi_1^2 + \Phi_2^2\right)\right] \\
&- r_{44}\left(u_4 \Phi_2 \Phi_3 + u_5 \Phi_3 \Phi_1 + u_6 \Phi_1 \Phi_2\right) - \\
&- \eta_{11}^u\left(\Phi_1^2 P_1^2 + \Phi_2^2 P_2^2 + \Phi_3^2 P_3^2\right) - \eta_{12}^u\left(\Phi_3^2\left(P_1^2 + P_2^2\right) + \Phi_1^2\left(P_2^2 + P_3^2\right) + \Phi_2^2\left(P_3^2 + P_1^2\right)\right) - \\
&- \eta_{44}^u\left(\Phi_2 \Phi_3 P_2 P_3 + \Phi_3 \Phi_1 P_3 P_1 + \Phi_1 \Phi_2 P_1 P_2\right) - \\
&+ \frac{1}{2} v_{ijkl}\left(\frac{\partial \Phi_i}{\partial x_j} \frac{\partial \Phi_k}{\partial x_l}\right) + \frac{f_{ijkl}}{2}\left(\frac{\partial P_k}{\partial x_l} u_{ij} - P_k \frac{\partial u_{ij}}{\partial x_l}\right)
\end{aligned} \tag{A.4}$$

Coefficients $\alpha_1(T)$ and $\beta_1(T)$ depend on temperature in accordance with Barrett law.

It is convinient to switch from the coefficients at given strain (with superscripts "u", for a clamped sample) to coefficients given stress (with superscripts "σ", for a free sample) for further consideration. Using Legendre transformation of (A.4) $G = F - \sigma_i u_i$, one could get the following relations:

$$a_{11}^{\sigma} = a_{11}^{u} - \frac{4}{3}\frac{(q_{11}-q_{12})^2}{c_{11}-c_{12}} - \frac{2}{3}\frac{(q_{11}+2q_{12})^2}{c_{11}+2c_{12}}, \quad a_{12}^{\sigma} = a_{12}^{u} + \frac{1}{3}\frac{(q_{11}-q_{12})^2}{c_{11}-c_{12}} - \frac{1}{3}\frac{(q_{11}+2q_{12})^2}{c_{11}+2c_{12}} - \frac{(q_{44})^2}{2c_{44}}. \quad \text{(A.5a)}$$

$$b_{11}^{\sigma} = b_{11}^{u} - \frac{4}{3}\frac{(r_{11}-r_{12})^2}{c_{11}-c_{12}} - \frac{2}{3}\frac{(r_{11}+2r_{12})^2}{c_{11}+2c_{12}}, \quad b_{12}^{\sigma} = b_{12}^{u} + \frac{1}{3}\frac{(r_{11}-r_{12})^2}{c_{11}-c_{12}} - \frac{1}{3}\frac{(r_{11}+2r_{12})^2}{c_{11}+2c_{12}} - \frac{(r_{44})^2}{2c_{44}} \quad \text{(A.5b)}$$

$$\eta_{11}^{\sigma} = \eta_{11}^{u} + \frac{c_{11}(q_{11}r_{11}+2q_{12}r_{12})+c_{12}(q_{11}r_{11}-2q_{12}r_{11}-2q_{11}r_{12})}{(c_{11}-c_{12})(c_{11}+2c_{12})} \quad \text{(A.5c)}$$

$$\eta_{12}^{\sigma} = \eta_{12}^{u} + \frac{c_{11}(q_{12}r_{11}+q_{11}r_{12}+q_{12}r_{12})-c_{12}(q_{11}r_{11}+2q_{12}r_{12})}{(c_{11}-c_{12})(c_{11}+2c_{12})} \quad \text{(A.5d)}$$

$$\eta_{44}^{\sigma} = \eta_{44}^{u} + \frac{q_{44}r_{44}}{c_{44}} \quad \text{(A.5e)}$$

Finally, free energy at given stresses could be decomposed into several parts:

$$G = G_{\Phi P} + G_{in\,homo} + G_{striction} + G_{flexo} \quad \text{(A.6)}$$

where

$$\begin{aligned}
G_{\Phi P} &= b_1(T)\left(\Phi_1^2 + \Phi_2^2 + \Phi_3^2\right) + b_{11}^{\sigma}\left(\Phi_1^4 + \tilde{\Phi}_2^4 + \Phi_3^4\right) + b_{12}^{\sigma}\left(\Phi_1^2\Phi_2^2 + \Phi_1^2\Phi_3^2 + \Phi_2^2\Phi_3^2\right) + \\
&+ a_1(T)\left(P_1^2 + P_2^2 + P_3^2\right) + a_{11}^{\sigma}\left(P_1^4 + P_2^4 + P_3^4\right) + a_{12}^{\sigma}\left(P_1^2P_2^2 + P_1^2P_3^2 + P_2^2P_3^2\right) + \\
&- \eta_{11}^{\sigma}\left(\Phi_1^2P_1^2 + \Phi_2^2P_2^2 + \Phi_3^2P_3^2\right) - \eta_{12}^{\sigma}\left(\Phi_1^2P_2^2 + \Phi_2^2P_1^2 + \Phi_3^2(P_1^2+P_2^2)+(\Phi_1^2+\Phi_2^2)P_3^2\right) - \\
&- \eta_{44}^{\sigma}\left(\Phi_1 P_1 + \Phi_2 P_2 + (\Phi_1 P_1 + \Phi_2 P_2)\Phi_3 P_3\right) +
\end{aligned} \quad \text{(A.7a)}$$

Gradient energy has the form:

$$\begin{aligned}
G_{inhomo} &= \frac{v_{11}}{2}\left(\left(\frac{\partial \Phi_1}{\partial x_1}\right)^2 + \left(\frac{\partial \Phi_2}{\partial x_2}\right)^2 + \left(\frac{\partial \Phi_3}{\partial x_3}\right)^2\right) + v_{12}\left(\frac{\partial \Phi_1}{\partial x_1}\frac{\partial \Phi_2}{\partial x_2} + \frac{\partial \Phi_1}{\partial x_1}\frac{\partial \Phi_3}{\partial x_3} + \frac{\partial \Phi_3}{\partial x_3}\frac{\partial \Phi_2}{\partial x_2}\right) + \\
&+ \frac{v_{44}}{2}\left(\left(\frac{\partial \Phi_1}{\partial x_2} + \frac{\partial \Phi_2}{\partial x_1}\right)^2 + \left(\frac{\partial \Phi_2}{\partial x_3} + \frac{\partial \Phi_3}{\partial x_2}\right)^2 + \left(\frac{\partial \Phi_3}{\partial x_1} + \frac{\partial \Phi_1}{\partial x_3}\right)^2\right) + \\
&+ \frac{g_{11}}{2}\left(\left(\frac{\partial P_1}{\partial x_1}\right)^2 + \left(\frac{\partial P_2}{\partial x_2}\right)^2 + \left(\frac{\partial P_3}{\partial x_3}\right)^2\right) + g_{12}\left(\frac{\partial P_1}{\partial x_1}\frac{\partial P_2}{\partial x_2} + \frac{\partial P_1}{\partial x_1}\frac{\partial P_3}{\partial x_3} + \frac{\partial P_3}{\partial x_3}\frac{\partial P_2}{\partial x_2}\right) + \\
&+ \frac{g_{44}}{2}\left(\left(\frac{\partial P_1}{\partial x_2} + \frac{\partial P_2}{\partial x_1}\right)^2 + \left(\frac{\partial P_2}{\partial x_3} + \frac{\partial P_3}{\partial x_2}\right)^2 + \left(\frac{\partial P_3}{\partial x_1} + \frac{\partial P_1}{\partial x_3}\right)^2\right)
\end{aligned} \quad \text{(A.7b)}$$

Quadratic coupling with elastic stress components $\sigma_i$ (electrostriction and rotostriction effects) gives the following contribution:

$$G_{striction} = -Q_{11}(\sigma_1 P_1^2 + \sigma_2 P_2^2 + \sigma_3 P_3^2) - Q_{12}[\sigma_1(P_2^2 + P_3^2) + \sigma_2(P_3^2 + P_1^2) + \sigma_3(P_1^2 + P_2^2)]$$
$$- Q_{44}(\sigma_4 P_2 P_3 + \sigma_5 P_3 P_1 + \sigma_6 P_1 P_2) -$$
$$- \frac{1}{2} s_{11}(\sigma_1^2 + \sigma_2^2 + \sigma_3^2) - s_{12}(\sigma_1 \sigma_2 + \sigma_2 \sigma_3 + \sigma_3 v_1) - \frac{1}{2} s_{44}(\sigma_4^2 + \sigma_5^2 + \sigma_6^2) -$$
$$- R_{11}(\sigma_1 \Phi_1^2 + \sigma_2 \Phi_2^2 + \sigma_3 \Phi_3^2) - R_{12}[\sigma_1(\Phi_2^2 + \Phi_3^2) + \sigma_2(\Phi_3^2 + \Phi_1^2) + \sigma_3(\Phi_1^2 + \Phi_2^2)]$$
$$- R_{44}(\sigma_4 \Phi_2 \Phi_3 + \sigma_5 \Phi_3 \Phi_1 + \sigma_6 \Phi_1 \Phi_2) -$$
(A.7c)

After integration in parts flexoelectric contribution could be rewritten as:

$$G_{flexo} = F_{11}\left(\sigma_1 \frac{\partial P_1}{\partial x_1} + \sigma_2 \frac{\partial P_2}{\partial x_2} + \sigma_3 \frac{\partial P_3}{\partial x_3}\right) +$$
$$F_{12}\left(\sigma_2 \frac{\partial P_1}{\partial x_1} + \sigma_1 \frac{\partial P_2}{\partial x_2} + \sigma_1 \frac{\partial P_3}{\partial x_3} + \sigma_3 \frac{\partial P_1}{\partial x_1} + \sigma_2 \frac{\partial P_3}{\partial x_3} + \sigma_3 \frac{\partial P_2}{\partial x_2}\right) +$$
$$+ F_{44}\left(\sigma_4 \frac{\partial P_3}{\partial x_2} + \sigma_4 \frac{\partial P_2}{\partial x_3} + \sigma_5 \frac{\partial P_1}{\partial x_3} + \sigma_5 \frac{\partial P_3}{\partial x_1} + \sigma_6 \frac{\partial P_2}{\partial x_1} + \sigma_6 \frac{\partial P_1}{\partial x_2}\right)$$
(A.7d)

Hereinafter we used Voigt matrix notations when appropriate. For instance, the stress and strain tensors in Voigt matrix notations have the view:

$$\sigma_{11} \equiv \sigma_1, \quad \sigma_{22} \equiv \sigma_2, \quad \sigma_{33} \equiv \sigma_3, \quad \sigma_{23} \equiv \sigma_4, \quad \sigma_{13} \equiv \sigma_5, \quad \sigma_{12} \equiv \sigma_6,$$

$$u_{11} \equiv u_1, \quad u_{22} \equiv u_2, \quad u_{33} \equiv u_3, \quad u_{23} \equiv \frac{1}{2} u_4, \quad u_{13} \equiv \frac{1}{2} u_5, \quad u_{12} \equiv \frac{1}{2} u_6,$$

## Appendix B. 180 degree DW (APB) in tilted perovskites

Elastic equation of state could be obtained by the minimization $\delta G / \delta \sigma_{jk} = -u_{jk}$. Subscripts 1, 2 and 3 denote Cartesian coordinates $x, y, z$ and Voigt's (matrix) notations are used.

Equations of state, including the flexoelectric contributions, have the form:

$$u_1 - u_1^{(F)} = s_{11}\sigma_1 + s_{12}(\sigma_2 + \sigma_3),$$
(B.1a)
$$u_2 - u_2^{(F)} = s_{11}\sigma_2 + s_{12}(\sigma_1 + \sigma_3),$$
(B.1b)
$$u_3 - u_3^{(F)} = s_{11}\sigma_3 + s_{12}(\sigma_1 + \sigma_2),$$
(B.1c)
$$u_4 - u_4^{(F)} = s_{44}\sigma_4,$$
(B.1d)
$$u_5 - u_5^{(F)} = s_{44}\sigma_5,$$
(B.1d)
$$u_6 - u_6^{(F)} = s_{44}\sigma_6.$$
(B.1e)

Here we introduced designations for the strains in the absence of any elastic stresses:

$$u_1^{(F)} = -F_{11}\frac{\partial P_1}{\partial x_1} - F_{12}\frac{\partial P_3}{\partial x_3} + Q_{11}P_1^2 + Q_{12}(P_2^2 + P_3^2) + R_{11}\Phi_1^2 + R_{12}(\Phi_2^2 + \Phi_3^2),$$
(B.2a)

$$u_2^{(F)} = -F_{12}\left(\frac{\partial P_1}{\partial x_1} + \frac{\partial P_3}{\partial x_3}\right) + Q_{11}P_2^2 + Q_{12}(P_1^2 + P_3^2) + R_{11}\Phi_2^2 + R_{12}(\Phi_1^2 + \Phi_3^2),$$
(B.2b)

$$u_3^{(F)} = -F_{11}\frac{\partial P_3}{\partial x_3} - F_{12}\frac{\partial P_1}{\partial x_1} + Q_{11}P_3^2 + Q_{12}\left(P_2^2 + P_1^2\right) + R_{11}\Phi_3^2 + R_{12}\left(\Phi_2^2 + \Phi_1^2\right), \quad \text{(B.2c)}$$

$$u_4^{(F)} = -F_{44}\frac{\partial P_2}{\partial x_3} + Q_{44}P_2P_3 + R_{44}\Phi_2\Phi_3, \quad \text{(B.2d)}$$

$$u_5^{(F)} = -F_{44}\left(\frac{\partial P_3}{\partial x_1} + \frac{\partial P_1}{\partial x_3}\right) + Q_{44}P_1P_3 + R_{44}\Phi_1\Phi_3, \quad \text{(B.2e)}$$

$$u_6^{(F)} = -F_{44}\frac{\partial P_2}{\partial x_1} + Q_{44}P_1P_2 + R_{44}\Phi_1\Phi_2. \quad \text{(B.2f)}$$

Below we consider the case of $x_1$- dependent 1D solution.

Note, that elastic contribution to the free energy could be written as

$$\Delta G_{elast} + \Delta G_{strict} + \Delta G_{flexo} = \begin{pmatrix} -\frac{1}{2}s_{11}\left(\sigma_1^2 + \sigma_2^2 + \sigma_3^2\right) - s_{12}\left(\sigma_1\sigma_2 + \sigma_2\sigma_3 + \sigma_3\sigma_1\right) \\ -\frac{1}{2}s_{44}\left(\sigma_4^2 + \sigma_5^2 + \sigma_6^2\right) - \\ -\sigma_1 u_1^{(F)} - \sigma_2 u_2^{(F)} - \sigma_3 u_3^{(F)} - \sigma_4 u_4^{(F)} - \sigma_5 u_5^{(F)} - \sigma_6 u_6^{(F)} \end{pmatrix} \quad \text{(B.3)}$$

In the case of $x_1$-dependent solution, compatibility relation leads to the conditions of constant strains $u_2 = const$, $u_3 = const$, $u_4 = const$, while the coordinate-dependent $u_1(x_1)$, $u_5(x_1)$ and $u_6(x_1)$ do not involved in compatibility relations. Mechanical equilibrium conditions could be written as $\partial\sigma_1/\partial x_1 = 0$, $\partial\sigma_5/\partial x_1 = 0$, $\partial\sigma_6/\partial x_1 = 0$. Since $\sigma_{ij}(x_1 \to \pm\infty) = 0$, one obtains $\sigma_1 = \sigma_5 = \sigma_6 = 0$.

The constant strains $u_{22}$, $u_{33}$ and $u_{23}$ are equal to values of spontaneous strains in stress free homogeneous system ($u_{ii}^{(S)} = R_{ij}^{(\Phi)}\left(\Phi_j^{(S)}\right)^2$ is the spontaneous strain far from the wall):

$$u_2 \equiv u_{22}^S = R_{11}\left(\Phi_2^{(S)}\right)^2 + R_{12}\left(\left(\Phi_3^{(S)}\right)^2 + \left(\Phi_1^{(S)}\right)^2\right), \quad \text{(B.4a)}$$

$$u_3 \equiv u_{33}^S = R_{11}\left(\Phi_3^{(S)}\right)^2 + R_{12}\left(\left(\Phi_2^{(S)}\right)^2 + \left(\Phi_1^{(S)}\right)^2\right), \quad \text{(B.4b)}$$

$$u_4 \equiv 2u_{23}^S = R_{44}\Phi_2^{(S)}\Phi_3^{(S)}. \quad \text{(B.4c)}$$

Here $\mathbf{\Phi}^{(S)}$ is spontaneous tilt vector of two adjacent domains.

Using (B.4) one can rewrite the system of equations (3) with respect to unknown stresses $\sigma_2$, $\sigma_3$ and $\sigma_4$ as:

$$s_{11}\sigma_2 + s_{12}\sigma_3 = F_{12}\frac{\partial P_1}{\partial x_1} + U_2, \quad \text{(B.5a)}$$

$$s_{12}\sigma_2 + s_{11}\sigma_3 = F_{12}\frac{\partial P_1}{\partial x_1} + U_3, \quad \text{(B.5b)}$$

$$s_{44}\sigma_4 = U_4. \quad \text{(B.5c)}$$

where.

$$U_2 = -Q_{11}P_2^2 - Q_{12}(P_1^2 + P_3^2) + R_{12}\left((\Phi_1^{(S)})^2 + (\Phi_3^{(S)})^2 - \Phi_1^2 - \Phi_3^2\right) + R_{11}\left((\Phi_2^{(S)})^2 - \Phi_2^2\right), \quad (B.6a)$$

$$U_3 = -Q_{11}P_3^2 - Q_{12}(P_2^2 + P_1^2) + R_{11}\left((\Phi_3^{(S)})^2 - \Phi_3^2\right) + R_{12}\left((\Phi_1^{(S)})^2 + (\Phi_2^{(S)})^2 - \Phi_1^2 - \Phi_2^2\right), \quad (B.6b)$$

$$U_4 = R_{44}\left(\Phi_2^{(S)}\Phi_3^{(S)} - \Phi_2\Phi_3\right) - Q_{44}P_2P_3. \quad (B.6c)$$

Finally, inhomogeneous stress components can be written in the form:

$$\sigma_2 = \frac{F_{12}}{s_{11} + s_{12}}\frac{\partial P_1}{\partial x_1} + \frac{s_{11}U_2 - s_{12}U_3}{s_{11}^2 - s_{12}^2}, \quad (B.7a)$$

$$\sigma_3 = \frac{F_{12}}{s_{11} + s_{12}}\frac{\partial P_1}{\partial x_1} + \frac{s_{11}U_3 - s_{12}U_2}{s_{11}^2 - s_{12}^2}, \quad (B.7b)$$

$$\sigma_4 = \frac{U_4}{s_{44}} \quad (B.7c)$$

$$\sigma_1 = \sigma_5 = \sigma_6 = 0, \quad (B.7d)$$

Then the strain $u_{11}$ given by Eq.(B.5c) could be expressed via nonzero stresses $\sigma_2$ and $\sigma_3$

Below let us consider typical hard APB, i.e. one-dimensional problem with periodically modulated $\Phi_1(x_1)$ and $\Phi_3(x_1)$, which induces $P_1(x_1)$ and $P_3(x_1)$, but $P_2(x_1) \equiv 0$. For the case Eqs.(B.2) reduces to the system of four coupled equations:

$$\begin{pmatrix} 2b_1\Phi_1 + 4b_{11}\Phi_1^3 + 2b_{12}\Phi_1\Phi_3^2 + 2(\eta_{11}P_1^2 + \eta_{12}P_3^2)\Phi_1 \\ + \eta_{44}P_1P_3\Phi_3 - v_{11}\frac{\partial^2\Phi_1}{\partial x_1 \partial x_1} - 2R_{12}(\sigma_2 + \sigma_3)\Phi_1 \end{pmatrix} = 0, \quad (B.8a)$$

$$\begin{pmatrix} 2b_1\Phi_3 + 4b_{11}\Phi_3^3 + 2b_{12}\Phi_3\Phi_1^2 + 2(\eta_{11}P_3^2 + \eta_{12}P_1^2)\Phi_3 \\ + \eta_{44}P_1P_3\Phi_1 - v_{44}\frac{\partial^2\Phi_3}{\partial x_1 \partial x_1} - 2(\sigma_3 R_{11} + \sigma_2 R_{12})\Phi_3 \end{pmatrix} = 0, \quad (B.8b)$$

$$\begin{pmatrix} 2a_1P_1 + 4a_{11}P_1^3 + 2a_{12}P_1P_3^2 + 2(\eta_{11}\Phi_1^2 + \eta_{12}\Phi_3^2)P_1 \\ + \eta_{44}\Phi_1\Phi_3P_3 - g_{11}\frac{\partial^2 P_1}{\partial x_1 \partial x_1} - 2Q_{12}(\sigma_2 + \sigma_3)P_1 \end{pmatrix} = E_1^d + F_{12}\left(\frac{\partial\sigma_2}{\partial x_1} + \frac{\partial\sigma_3}{\partial x_1}\right), \quad (B.8c)$$

$$\begin{pmatrix} 2a_1P_3 + 4a_{11}P_3^3 + 2a_{12}P_3P_1^2 + 2(\eta_{11}\Phi_3^2 + \eta_{12}\Phi_1^2)P_3 \\ + \eta_{44}\Phi_1\Phi_3P_1 - g_{44}\frac{\partial^2 P_3}{\partial x_1 \partial x_1} - 2(\sigma_3 Q_{11} + \sigma_2 Q_{12})P_3 \end{pmatrix} = 0. \quad (B.8d)$$

It is seen from Eqs.(B.8c) that nonzero "flexoelectric" field $F_{12}\left(\frac{\partial\sigma_2}{\partial x_1} + \frac{\partial\sigma_3}{\partial x_1}\right)$ exists for the component $P_1$.

Inhomogeneous stress components are given by Eq.(B.7). Substituting of Eqs.(B.7) into Eq.(B.8) leads to the closed system for the polarization and tilt vector components. The system (B.8) should be supplemented by the boundary conditions. $\Phi_3$ is zero on the domain walls and at infinity. Polarization component $P_3$ is zero at the walls.

**Appendix C. 90°-domain wall (TB).**

Let us consider domains "1" and "2" with different orientation of structural order parameters, $\mathbf{\Phi}^{(1)} = (\Phi_S, 0, 0)$ and $\mathbf{\Phi}^{(2)} = (0, \pm \Phi_S, 0)$, different signs correspond to head-to-tail and head-to-head domain walls. For tetragonal ferroelastics corresponding spontaneous strain tensors are

$$\hat{u}^{(1)} = \begin{pmatrix} R_{11}\Phi_S^2 & 0 & 0 \\ 0 & R_{12}\Phi_S^2 & 0 \\ 0 & 0 & R_{12}\Phi_S^2 \end{pmatrix} \text{ and } \hat{u}^{(2)} = \begin{pmatrix} R_{12}\Phi_S^2 & 0 & 0 \\ 0 & R_{11}\Phi_S^2 & 0 \\ 0 & 0 & R_{12}\Phi_S^2 \end{pmatrix} \quad \text{(C.1a)}$$

It is well known [1], that mechanical compatibility of domains with spontaneous strain (C.1) is possible only for the specific orientation of domain wall (DW), which could be determined from the condition

$$e_{lik} n_k \left( u_{ij}^{(1)} - u_{ij}^{(2)} \right) e_{mjp} n_p = 0 \quad \text{(C.1b)}$$

Where $e_{lik}$ is the fully antisymmetric tensor, $n_k$ is the components of the normal to the DW plane. For the specific case of spontaneous strains (C.1a) system of equations (C.1b) has the solutions corresponding to plane of DW should be perpendicular to $[110]$ or $[\bar{1}10]$ directions (see Fig. C1).

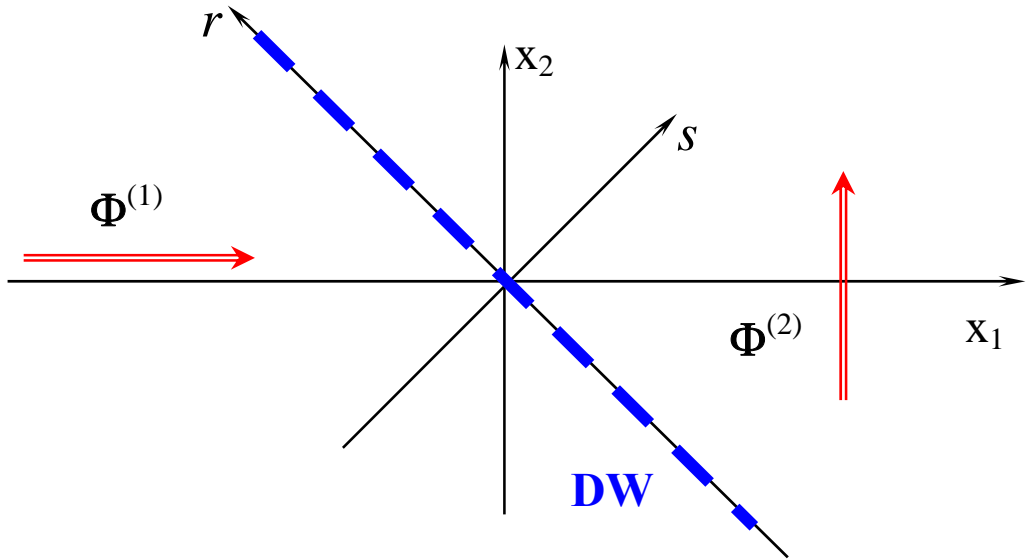

Fig. C1. Coordinate systems related to DW between 90-degree domains (twins).

Relations between the coordinates in crystallographic system and rotated one, with one axis, perpendicular to domain wall:

$$\tilde{x}_1 = \frac{x_1 + x_2}{\sqrt{2}}, \quad \tilde{x}_2 = \frac{-x_1 + x_2}{\sqrt{2}}, \quad \tilde{x}_3 \equiv x_3 \qquad (C.2a)$$

The relations between the vector components are

$$\tilde{P}_1 = \frac{P_1 + P_2}{\sqrt{2}}, \quad \tilde{P}_2 = \frac{-P_1 + P_2}{\sqrt{2}}, \quad \tilde{P}_3 \equiv P_3 \qquad (C.2b)$$

$$P_1 = \frac{\tilde{P}_1 - \tilde{P}_2}{\sqrt{2}}, \quad P_2 = \frac{\tilde{P}_1 + \tilde{P}_2}{\sqrt{2}}, \quad \tilde{P}_3 \equiv P_3 \qquad (C.2c)$$

For the symmetric tensors of second rank (e.g. strain of stress) the relations between the tensor components in two coordinate system have the view:

$$\tilde{u}_{11} = \frac{u_{11} + u_{22}}{2} - u_{12}, \quad \tilde{u}_{22} = \frac{u_{11} + u_{22}}{2} + u_{12}, \quad \tilde{u}_{33} = u_{33};$$

$$\tilde{u}_{12} = \frac{u_{11} - u_{22}}{2}, \quad \tilde{u}_{13} = \frac{u_{13} - u_{23}}{\sqrt{2}}, \quad \tilde{u}_{23} = \frac{u_{13} + u_{23}}{\sqrt{2}}. \qquad (C.2d)$$

For instance, spontaneous strain (C.1) far from the DW have the view in the new coordinate system (C.2):

$$\hat{u}^{(1)}\Big|_{\tilde{x}} = \begin{pmatrix} \frac{R_{11} + R_{12}}{2}\Phi_S^2 & \frac{R_{11} - R_{12}}{2}\Phi_S^2 & 0 \\ \frac{R_{11} - R_{12}}{2}\Phi_S^2 & \frac{R_{11} + R_{12}}{2}\Phi_S^2 & 0 \\ 0 & 0 & R_{12}\Phi_S^2 \end{pmatrix} \qquad (C.3a)$$

$$\hat{u}^{(2)}\Big|_{\tilde{x}} = \begin{pmatrix} \frac{R_{11} + R_{12}}{2}\Phi_S^2 & \frac{-R_{11} + R_{12}}{2}\Phi_S^2 & 0 \\ \frac{-R_{11} + R_{12}}{2}\Phi_S^2 & \frac{R_{11} + R_{12}}{2}\Phi_S^2 & 0 \\ 0 & 0 & R_{12}\Phi_S^2 \end{pmatrix} \qquad (C.3b)$$

Different types of twin boundaries (TB), or 90 degree DW, are depicted in Fig. C2.

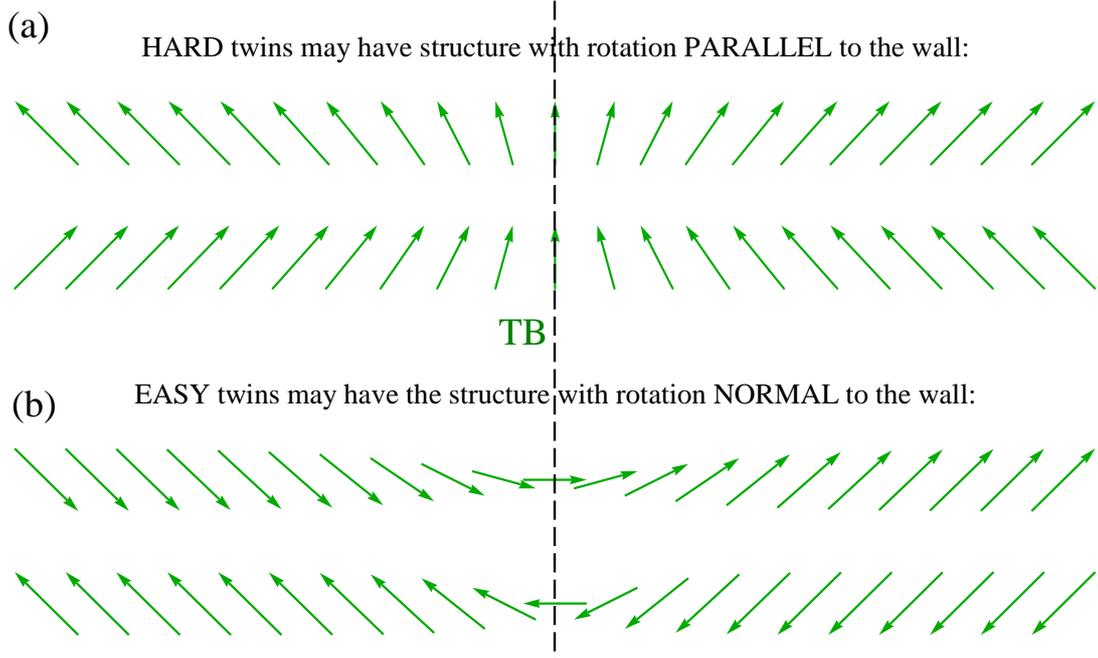

Fig. C2. Twin boundaries structures. Fields are plotted using the approximate solutions proposed by [2] (Twin plane is in the center).

An EASY DW structure usually have lower energy than those of HARD DW.

For the ferroelectrics head-to-tail twins are preferable since they do not create internal (depolarization) electric field, while for the structural order parameter there are no basically field conjugate to the structural order parameter, so the energetics of the twins are determined by the other factors.

Elastic equation of state could be obtained by the minimization $\delta G/\delta \tilde{\sigma}_{jk} = -\tilde{u}_{jk}$, in our case, including the flexoelectric contributions, could be written in the form:

$$\tilde{u}_1 - u_1^{(F)} = \tilde{s}_{11}\tilde{\sigma}_1 + (\tilde{s}_{12}\tilde{\sigma}_2 + s_{12}\tilde{\sigma}_3), \qquad (C.4a)$$

$$\tilde{u}_2 - u_2^{(F)} = \tilde{s}_{11}\tilde{\sigma}_2 + \tilde{s}_{12}\tilde{\sigma}_1 + s_{12}\tilde{\sigma}_3, \qquad (C.4b)$$

$$\tilde{u}_3 - u_3^{(F)} = s_{11}\tilde{\sigma}_3 + s_{12}(\tilde{\sigma}_1 + \tilde{\sigma}_2), \qquad (C.4c)$$

$$\tilde{u}_4 - u_4^{(F)} = s_{44}\tilde{\sigma}_4, \qquad (C.4d)$$

$$\tilde{u}_5 - u_5^{(F)} = s_{44}\tilde{\sigma}_5, \qquad (C.4d)$$

$$\tilde{u}_6 - u_6^{(F)} = \tilde{s}_{66}\tilde{\sigma}_6. \qquad (C.4e)$$

In Eq. (C.4) we introduced designations for the strains in the absence of any elastic stresses:

$$u_1^{(F)} = -\tilde{F}_{11}\frac{\partial \tilde{P}_1}{\partial \tilde{x}_1} + \tilde{Q}_{11}\tilde{P}_1^2 + \tilde{Q}_{12}\tilde{P}_2^2 + Q_{12}P_3^2 + \tilde{R}_{11}\tilde{\Phi}_1^2 + \tilde{R}_{12}\tilde{\Phi}_2^2 + R_{12}\Phi_3^2, \qquad (C.5a)$$

$$u_2^{(F)} = -\tilde{F}_{12}\frac{\partial \tilde{P}_1}{\partial \tilde{x}_1} + \tilde{Q}_{11}\tilde{P}_2^2 + \tilde{Q}_{12}\tilde{P}_1^2 + Q_{12}P_3^2 + \tilde{R}_{11}\tilde{\Phi}_2^2 + \tilde{R}_{12}\tilde{\Phi}_1^2 + R_{12}\Phi_3^2, \quad \text{(C.5b)}$$

$$u_3^{(F)} = -F_{12}\frac{\partial \tilde{P}_1}{\partial \tilde{x}_1} + Q_{11}P_3^2 + Q_{12}(\tilde{P}_2^2 + \tilde{P}_1^2) + R_{11}\Phi_3^2 + R_{12}(\tilde{\Phi}_2^2 + \tilde{\Phi}_1^2), \quad \text{(C.5c)}$$

$$u_4^{(F)} = Q_{44}\tilde{P}_2 P_3 + R_{44}\tilde{\Phi}_2 \Phi_3, \quad \text{(C.5d)}$$

$$u_5^{(F)} = -F_{44}\frac{\partial P_3}{\partial \tilde{x}_1} + Q_{44}\tilde{P}_1 P_3 + R_{44}\tilde{\Phi}_1 \Phi_3, \quad \text{(C.5e)}$$

$$u_6^{(F)} = -\tilde{F}_{66}\frac{\partial \tilde{P}_2}{\partial \tilde{x}_1} + \tilde{Q}_{66}\tilde{P}_1\tilde{P}_2 + \tilde{R}_{66}\tilde{\Phi}_1\tilde{\Phi}_2. \quad \text{(C.5f)}$$

Here we used compliance tensor components in the new reference frame:

$\tilde{s}_{11} = (s_{11} + s_{12} + s_{44}/2)/2$, $\tilde{s}_{12} = (s_{11} + s_{12} - s_{44}/2)/2$, $\tilde{s}_{66} = 2(s_{11} - s_{12})$;

$\tilde{R}_{11} = (R_{11} + R_{12} + R_{44}/2)/2$, $\tilde{R}_{12} = (R_{11} + R_{12} - R_{44}/2)/2$, $\tilde{R}_{66} = 2(R_{11} - R_{12})$;

$\tilde{Q}_{11} = (Q_{11} + Q_{12} + Q_{44}/2)/2$, $\tilde{Q}_{12} = (Q_{11} + Q_{12} - Q_{44}/2)/2$, $\tilde{Q}_{66} = 2(Q_{11} - Q_{12})$;

$\tilde{F}_{11} = (F_{11} + F_{12} + F_{44})/2$, $\tilde{F}_{12} = (F_{11} + F_{12} - F_{44})/2$, $\tilde{F}_{66} = F_{11} - F_{12}$.

Using compatibility relation $e_{ikl}e_{jmn}(\partial^2 \tilde{u}_{ln}/\partial \tilde{x}_k \partial \tilde{x}_m) = 0$ and mechanical equilibrium conditions $\partial \tilde{\sigma}_{ij}/\partial \tilde{x}_i = 0$, we could easily get the evident form of elastic strains and stresses. In the case of $\tilde{x}_1$-dependent solution, compatibility relation leads to the conditions of constant strains $\tilde{u}_2 = const$, $\tilde{u}_3 = const$, $\tilde{u}_4 = const$, while general form dependences like $\tilde{u}_1 = \tilde{u}_1(\tilde{x}_1)$, $\tilde{u}_5 = \tilde{u}_5(\tilde{x}_1)$ and $\tilde{u}_6 = \tilde{u}_6(\tilde{x}_1)$ do not contradict to these relations. Mechanical equilibrium conditions could be written as $\partial \tilde{\sigma}_1/\partial \tilde{x}_1 = 0$, $\partial \tilde{\sigma}_5/\partial \tilde{x}_1 = 0$, $\partial \tilde{\sigma}_6/\partial \tilde{x}_1 = 0$. Since $\tilde{\sigma}_{ij}(\tilde{x}_1 \to \pm\infty) = 0$, one obtains $\tilde{\sigma}_1 = \tilde{\sigma}_5 = \tilde{\sigma}_6 = 0$. In this case some of the shear strains are trivial, $\tilde{u}_5 = u_5^{(F)}$, $\tilde{u}_6 = u_6^{(F)}$.

Constant strains $\tilde{u}_2$, $\tilde{u}_3$ and $\tilde{u}_4$ are equal to values of spontaneous strains in stress free, homogeneous system:

$$\tilde{u}_2 = \tilde{u}_2^{(S)} \equiv \tilde{R}_{11}(\tilde{\Phi}_2^{(S)})^2 + \tilde{R}_{12}(\tilde{\Phi}_1^{(S)})^2, \quad \text{(C.6a)}$$

$$\tilde{u}_3 = \tilde{u}_3^{(S)} \equiv R_{12}((\tilde{\Phi}_2^{(S)})^2 + (\tilde{\Phi}_1^{(S)})^2), \quad \text{(C.6b)}$$

$$\tilde{u}_4 = \tilde{u}_4^{(S)} \equiv 0. \quad \text{(C.6c)}$$

Here $\tilde{\mathbf{\Phi}}^{(S)} = \pm(\Phi_S, \pm\Phi_S, 0)/\sqrt{2}$ are spontaneous tilt vectors of two adjacent twins. Since strains (C.6) depend only on squares of spontaneous tilt, strains $\tilde{u}_2$, $\tilde{u}_3$ and $\tilde{u}_4$ are the same for both the domains (as it should be expected from compatibility conditions).

Using (C.6) one can rewrite the system of equations (C.4) with respect to unknown stresses $(\tilde{\sigma}_2, \tilde{\sigma}_3)$:

$$\tilde{s}_{11}\tilde{\sigma}_2 + s_{12}\tilde{\sigma}_3 = \tilde{F}_{12}\frac{\partial \tilde{P}_1}{\partial \tilde{x}_1} + U_2 \tag{C.7a}$$

$$s_{12}\tilde{\sigma}_2 + s_{11}\tilde{\sigma}_3 = F_{12}\frac{\partial \tilde{P}_1}{\partial \tilde{x}_1} + U_3 \tag{C.7b}$$

$$s_{44}\tilde{\sigma}_4 = U_4 \tag{C.7c}$$

while strain $u_1$ could be expressed via $\tilde{\sigma}_2$ and $\tilde{\sigma}_3$. Finally, inhomogeneous stress and strain components can be written in the form:

$$\tilde{\sigma}_2 = \frac{s_{11}\tilde{F}_{12} - s_{12}F_{12}}{s_{11}\tilde{s}_{11} - s_{12}^2}\frac{\partial \tilde{P}_1}{\partial \tilde{x}_1} + \frac{s_{11}U_2 - s_{12}U_3}{s_{11}\tilde{s}_{11} - s_{12}^2}, \tag{C.8a}$$

$$\tilde{\sigma}_3 = \frac{\tilde{s}_{11}F_{12} - s_{12}\tilde{F}_{12}}{s_{11}\tilde{s}_{11} - s_{12}^2}\frac{\partial \tilde{P}_1}{\partial \tilde{x}_1} + \frac{\tilde{s}_{11}U_3 - s_{12}U_2}{s_{11}\tilde{s}_{11} - s_{12}^2}, \tag{C.8b}$$

$$\tilde{\sigma}_4 = \frac{U_4}{s_{44}}, \tag{C.8c}$$

$$\tilde{\sigma}_1 = \tilde{\sigma}_5 = \tilde{\sigma}_6 = 0 \tag{C.8d}$$

Here we introduced the following designations

$$U_2 = \tilde{R}_{11}\left(\frac{\Phi_S^2}{2} - \tilde{\Phi}_2^2\right) + \tilde{R}_{12}\left(\frac{\Phi_S^2}{2} - \tilde{\Phi}_1^2\right) - R_{12}\Phi_3^2 - \tilde{Q}_{12}\tilde{P}_1^2 - \tilde{Q}_{11}\tilde{P}_2^2 - Q_{12}P_3^2 \tag{C.9a}$$

$$U_3 = R_{12}\left(\Phi_S^2 - \tilde{\Phi}_2^2 - \tilde{\Phi}_1^2\right) - R_{11}\Phi_3^2 - Q_{12}\left(\tilde{P}_2^2 + \tilde{P}_1^2\right) - Q_{11}P_3^2 \tag{C.9a}$$

$$U_4 = -Q_{44}\tilde{P}_2 P_3 - R_{44}\tilde{\Phi}_2\Phi_3 \tag{C.9a}$$

Since the elastic stresses are accommodated near the DW, the state with minimal energy corresponds to another thermodynamic potential:

$$F = G + \tilde{\sigma}_i \tilde{u}_i = $$
$$= G_{\Phi P} + G_{in\,homo} + G_{striction} + G_{flexo} + \tilde{\sigma}_1\tilde{u}_1 + \tilde{\sigma}_2\tilde{u}_2 + \tilde{\sigma}_3\tilde{u}_3 + \tilde{\sigma}_4\tilde{u}_4 + \tilde{\sigma}_5\tilde{u}_5 + \tilde{\sigma}_6\tilde{u}_6 \tag{C.10}$$

Note, that elastic contribution to the free energy (C.3) could be written as

$$G_{striction} + G_{flexo} = $$
$$= -\frac{1}{2}\left(\tilde{s}_{11}\left(\tilde{\sigma}_1^2 + \tilde{\sigma}_2^2\right) + s_{11}\tilde{\sigma}_3^2\right) - \left(\tilde{s}_{12}\tilde{\sigma}_1\tilde{\sigma}_2 + s_{12}\left(\tilde{\sigma}_2\tilde{\sigma}_3 + \tilde{\sigma}_3\tilde{\sigma}_1\right)\right) - \frac{1}{2}\left(s_{44}\left(\tilde{\sigma}_4^2 + \tilde{\sigma}_5^2\right) + \tilde{s}_{66}\tilde{\sigma}_6^2\right) - \tag{C.11}$$
$$- \tilde{\sigma}_1 u_1^{(F)} - \tilde{\sigma}_2 u_2^{(F)} - \tilde{\sigma}_3 u_3^{(F)} - \tilde{\sigma}_4 u_4^{(F)} - \tilde{\sigma}_5 u_5^{(F)} - \tilde{\sigma}_6 u_6^{(F)}$$

Using the expression (C.10) and solution (C.8), it is easy to show that

$$G_{striction} + G_{flexo} + \tilde{\sigma}_i \tilde{u}_i =$$

$$-\frac{1}{2}\left(\tilde{s}_{11}\tilde{\sigma}_2^2 + s_{11}\tilde{\sigma}_3^2\right) - s_{12}\tilde{\sigma}_2\tilde{\sigma}_3 - \frac{1}{2}s_{44}\tilde{\sigma}_4^2 + \tilde{\sigma}_2\left(\tilde{u}_2^{(S)} - u_2^{(F)}\right) + \tilde{\sigma}_3\left(\tilde{u}_3^{(S)} - u_3^{(F)}\right) + \tilde{\sigma}_4\left(\tilde{u}_4^{(S)} - u_4^{(F)}\right) =$$

$$= \frac{s_{11}\left(\tilde{u}_2^{(S)} - u_2^{(F)}\right)^2}{2\left(s_{11}\tilde{s}_{11} - s_{12}^2\right)} + \frac{\tilde{s}_{11}\left(\tilde{u}_3^{(S)} - u_3^{(F)}\right)^2}{2\left(s_{11}\tilde{s}_{11} - s_{12}^2\right)} - \frac{s_{12}\left(\tilde{u}_2^{(S)} - u_2^{(F)}\right)\left(\tilde{u}_3^{(S)} - u_3^{(F)}\right)}{s_{11}\tilde{s}_{11} - s_{12}^2} + \frac{\left(\tilde{u}_4^{(S)} - u_4^{(F)}\right)^2}{2s_{44}} =$$

$$= \frac{s_{11}U_2^2 + \tilde{s}_{11}U_3^2 - 2s_{12}U_2U_3}{2\left(s_{11}\tilde{s}_{11} - s_{12}^2\right)} + \frac{U_4^2}{2s_{44}} + \frac{1}{2}\left(\frac{s_{11}\tilde{F}_{12} - s_{12}F_{12}}{s_{11}\tilde{s}_{11} - s_{12}^2}\tilde{F}_{12} + \frac{\tilde{s}_{11}F_{12} - s_{12}\tilde{F}_{12}}{s_{11}\tilde{s}_{11} - s_{12}^2}F_{12}\right)\left(\frac{\partial \tilde{P}_1}{\partial \tilde{x}_1}\right)^2 +$$

$$\left(\frac{\left(s_{11}\tilde{F}_{12} - s_{12}F_{12}\right)U_2}{\left(s_{11}\tilde{s}_{11} - s_{12}^2\right)} + \frac{\left(\tilde{s}_{11}F_{12} - s_{12}\tilde{F}_{12}\right)U_3}{\left(s_{11}\tilde{s}_{11} - s_{12}^2\right)}\right)\frac{\partial \tilde{P}_1}{\partial \tilde{x}_1}$$

(C.12)

Free energy in the rotated reference frame has the view:

$$G = b_1(T)\left(\tilde{\Phi}_1^2 + \tilde{\Phi}_2^2 + \Phi_3^2\right) + \tilde{b}_{11}\left(\tilde{\Phi}_1^4 + \tilde{\Phi}_2^4\right) + b_{11}\Phi_3^4 + \tilde{b}_{12}\tilde{\Phi}_1^2\tilde{\Phi}_2^2 + b_{12}\left(\tilde{\Phi}_1^2 + \tilde{\Phi}_2^2\right)\Phi_3^2 +$$
$$+ a_1(T)\left(\tilde{P}_1^2 + \tilde{P}_2^2 + P_3^2\right) + \tilde{a}_{11}\left(\tilde{P}_1^4 + \tilde{P}_2^4\right) + a_{11}P_3^4 + \tilde{a}_{12}\tilde{P}_1^2\tilde{P}_2^2 + a_{12}\left(\tilde{P}_1^2 + \tilde{P}_2^2\right)P_3^2 +$$
$$- \tilde{\eta}_{11}\left(\tilde{\Phi}_1^2\tilde{P}_1^2 + \tilde{\Phi}_2^2\tilde{P}_2^2\right) - \eta_{11}\Phi_3^2 P_3^2 - \tilde{\eta}_{12}\left(\tilde{\Phi}_1^2\tilde{P}_2^2 + \tilde{\Phi}_2^2\tilde{P}_1^2\right) -$$
$$- \eta_{12}\left(\Phi_3^2\left(\tilde{P}_1^2 + \tilde{P}_2^2\right) + \left(\tilde{\Phi}_1^2 + \tilde{\Phi}_2^2\right)P_3^2\right) - \tilde{\eta}_{66}\tilde{\Phi}_1\tilde{\Phi}_2\tilde{P}_1\tilde{P}_2 - \eta_{44}\left(\tilde{\Phi}_1\tilde{P}_1 + \tilde{\Phi}_2\tilde{P}_2\right)\Phi_3 P_3 +$$
$$- \tilde{Q}_{11}\tilde{\sigma}_2\tilde{P}_2^2 - Q_{11}\tilde{\sigma}_3 P_3^2 - Q_{12}\tilde{\sigma}_2 P_3^2 - \tilde{Q}_{12}\tilde{\sigma}_2\tilde{P}_1^2 - Q_{12}\tilde{\sigma}_3\left(\tilde{P}_1^2 + \tilde{P}_2^2\right) - Q_{44}\tilde{\sigma}_4\tilde{P}_2 P_3 -$$
$$- \tilde{R}_{11}\tilde{\sigma}_2\tilde{\Phi}_2^2 - R_{11}\tilde{\sigma}_3\Phi_3^2 - R_{12}\tilde{\sigma}_2\Phi_3^2 - \tilde{R}_{12}\tilde{\sigma}_2\tilde{\Phi}_1^2 - R_{12}\tilde{\sigma}_3\left(\tilde{\Phi}_1^2 + \tilde{\Phi}_2^2\right) - R_{44}\tilde{\sigma}_4\tilde{\Phi}_2\Phi_3 -$$
$$- \frac{\tilde{s}_{11}}{2}\tilde{\sigma}_2^2 - \frac{s_{11}}{2}\tilde{\sigma}_3^2 - s_{12}\tilde{\sigma}_3\tilde{\sigma}_2 - \frac{1}{2}s_{44}\tilde{\sigma}_4^2 + \tilde{F}_{12}\tilde{\sigma}_2\frac{\partial \tilde{P}_1}{\partial \tilde{x}_1} + F_{12}\tilde{\sigma}_3\frac{\partial \tilde{P}_1}{\partial \tilde{x}_1} +$$
$$+ \frac{\tilde{v}_{11}}{2}\left(\frac{\partial \tilde{\Phi}_1}{\partial \tilde{x}_1}\right)^2 + \frac{\tilde{v}_{66}}{2}\left(\frac{\partial \tilde{\Phi}_2}{\partial \tilde{x}_1}\right)^2 + \frac{v_{44}}{2}\left(\frac{\partial \Phi_3}{\partial \tilde{x}_1}\right)^2 + \frac{\tilde{g}_{11}}{2}\left(\frac{\partial \tilde{P}_1}{\partial \tilde{x}_1}\right)^2 + \frac{\tilde{g}_{66}}{2}\left(\frac{\partial \tilde{P}_2}{\partial \tilde{x}_1}\right)^2 + \frac{g_{44}}{2}\left(\frac{\partial P_3}{\partial \tilde{x}_1}\right)^2$$

(C.13a)

Here we introduced tensor components in the new reference frame

$$\tilde{b}_{11} = \frac{1}{4}(2b_{11} + b_{12}), \quad \tilde{b}_{12} = \frac{1}{2}(6b_{11} - b_{12}) \quad \tilde{a}_{11} = \frac{1}{4}(2a_{11} + a_{12}), \quad \tilde{a}_{12} = \frac{1}{2}(6a_{11} - a_{12})$$

$$\tilde{\eta}_{11} = (\eta_{11} + \eta_{12} + \eta_{44}/2)/2, \quad \tilde{\eta}_{12} = (\eta_{11} + \eta_{12} - \eta_{44}/2)/2, \quad \tilde{\eta}_{66} = 2(\eta_{11} - \eta_{12})$$

$$\tilde{v}_{11} = (v_{11} + v_{12} + 2v_{44})/2, \quad \tilde{v}_{12} = (v_{11} + v_{12} - 2v_{44})/2, \quad \tilde{v}_{66} = (v_{11} - v_{12})/2$$

$$\tilde{g}_{11} = (g_{11} + g_{12} + 2g_{44})/2, \quad \tilde{g}_{12} = (g_{11} + g_{12} - 2g_{44})/2, \quad \tilde{g}_{66} = (g_{11} - g_{12})/2$$

Finally, using Eq.(C.12) and (C.13) one could get the thermodynamic potential with stresses excluded

$$G + \tilde{\sigma}_i \tilde{u}_i = b_1(T)(\tilde{\Phi}_1^2 + \tilde{\Phi}_2^2 + \Phi_3^2) + \tilde{b}_{11}(\tilde{\Phi}_1^4 + \tilde{\Phi}_2^4) + b_{11}\Phi_3^4 + \tilde{b}_{12}\tilde{\Phi}_1^2\tilde{\Phi}_2^2 + b_{12}(\tilde{\Phi}_1^2 + \tilde{\Phi}_2^2)\Phi_3^2 +$$
$$+ a_1(T)(\tilde{P}_1^2 + \tilde{P}_2^2 + P_3^2) + \tilde{a}_{11}(\tilde{P}_1^4 + \tilde{P}_2^4) + a_{11}P_3^4 + \tilde{a}_{12}\tilde{P}_1^2\tilde{P}_2^2 + a_{12}(\tilde{P}_1^2 + \tilde{P}_2^2)P_3^2 +$$
$$- \tilde{\eta}_{11}(\tilde{\Phi}_1^2\tilde{P}_1^2 + \tilde{\Phi}_2^2\tilde{P}_2^2) - \eta_{11}\Phi_3^2 P_3^2 - \tilde{\eta}_{12}(\tilde{\Phi}_1^2\tilde{P}_2^2 + \tilde{\Phi}_2^2\tilde{P}_1^2) -$$
$$- \eta_{12}(\Phi_3^2(\tilde{P}_1^2 + \tilde{P}_2^2) + (\tilde{\Phi}_1^2 + \tilde{\Phi}_2^2)P_3^2) - \tilde{\eta}_{66}\tilde{\Phi}_1\tilde{\Phi}_2\tilde{P}_1\tilde{P}_2 - \eta_{44}(\tilde{\Phi}_1\tilde{P}_1 + \tilde{\Phi}_2\tilde{P}_2)\Phi_3 P_3 +$$
$$+ \frac{\tilde{v}_{11}}{2}\left(\frac{\partial\tilde{\Phi}_1}{\partial\tilde{x}_1}\right)^2 + \frac{\tilde{v}_{66}}{2}\left(\frac{\partial\tilde{\Phi}_2}{\partial\tilde{x}_1}\right)^2 + \frac{v_{44}}{2}\left(\frac{\partial\Phi_3}{\partial\tilde{x}_1}\right)^2 + \frac{\tilde{g}_{11}}{2}\left(\frac{\partial\tilde{P}_1}{\partial\tilde{x}_1}\right)^2 + \frac{\tilde{g}_{66}}{2}\left(\frac{\partial\tilde{P}_2}{\partial\tilde{x}_1}\right)^2 + \frac{g_{44}}{2}\left(\frac{\partial P_3}{\partial\tilde{x}_1}\right)^2 +$$
$$+ \frac{s_{11}U_2^2 + \tilde{s}_{11}U_3^2 - 2s_{12}U_2 U_3}{2(s_{11}\tilde{s}_{11} - s_{12}^2)} + \frac{U_4^2}{2s_{44}} + \frac{1}{2}\left(\frac{s_{11}\tilde{F}_{12} - s_{12}F_{12}}{s_{11}\tilde{s}_{11} - s_{12}^2}\tilde{F}_{12} + \frac{\tilde{s}_{11}F_{12} - s_{12}\tilde{F}_{12}}{s_{11}\tilde{s}_{11} - s_{12}^2}F_{12}\right)\left(\frac{\partial\tilde{P}_1}{\partial\tilde{x}_1}\right)^2 +$$
$$+ \left(\frac{(s_{11}\tilde{F}_{12} - s_{12}F_{12})U_2}{(s_{11}\tilde{s}_{11} - s_{12}^2)} + \frac{(\tilde{s}_{11}F_{12} - s_{12}\tilde{F}_{12})U_3}{(s_{11}\tilde{s}_{11} - s_{12}^2)}\right)\frac{\partial\tilde{P}_1}{\partial\tilde{x}_1}$$

Equations of state for rotation vector components could be found from minimization of (C.13a):

$$2b_1\tilde{\Phi}_1 + 4\tilde{b}_{11}\tilde{\Phi}_1^3 + 2\tilde{b}_{12}\tilde{\Phi}_1\tilde{\Phi}_2^2 + 2b_{12}\tilde{\Phi}_1\Phi_3^2 - \tilde{v}_{11}\frac{\partial^2\tilde{\Phi}_1}{\partial\tilde{x}_1\partial\tilde{x}_1} -$$
$$- 2(\tilde{\eta}_{11}\tilde{P}_1^2 + \tilde{\eta}_{12}\tilde{P}_2^2 + \eta_{12}P_3^2)\tilde{\Phi}_1 - \tilde{\eta}_{66}\tilde{P}_1\tilde{P}_2\tilde{\Phi}_2 - \eta_{44}\tilde{P}_1 P_3 \Phi_3 - 2(\tilde{R}_{12}\tilde{\sigma}_2 + R_{12}\tilde{\sigma}_3)\tilde{\Phi}_1 = 0 \quad \text{(C.14a)}$$

$$2b_1\tilde{\Phi}_2 + 4\tilde{b}_{11}\tilde{\Phi}_2^3 + 2(\tilde{b}_{12}\tilde{\Phi}_1^2 + b_{12}\Phi_3^2)\tilde{\Phi}_2 - \tilde{v}_{66}\frac{\partial^2\tilde{\Phi}_2}{\partial\tilde{x}_1\partial\tilde{x}_1} -$$
$$- 2(\tilde{\eta}_{11}\tilde{P}_2^2 + \tilde{\eta}_{12}\tilde{P}_1^2 + \eta_{12}P_3^2)\tilde{\Phi}_2 - \tilde{\eta}_{66}\tilde{P}_1\tilde{P}_2\tilde{\Phi}_1 - \eta_{44}\tilde{P}_2 P_3 \Phi_3 - 2(\tilde{\sigma}_2\tilde{R}_{11} + \tilde{\sigma}_3 R_{12})\tilde{\Phi}_2 - R_{44}\tilde{\sigma}_4\Phi_3 = 0$$

(C.14b)

$$2b_1\Phi_3 + 4b_{11}\Phi_3^3 + 2b_{12}\Phi_3(\tilde{\Phi}_1^2 + \tilde{\Phi}_2^2) - v_{44}\frac{\partial^2\Phi_3}{\partial\tilde{x}_1\partial\tilde{x}_1} -$$
$$- 2(\eta_{11}P_3^2 + \eta_{12}(\tilde{P}_1^2 + \tilde{P}_2^2))\Phi_3 - \eta_{44}(\tilde{P}_1\tilde{\Phi}_1 + \tilde{P}_2\tilde{\Phi}_2)P_3 - 2(R_{11}\tilde{\sigma}_3 + R_{12}\tilde{\sigma}_2)\Phi_3 - R_{44}\tilde{\sigma}_4\tilde{\Phi}_2 = 0$$

(C.14c)

For the polarization components

$$2a_1\tilde{P}_1 + 4\tilde{a}_{11}\tilde{P}_1^3 + 2(a_{12}P_3^2 + \tilde{a}_{12}\tilde{P}_2^2)\tilde{P}_1 - 2(\tilde{\eta}_{11}\tilde{\Phi}_1^2 + \eta_{12}\Phi_3^2 + \tilde{\eta}_{12}\tilde{\Phi}_2^2)\tilde{P}_1 - (\eta_{44}\Phi_3 P_3 + \tilde{\eta}_{66}\tilde{\Phi}_2\tilde{P}_2)\tilde{\Phi}_1 -$$
$$- \tilde{g}_{11}\frac{\partial^2\tilde{P}_1}{\partial\tilde{x}_1\partial\tilde{x}_1} - 2(\tilde{Q}_{12}\tilde{\sigma}_2 + Q_{12}\tilde{\sigma}_3)\tilde{P}_1 + \frac{\tilde{P}_1}{\varepsilon_0\varepsilon_b} - \tilde{F}_{12}\frac{\partial\tilde{\sigma}_2}{\partial\tilde{x}_1} - F_{12}\frac{\partial\tilde{\sigma}_3}{\partial\tilde{x}_1} = 0$$

(C.15a)

$$2a_1\tilde{P}_2 + 4\tilde{a}_{11}\tilde{P}_2^3 + 2(a_{12}P_3^2 + \tilde{a}_{12}\tilde{P}_1^2)\tilde{P}_2 - 2(\tilde{\eta}_{11}\tilde{\Phi}_2^2 + \eta_{12}\Phi_3^2 + \tilde{\eta}_{12}\tilde{\Phi}_1^2)\tilde{P}_2 - (\eta_{44}\Phi_3 P_3 + \tilde{\eta}_{66}\tilde{\Phi}_1\tilde{P}_1)\tilde{\Phi}_2 -$$
$$- \tilde{g}_{66}\frac{\partial^2\tilde{P}_2}{\partial\tilde{x}_1\partial\tilde{x}_1} - 2(\tilde{Q}_{11}\tilde{\sigma}_2 + Q_{12}\tilde{\sigma}_3)\tilde{P}_2 - \tilde{\sigma}_4 Q_{44}P_3 = 0$$

(C.15b)

$$2a_1 P_3 + 4a_{11}P_3^3 + 2a_{12}P_3(\tilde{P}_1^2 + \tilde{P}_2^2) - 2(\eta_{11}\Phi_3^2 + \eta_{12}(\tilde{\Phi}_1^2 + \tilde{\Phi}_2^2))P_3 - \eta_{44}(\tilde{\Phi}_1\tilde{P}_1 + \tilde{\Phi}_2\tilde{P}_2)\Phi_3 -$$
$$- g_{44}\frac{\partial^2 P_3}{\partial\tilde{x}_1\partial\tilde{x}_1} - 2(\tilde{\sigma}_3 Q_{11} + \tilde{\sigma}_2 Q_{12})P_3 - \tilde{\sigma}_4 Q_{44}\tilde{P}_2 = 0$$

(C.15c)

After substitution of stress components (C.8) into Eqs.(C.12-B13) one could get the evident form equations for tilt components:

$$2b_1\tilde{\Phi}_1 + 4\tilde{b}_{11}\tilde{\Phi}_1^3 + 2\tilde{b}_{12}\tilde{\Phi}_1\tilde{\Phi}_2^2 + 2b_{12}\tilde{\Phi}_1\Phi_3^2 - \tilde{v}_{11}\frac{\partial^2\tilde{\Phi}_1}{\partial\tilde{x}_1\partial\tilde{x}_1} -$$
$$- 2(\tilde{\eta}_{11}\tilde{P}_1^2 + \tilde{\eta}_{12}\tilde{P}_2^2 + \eta_{12}P_3^2)\tilde{\Phi}_1 - \tilde{\eta}_{66}\tilde{P}_1\tilde{P}_2\tilde{\Phi}_2 - \eta_{44}\tilde{P}_1P_3\Phi_3 -$$
$$- 2\left\{\begin{array}{l}\frac{(\tilde{R}_{12}s_{11} - R_{12}s_{12})}{s_{11}\tilde{s}_{11} - s_{12}^2}\left(\tilde{R}_{11}\left(\frac{\Phi_S^2}{2} - \tilde{\Phi}_2^2\right) + \tilde{R}_{12}\left(\frac{\Phi_S^2}{2} - \tilde{\Phi}_1^2\right) - R_{12}\Phi_3^2 - \tilde{Q}_{12}\tilde{P}_1^2 - \tilde{Q}_{11}\tilde{P}_2^2 - Q_{12}P_3^2\right) + \\ + \frac{(R_{12}\tilde{s}_{11} - \tilde{R}_{12}s_{12})}{s_{11}\tilde{s}_{11} - s_{12}^2}\left(R_{12}(\Phi_S^2 - \tilde{\Phi}_2^2 - \tilde{\Phi}_1^2) - R_{11}\Phi_3^2 - Q_{12}(\tilde{P}_2^2 + \tilde{P}_1^2) - Q_{11}P_3^2\right)\end{array}\right\}\tilde{\Phi}_1 +$$
$$- 2\left(\frac{s_{11}\tilde{F}_{12} - s_{12}F_{12}}{s_{11}\tilde{s}_{11} - s_{12}^2}\tilde{R}_{12} + \frac{\tilde{s}_{11}F_{12} - s_{12}\tilde{F}_{12}}{s_{11}\tilde{s}_{11} - s_{12}^2}R_{12}\right)\tilde{\Phi}_1\frac{\partial\tilde{P}_1}{\partial\tilde{x}_1} = 0$$

(C.16a)

$$2b_1\tilde{\Phi}_2 + 4\tilde{b}_{11}\tilde{\Phi}_2^3 + 2(\tilde{b}_{12}\tilde{\Phi}_1^2 + b_{12}\Phi_3^2)\tilde{\Phi}_2 - \tilde{v}_{66}\frac{\partial^2\tilde{\Phi}_2}{\partial\tilde{x}_1\partial\tilde{x}_1} -$$
$$- 2(\tilde{\eta}_{11}\tilde{P}_2^2 + \tilde{\eta}_{12}\tilde{P}_1^2 + \eta_{12}P_3^2)\tilde{\Phi}_2 - \tilde{\eta}_{66}\tilde{P}_1\tilde{P}_2\tilde{\Phi}_1 - \eta_{44}\tilde{P}_2P_3\Phi_3 +$$
$$+ R_{44}\frac{Q_{44}\tilde{P}_2P_3 + R_{44}\tilde{\Phi}_2\Phi_3}{s_{44}}\Phi_3 - 2\left(\frac{s_{11}\tilde{F}_{12} - s_{12}F_{12}}{s_{11}\tilde{s}_{11} - s_{12}^2}\tilde{R}_{11} + \frac{\tilde{s}_{11}F_{12} - s_{12}\tilde{F}_{12}}{s_{11}\tilde{s}_{11} - s_{12}^2}R_{12}\right)\tilde{\Phi}_2\frac{\partial\tilde{P}_1}{\partial\tilde{x}_1} -$$
$$- 2\left\{\begin{array}{l}\frac{(\tilde{R}_{11}s_{11} - R_{12}s_{12})}{s_{11}\tilde{s}_{11} - s_{12}^2}\left(\tilde{R}_{11}\left(\frac{\Phi_S^2}{2} - \tilde{\Phi}_2^2\right) + \tilde{R}_{12}\left(\frac{\Phi_S^2}{2} - \tilde{\Phi}_1^2\right) - R_{12}\Phi_3^2 - \tilde{Q}_{12}\tilde{P}_1^2 - \tilde{Q}_{11}\tilde{P}_2^2 - Q_{12}P_3^2\right) + \\ + \frac{(R_{12}\tilde{s}_{11} - \tilde{R}_{11}s_{12})}{s_{11}\tilde{s}_{11} - s_{12}^2}\left(R_{12}(\Phi_S^2 - \tilde{\Phi}_2^2 - \tilde{\Phi}_1^2) - R_{11}\Phi_3^2 - Q_{12}(\tilde{P}_2^2 + \tilde{P}_1^2) - Q_{11}P_3^2\right)\end{array}\right\}\tilde{\Phi}_2 = 0$$

(C.16b)

$$2b_1\Phi_3 + 4b_{11}\Phi_3^3 + 2b_{12}\Phi_3(\tilde{\Phi}_1^2 + \tilde{\Phi}_2^2) - v_{44}\frac{\partial^2\Phi_3}{\partial\tilde{x}_1\partial\tilde{x}_1} - 2\left(\frac{s_{11}\tilde{F}_{12} - s_{12}F_{12}}{s_{11}\tilde{s}_{11} - s_{12}^2}R_{12} + \frac{\tilde{s}_{11}F_{12} - s_{12}\tilde{F}_{12}}{s_{11}\tilde{s}_{11} - s_{12}^2}R_{11}\right)\tilde{\Phi}_3\frac{\partial\tilde{P}_1}{\partial\tilde{x}_1} -$$
$$- 2(\eta_{11}P_3^2 + \eta_{12}(\tilde{P}_1^2 + \tilde{P}_2^2))\Phi_3 - \eta_{44}(\tilde{P}_1P_3\tilde{\Phi}_1 + \tilde{P}_2P_3\tilde{\Phi}_2) + \frac{Q_{44}\tilde{P}_2P_3 + R_{44}\tilde{\Phi}_2\Phi_3}{s_{44}}R_{44}\tilde{\Phi}_2 -$$
$$- 2\left\{\begin{array}{l}\frac{(R_{12}s_{11} - R_{11}s_{12})}{s_{11}\tilde{s}_{11} - s_{12}^2}\left(\tilde{R}_{11}\left(\frac{\Phi_S^2}{2} - \tilde{\Phi}_2^2\right) + \tilde{R}_{12}\left(\frac{\Phi_S^2}{2} - \tilde{\Phi}_1^2\right) - R_{12}\Phi_3^2 - \tilde{Q}_{12}\tilde{P}_1^2 - \tilde{Q}_{11}\tilde{P}_2^2 - Q_{12}P_3^2\right) + \\ + \frac{(R_{11}\tilde{s}_{11} - R_{12}s_{12})}{s_{11}\tilde{s}_{11} - s_{12}^2}\left(R_{12}(\Phi_S^2 - \tilde{\Phi}_2^2 - \tilde{\Phi}_1^2) - R_{11}\Phi_3^2 - Q_{12}(\tilde{P}_2^2 + \tilde{P}_1^2) - Q_{11}P_3^2\right)\end{array}\right\}\Phi_3 = 0$$

(C.16c)

The same for polarization components:

$$2a_1\tilde{P}_1 + 4\tilde{a}_{11}\tilde{P}_1^3 + 2(a_{12}P_3^2 + \tilde{a}_{12}\tilde{P}_2^2)\tilde{P}_1 + \frac{\tilde{P}_1}{\varepsilon_0\varepsilon_b} - \left(\tilde{g}_{11} + \frac{s_{11}\tilde{F}_{12} - s_{12}F_{12}}{s_{11}\tilde{s}_{11} - s_{12}^2}\tilde{F}_{12} + \frac{\tilde{s}_{11}F_{12} - s_{12}\tilde{F}_{12}}{s_{11}\tilde{s}_{11} - s_{12}^2}F_{12}\right)\frac{\partial^2\tilde{P}_1}{\partial\tilde{x}_1\partial\tilde{x}_1} -$$

$$- 2(\tilde{\eta}_{11}\tilde{\Phi}_1^2 + \eta_{12}\Phi_3^2 + \tilde{\eta}_{12}\tilde{\Phi}_2^2)\tilde{P}_1 - (\eta_{44}\Phi_3 P_3 + \tilde{\eta}_{66}\tilde{\Phi}_2\tilde{P}_2)\tilde{\Phi}_1 -$$

$$- 2\left(\begin{array}{l}\dfrac{(\tilde{Q}_{12}s_{11} - Q_{12}s_{12})}{s_{11}\tilde{s}_{11} - s_{12}^2}\left(\tilde{R}_{11}\left(\dfrac{\Phi_S^2}{2} - \tilde{\Phi}_2^2\right) + \tilde{R}_{12}\left(\dfrac{\Phi_S^2}{2} - \tilde{\Phi}_1^2\right) - R_{12}\Phi_3^2 - \tilde{Q}_{12}\tilde{P}_1^2 - \tilde{Q}_{11}\tilde{P}_2^2 - Q_{12}P_3^2\right) + \\ + \dfrac{(Q_{12}\tilde{s}_{11} - \tilde{Q}_{12}s_{12})}{s_{11}\tilde{s}_{11} - s_{12}^2}\left(R_{12}(\Phi_S^2 - \tilde{\Phi}_2^2 - \tilde{\Phi}_1^2) - R_{11}\Phi_3^2 - Q_{12}(\tilde{P}_2^2 + \tilde{P}_1^2) - Q_{11}P_3^2\right)\end{array}\right)\tilde{P}_1 +$$

$$+ \frac{\partial}{\partial\tilde{x}_1}\left(\begin{array}{l}\dfrac{(\tilde{F}_{12}s_{11} - F_{12}s_{12})}{s_{11}\tilde{s}_{11} - s_{12}^2}(\tilde{Q}_{11}\tilde{P}_2^2 + Q_{12}P_3^2 + \tilde{R}_{11}\tilde{\Phi}_2^2 + \tilde{R}_{12}\tilde{\Phi}_1^2 + R_{12}\Phi_3^2) + \\ + \dfrac{(F_{12}\tilde{s}_{11} - \tilde{F}_{12}s_{12})}{s_{11}\tilde{s}_{11} - s_{12}^2}(Q_{11}P_3^2 + Q_{12}\tilde{P}_2^2 + R_{11}\Phi_3^2 + R_{12}(\tilde{\Phi}_2^2 + \tilde{\Phi}_1^2))\end{array}\right) = 0$$

(C.17a)

$$2a_1\tilde{P}_2 + 4\tilde{a}_{11}\tilde{P}_2^3 + 2(a_{12}P_3^2 + \tilde{a}_{12}\tilde{P}_1^2)\tilde{P}_2 - 2(\tilde{\eta}_{11}\tilde{\Phi}_2^2 + \eta_{12}\Phi_3^2 + \tilde{\eta}_{12}\tilde{\Phi}_1^2)\tilde{P}_2 - (\eta_{44}\Phi_3 P_3 + \tilde{\eta}_{66}\tilde{\Phi}_1\tilde{P}_1)\tilde{\Phi}_2 -$$

$$+ \frac{Q_{44}^2\tilde{P}_2 P_3 + Q_{44}R_{44}\tilde{\Phi}_2\Phi_3}{s_{44}}P_3 - \tilde{g}_{66}\frac{\partial^2\tilde{P}_2}{\partial\tilde{x}_1\partial\tilde{x}_1} - 2\left(\frac{s_{11}\tilde{F}_{12} - s_{12}F_{12}}{s_{11}\tilde{s}_{11} - s_{12}^2}\tilde{Q}_{11} + \frac{\tilde{s}_{11}F_{12} - s_{12}\tilde{F}_{12}}{s_{11}\tilde{s}_{11} - s_{12}^2}Q_{12}\right)\tilde{P}_2\frac{\partial\tilde{P}_1}{\partial\tilde{x}_1} -$$

$$- 2\left(\frac{(s_{11}\tilde{Q}_{11} - s_{12}Q_{12})U_2}{s_{11}\tilde{s}_{11} - s_{12}^2} + \frac{(\tilde{s}_{11}Q_{12} - s_{12}\tilde{Q}_{11})U_3}{s_{11}\tilde{s}_{11} - s_{12}^2}\right)\tilde{P}_2 = 0$$

(C.17b)

$$2a_1 P_3 + 4a_{11}P_3^3 + 2a_{12}P_3(\tilde{P}_1^2 + \tilde{P}_2^2) - 2(\eta_{11}\Phi_3^2 + \eta_{12}(\tilde{\Phi}_1^2 + \tilde{\Phi}_2^2))P_3 - \eta_{44}(\tilde{\Phi}_1\tilde{P}_1 + \tilde{\Phi}_2\tilde{P}_2)\Phi_3 +$$

$$+ \frac{Q_{44}^2\tilde{P}_2 P_3 + Q_{44}R_{44}\tilde{\Phi}_2\Phi_3}{s_{44}}\tilde{P}_2 - g_{44}\frac{\partial^2 P_3}{\partial\tilde{x}_1\partial\tilde{x}_1} - 2\left(\frac{\tilde{s}_{11}F_{12} - s_{12}\tilde{F}_{12}}{s_{11}\tilde{s}_{11} - s_{12}^2}Q_{11} + \frac{s_{11}\tilde{F}_{12} - s_{12}F_{12}}{s_{11}\tilde{s}_{11} - s_{12}^2}Q_{12}\right)P_3\frac{\partial\tilde{P}_1}{\partial\tilde{x}_1} -$$

$$- 2\left(\frac{(s_{11}Q_{12} - s_{12}Q_{11})U_2}{s_{11}\tilde{s}_{11} - s_{12}^2} + \frac{(\tilde{s}_{11}Q_{11} - s_{12}Q_{12})U_3}{s_{11}\tilde{s}_{11} - s_{12}^2}\right)P_3 = 0$$

(C.17c)

These equations should be accompanied by the following boundary conditions:

Polarization vector components are zero far from the TB:

$$\tilde{P}_1(\tilde{x}_1 \to \pm\infty) = 0, \quad \tilde{P}_2(\tilde{x}_1 \to \pm\infty) = 0 \qquad (C.18)$$

Polarization component $\tilde{P}_2$ has zero derivatives at the TB (even type solution). Polarization component $\tilde{P}_1$ either has zero derivatives at the TB (even type solution) or is zero at the TB (odd type solution):

$$\frac{\partial\tilde{P}_1}{\partial\tilde{x}_1}(\tilde{x}_1 = 0) = 0, \quad \frac{\partial\tilde{P}_2}{\partial\tilde{x}_1}(\tilde{x}_1 = 0) = 0, \qquad (\tilde{P}_1\text{-even and } \tilde{P}_2\text{-even}) \qquad (C.19b)$$

$$\tilde{P}_1(\tilde{x}_1 = 0) = 0, \quad \frac{\partial\tilde{P}_2}{\partial\tilde{x}_1}(\tilde{x}_1 = 0) = 0. \qquad (\tilde{P}_1\text{-odd and } \tilde{P}_2\text{-even}) \qquad (C.19c)$$

Note that $\tilde{P}_2$-odd solution appeared unstable.

Boundary conditions for the tilt vector at **hard** twins (with rotation vector parallel to the wall plane as shown in **Fig. C2a**) are $\tilde{\Phi}_1(\tilde{x}_1 = 0) = 0$, $\frac{\partial \tilde{\Phi}_2}{\partial \tilde{x}_1}(\tilde{x}_1 = 0) = 0$ at the TB and very far from TB:

$$\tilde{\Phi}_1(\tilde{x}_1 \to +\infty) = \mp \frac{\Phi_S}{\sqrt{2}}, \quad \tilde{\Phi}_1(\tilde{x}_1 \to -\infty) = \pm \frac{\Phi_S}{\sqrt{2}}, \quad \tilde{\Phi}_2(\tilde{x}_1 \to +\infty) = \frac{\Phi_S}{\sqrt{2}}, \quad \tilde{\Phi}_2(\tilde{x}_1 \to -\infty) = \frac{\Phi_S}{\sqrt{2}},$$

(C.20a)

Boundary conditions for **easy** twins (with rotation vector perpendicular to the wall plane as shown in **Fig. C2b**) are $\frac{\partial \tilde{\Phi}_1}{\partial \tilde{x}_1}(\tilde{x}_1 = 0) = 0$, $\tilde{\Phi}_2(\tilde{x}_1 = 0) = 0$ at the TB and very far from for TB:

$$\tilde{\Phi}_1(\tilde{x}_1 \to +\infty) = \tilde{\Phi}_1(\tilde{x}_1 \to -\infty) = \pm \frac{\Phi_S}{\sqrt{2}}, \quad \tilde{\Phi}_2(\tilde{x}_1 \to +\infty) = \pm \frac{\Phi_S}{\sqrt{2}}, \quad \tilde{\Phi}_2(\tilde{x}_1 \to -\infty) = \mp \frac{\Phi_S}{\sqrt{2}} \quad \text{(C.20b)}$$